\documentclass{elsart}
\usepackage{setspace}
\usepackage{amssymb}
\usepackage{amsmath}
\usepackage{graphicx}
\usepackage{amsfonts}
\makeindex

\begin{document}

\begin{frontmatter}

\title{Frontiers in Nuclear Astrophysics}
\author[TAMUC,TAMU]{C.\ A.\ Bertulani}
\ead{carlos.bertulani@tamuc.edu} and
\author[NAO,UT] {T. Kajino}
\ead{kajino@nao.ac.jp}
\address[TAMUC]{Department of Physics and Astronomy, Texas A \& M University-Commerce, Commerce, Texas  75429, USA}
\address[TAMU]{Department of Physics and Astronomy, Texas A\&M University, College Station, Texas 77843, USA}
\address[NAO]{National Astronomical Observatory of Japan, 2-21-1 Osawa, Mitaka, Tokyo 181-8588, Japan}
\address[UT]{Department of Astronomy, Graduate School of Science, The University of Tokyo, 7-3-1 Hongo, Bunkyo-ku, Tokyo, 113-0033, Japan}
\begin{abstract}
The synthesis of nuclei in diverse cosmic scenarios is reviewed, with a summary  of the basic concepts involved before a discussion of the current status in each case is made. We review the physics of the early universe, the proton to neutron ratio influence in the observed helium abundance, reaction networks, the formation of elements up to beryllium, the inhomogeneous Big Bang model, and the Big Bang nucleosynthesis constraints on cosmological models. Attention is paid to element production in stars, together with the details of the pp chain, the  pp reaction, $^3$He formation and destruction, electron capture on $^7$Be, the importance of $^8$B formation and its relation to solar neutrinos, and neutrino oscillations.   Nucleosynthesis in massive stars is also reviewed, with focus on the CNO cycle and its hot companion cycle, the rp-process,  triple-$\alpha$ capture,   and red giants and AGB stars. The stellar burning of carbon, neon, oxygen, and silicon is presented in a separate section, as well as the slow and rapid nucleon capture processes and the importance of medium modifications due to electrons also for pycnonuclear reactions. The nucleosynthesis in cataclysmic events such as in novae, X-ray bursters and in core-collapse supernovae, the role of neutrinos, and the supernova radioactivity and light-curve is further discussed, as well as the structure of neutron stars and its equation of state. A brief review of the element composition found in cosmic rays is made in the end. 

\end{abstract}

\begin{keyword}
nuclear astrophysics, big bang, stellar nucleosynthesis, nuclear matter
\PACS 26.20.-f, 26.35.+c, 26.40.+r, 26.50.+x, 26.60.-c, 26.65.+t 
\end{keyword}

\end{frontmatter}

\tableofcontents

\pagebreak
 
  \maketitle

\pagenumbering{roman}

\setcounter{page}{5}

\pagenumbering{arabic} \setcounter{page}{1}

\section{Introduction}
Nuclear astrophysics is the science behind the synthesis of nuclei in temperature and pressure conditions existing within stars. It encompasses the synthesis and time evolution of nuclear abundances occurring through thermonuclear reactions from the Big Bang to the present date. Although the temperatures  during the Big Bang were very high, the densities were  low compared to typical stellar environments and only elements up to helium were produced appreciably. Hydrogen ($75\%$) and helium ($25\%$) remained as the source elements from which stars were formed long after the Big Bang Nucleosynthesis (BBN) stopped. Small traces of other elements such as Be and Li isotopes were also produced in the BBN. The oldest stars (population III stars) were formed from these and few other light primordial nuclei.  One theorizes that population III stars with large masses exhausted their nuclear fuel quickly and ejected heavier elements synthesized in their cores in very energetic supernovae explosions. Population II stars, found in the bulge and halo of galaxies, are the next generation of stars and very metal poor stars,  but can form elements such as carbon, oxygen, calcium and iron. These elements were dredged up from the core of stars and ejected by means of stellar winds to the interstellar medium.  A few of these stars, much heavier than our Sun, have also exploded, ejecting heavier elements. This stardust gave rise to a new generation of stars, population I stars, usually found in the disk of galaxies and with the highest metallicity content of all three populations. 

BBN is of impressive success, and has been used as a test of many new theories in physics, ranging from cosmology to particle and nuclear physics. In this review article we will discuss the state-of-the-art understanding of the physics of element production in the Universe and the challenges we face in nuclear astrophysics. In section \ref{bbnsect} we  discuss the physics of the Big Bang  both from the point of view on how particles, matter, and space evolve, and also on how nuclei are formed from the pre-existing nucleons during the BBN. In this section we discuss the physics of the early Universe, the proton to neutron ratio influence in the observed helium abundance, reaction networks, the formation of elements up to beryllium, the inhomogeneous Big Bang model, and the BBN constraints on cosmological models. In the following section \ref{stelarsynt} we discuss the physics of element production in stars, starting with the Sun which belongs to population I stars.  The Sun is burning hydrogen into helium slowly for about 4.5 Gyr and will continue to do it for 5 Gyr more. Nuclear reactions are responsible for the evolution and life-cycle of the stars, not only yielding internal heat, but also synthesizing  heavier elements than those manufactured in the early Universe. Stars with masses small compared with the solar mass did not evolve and may be still burning hydrogen, and some have already ended up as dwarf stars. In section \ref{stelarsynt} we present the general concepts in stellar nucleosynthesis, the pp chain, the  pp reaction, $^3$He formation and destruction, electron capture on $^7$Be, the importance of $^8$B formation and its relation to solar neutrinos, and neutrino oscillations.

The metallicity of a  star is defined as the proportion of chemical elements heavier than helium. E.g., the Sun has a metallicity of about 1.8 percent of its mass. Metallicity is often denoted by ``[Fe/H]",  the logarithm of the ratio of  iron abundance compared to the iron abundance in the Sun. Iron is used a reference because  it is easy to identify with spectral data in the visible spectrum. The gas from which Population I stars formed has been seeded with heavy elements produced in ancient giant stars. Such ancient stars still reside within star clusters.  Globular clusters are spherical star clusters with about 10 parsecs of radius and are  strongly concentrated in the galactic center. They usually contain  $\sim 10^{5-6}$, mostly old stars with the age of about 14 Gyr,  as old as the Universe. Population II stars are found in globular clusters, and are older, less bright and cooler than population I stars, with fewer heavy elements. Populations I and II stars differ in their age and their metallicity. Population III stars, with very low metallicity, have been formed within $10^6 - 10^7$ years after the Big Bang.  In section \ref{nuclsynt},  nucleosynthesis in massive stars is reviewed, with focus on the CNO cycle and its hot companion cycle, the rp-process,  triple-$\alpha$ capture, and red giants and AGB stars. The stellar burning of carbon, neon, oxygen, and silicon is discussed in section \ref{COSburn}. The slow and rapid nucleon capture processes are reviewed in section \ref{slowrapid}. The important medium modifications due to electrons and pycnonuclear reactions are discussed in section \ref{elctscr}.

One of the most astounding phenomena in the Universe are stellar runaway events, such as novae and supernovae stars. The nucleosynthesis in stellar scenarios such as in novae, X-ray bursters and in core-collapse supernovae, the role of neutrinos, and the supernovae radioactivity and light-curve is discussed in section \ref{cataclysm}. The structure of neutron stars, the search for a good equation of state to reproduce its characteristics such as mass versus radius relation and  a possible quark transition phase is introduced in section \ref{nstars}. Finally, in section \ref{cosmicr} we briefly discuss the element composition found in cosmic rays. 

It is impossible to make proper references to all seminal works in this vast field of science. Therefore, we apologize before hand to all authors whose valuable work were not referenced in this article.

\section{Big Bang Nucleosynthesis}\label{bbnsect}

The observed $^4$He abundance is one of the major predictions of the {\it Big Bang Nucleosynthesis} (BBN) model. $^4$He is  produced in nuclear fusion reactions in stars, but most of its presence in the Sun and in the Universe is due to BBN, and that is why it is often called by {\it primordial helium}.  In order to explain the primordial helium abundance,  the neutron to proton ratio at the moment when nucleosynthesis started should have been very close to ${\rm n/p} = 1/7$. This is because hydrogen and helium comprise almost 100\% of all elements in the Universe, and thus all neutrons existing in the primordial epoch must have in ended up inside helium. If ${\rm n/p} = 1/7$ then for every 14 protons there were 2 neutrons and they combined with 2 protons to form helium, with 12 protons remaining. In this  simple algebra, for every $^4$He nucleus, with 4 nucleons, there are a total of 16 nucleons in the Universe and the mass ratio of helium to all mass in the Universe is 25\%, whereas 75\% ($12/16 = 3/4$) is the mass ratio for hydrogen. This is indeed what is observed in the visible Universe, with a very small trace of heavy elements. 

\begin{figure}
\begin{center}
\includegraphics[
width=3.2in
]
{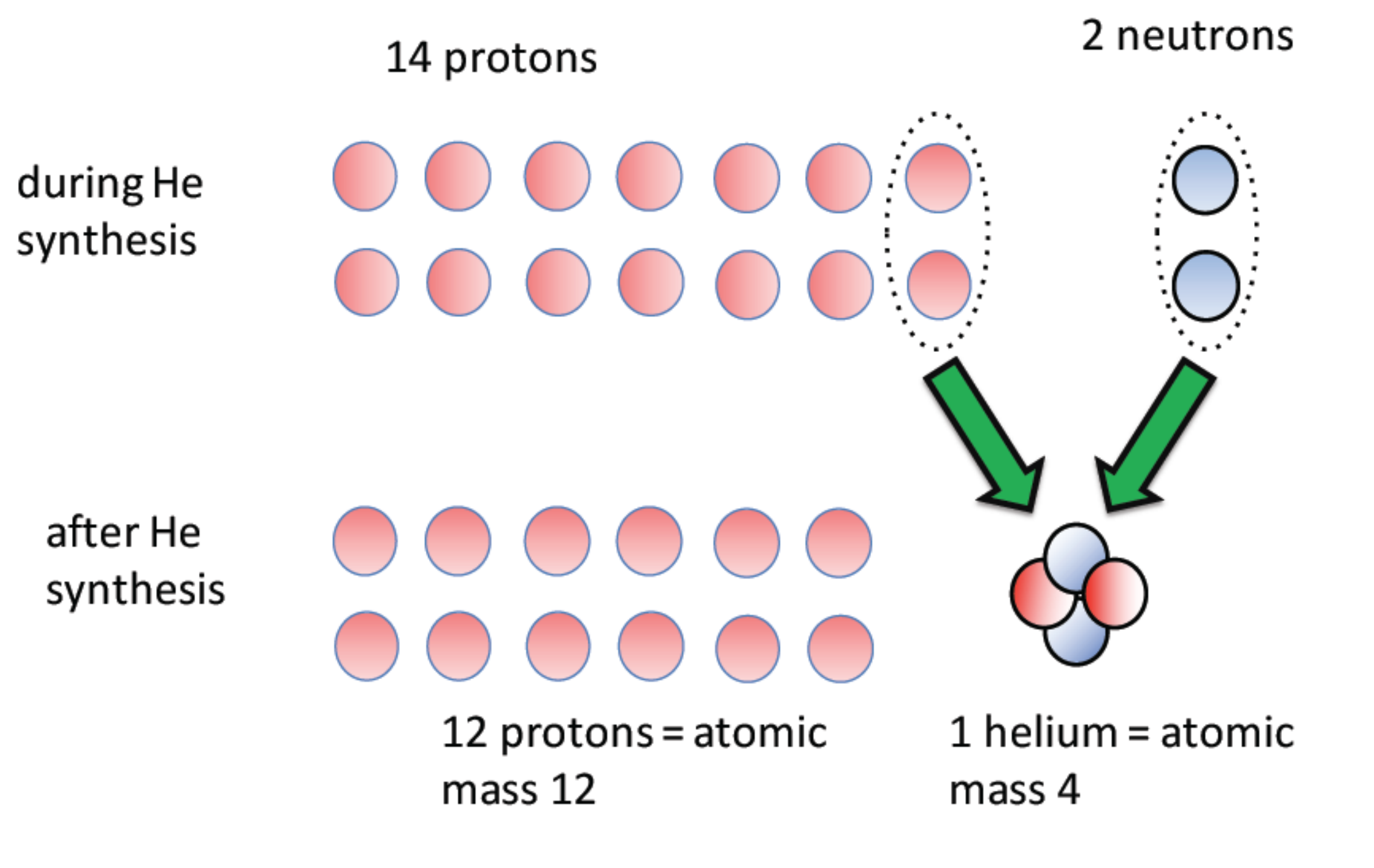} \ \ \ \ \ \ \ \ \ 
\includegraphics[
width=2.8in,
]
{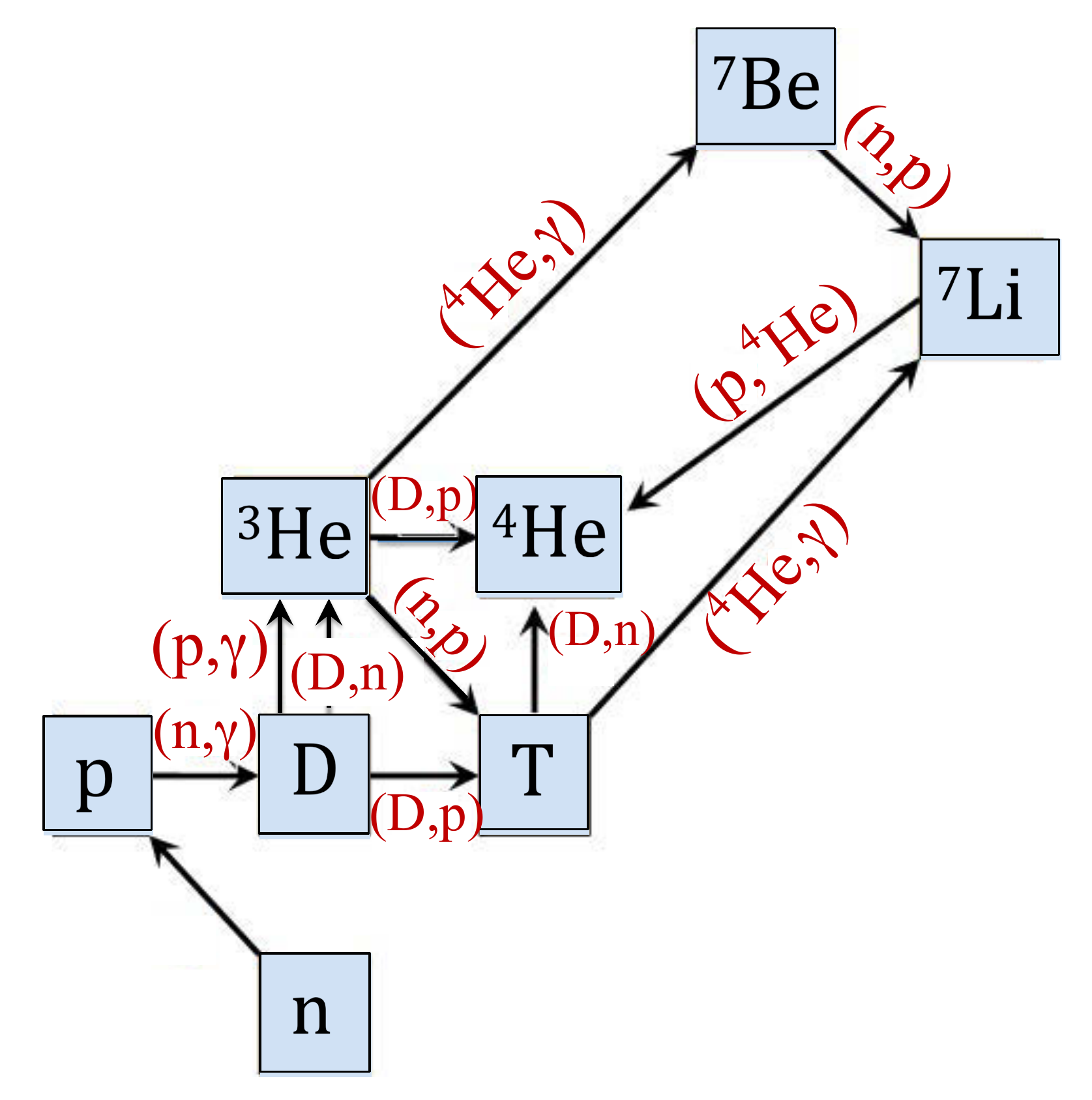}
\caption{\small {\it Left:} At  $t \lesssim$  1 min the Universe was very hot, $T \gtrsim 10^9$ K, and protons were more abundant that neutrons in the proportion  7:1. When H and He nuclei were formed, almost all neutrons ended up inside He nuclei yielding 1 helium for every 12 hydrogen nuclei. This explains why 75\% of the Universe's visible mass is in hydrogen nuclei and 25\% is in helium nuclei.  {\it Right:} The 12 most important nuclear reactions (arrows)  for Big Bang nucleosynthesis and the nuclei involved.}
\label{hehratio}
\end{center}
\end{figure}

In the following we will make a short survey of the physics of the early Universe and how one can explain the ratio  ${\rm n/p} = 1/7$ at the beginning of nucleosynthesis era after the Big Bang. This is also shown schematically in Figure \ref{hehratio}. Although they are very small, the abundances of other light elements can be reasonably well explained with the  BBN model. One exception maybe the lithium abundance.

\subsection{Physics of the Early Universe}

The physics of the early Universe can be formulated in terms of the laws of thermodynamics and statistics and with the assumption that particles are relativistic (see, e.g., the textbooks on cosmology in Refs. \cite{Wei93,KT94,Pea99,LL00,Dod03,Wei08}). It also relies on the {\it Cosmological Principle} of a homogeneous and isotropic Universe. The number density, $n$, energy density, $\rho$, and entropy density, $s$, of particles in the early Universe are found to be related to the temperature as
\begin{equation}
n = {\xi({3}) \over \pi^2} g'(T) T^3, \ \ \ \ \rho=3p={\pi^2\over 30} g(T) T^4, \ \ \ \ s={p+\rho \over T} = {2\pi^2 \over 45} g(T) T^3, \label{peu:n}
\end{equation}
where $\xi({s})=\sum_{n=1}^\infty  1/n^s$ is the Riemman Zeta function\footnote{$\xi({3})=1.202\dots$ is the Ap\'ery's constant.},  $p$ is the pressure and $T$ the temperature in units of the Boltzmann constant $k_B$\footnote{Here we use units $\hbar = c = k_B = 1$.}.  The temperature dependence in these relations is easily understood. At  a temperature $T$, the particle density  $n$ increases with $T^3$  as
their momentum states ($k_x,\ k_y,\ k_z$) will roughly fill up  to energy ($T,\ T,\ T$) and $n \propto k^3$. For the energy density $\rho$, an extra power of $T$  appears because each particle has energy  $T$. Therefore, the energy density in the early Universe is governed by the $T^4$ law and is due to relativistic particles, namely, photons, electrons, positrons, and the three neutrinos.

In Eq. \eqref{peu:n}, the factors $g$ and $g'$ account for the number of degrees of freedom for bosons, $b$, and fermions, $f$, with the appropriate spin statistics weights,
\begin{equation}
g'(T)=g_b(T) + {3\over 4} g_f(T), \ \ \ \ {\rm and} \ \ \ \ g(T)=g_b(T) +{7\over 8} g_f (T) .
\end{equation}
The factors $3/4$ and $7/8$ arise due to the different  fermionic and bosonic energy distributions in the integrals used to obtain $n$, $\rho$ and $s$ \cite{KT94}. At the early stages, the Universe is also assumed to be free from dissipative processes such as phase-transitions which can change the overall entropy. Then the following equations for the ``{\it comoving entropy}" holds \cite{Wei08},
\begin{equation}
{d(sa^3) \over dt}=0, \ \ \ \ {\rm so \ that} \ \ \ \ s \propto {1\over a^3}, \ \ \ \ \ {\rm and} \ \ \ \ \ T \propto {1\over a} ,
\end{equation}
where $a$ is a scale factor such that the distances in the Universe are measured as $r=a\zeta$, where $r$ is the distance to a point and $\zeta$ is a dimensionless measure of distance.   Thus, one predicts that in the early Universe the temperature decreased with the inverse of its ``size". 

The dynamics of the early Universe is governed by the Friedmann equation, which can be easily obtained from Newton's law of an expanding gas interacting gravitationally \cite{KT94,Pea99,LL00,Dod03,Wei08}. It reads
\begin{equation}
\left( {\dot a \over a } \right)^2 = {8\pi G \rho \over 3}. \label{peu:friedmann}
\end{equation}
Thus the ratio of expansion of the early Universe (known as the {\it Hubble constant}), $H=\dot a /a$,  is proportional to the square root of its energy density, $H\propto \sqrt{\rho}$.  A more general form of the Friedmann equation is given by
$
\Omega_m +\Omega_k +\Omega_\Lambda =1,
$
where $\Omega_m = \rho/\rho_{crit}$ is the {\it matter density parameter} of the Universe, $\rho_{crit} = 3H^2/8\pi G$ is the {\it critical density}, $\Omega_k$ is the {\it curvature density parameter} (for a flat Universe, $\Omega_k =0$), and $\Omega_\Lambda = \Lambda /3H^2$ is the {\it vacuum energy parameter}, or {\it dark energy}, density. Here,  $\Lambda$ is Einstein's cosmological constant.  Table \ref{nuc3:tab1} lists the ratio of the scale factor at the different epochs with that at the present time (denoted by index ``0")\footnote{To convert Kelvins to eV: 1 eV = 11,600 K, or 1 K = $8.6 \times 10^{-5}$ eV.}.

\begin{table}
\begin{center}
\begin{tabular}
[c]{|l|l|l|l|}\hline
Energy & T(K) & $a/a_0$ & $t$(s) \\\hline\hline
$\sim 10$ MeV & $10^{11}$ & $1.9 \times 10^{-11}$ & .01 \\\hline
$\sim 1$ MeV & $10^{10}$ & $1.9 \times 10^{-10}$ & 1.1 \\\hline
$\sim 100$ keV & $10^9$ & $2.6 \times 10^{-9}$ & 180 \\\hline
$\sim 10$ keV & $10^8$ & $2.7 \times 10^{-8}$ & 19,000 \\\hline
\end{tabular}
\caption{\small Properties of the early Universe according to Friedmann cosmology: particle energy, temperature, radius scale  and elapsed time. The subindex 0 means present time \cite{KT94,Pea99,LL00,Dod03,Wei08}.\label{nuc3:tab1}}
\end{center}
\end{table}

Equation  \eqref{peu:friedmann} can be readily integrated, using Eq. \eqref{peu:n}, yielding the time-temperature relationship \cite{OP11}
\begin{equation}
t(s) \simeq {2.4 \over T_{10}^2}, \label{peu:t}
\end{equation}
where $T_{X}$ is the temperature in units of $10^{X}$ K.
But  in order to get the time evolution right one needs to account for temperatures when particle energies changed enough so that phase transitions occurred. For example, at about 150 to 400 MeV quarks and gluons combined to form nucleons, thus changing the value of the available degrees of freedom $g$ of relativistic particles. This will change the time-temperature relationship, Eq. \eqref{peu:t}, accordingly. The observation of a perfect Planck spectrum for the {\it Cosmic Microwave Background} (CMB) is a testimony of our hot and dense past and directly demonstrates that the expansion of the Universe was adiabatic (with negligible energy release) back to at least  $t \sim 1$ d. With the proper inclusion of nucleosynthesis processes, we can even go back to obtain the proper temperature when $t \sim 1$ s \cite{Wei08}.

The several evolution steps of the Universe since the Big Bang to the present epoch are summarized in the following list (for more details, see Refs. \cite{KT94,Pea99,LL00,Dod03,Wei08}):

\begin{enumerate}

\item {\it Planck era -- from zero to $10^{-43}$ s (Planck time)}:    During this period,  all four known fundamental forces were unified, or equally strong. The distance light travels within the Planck time is $1.6 \times 10^{-35}$ m and this value should be the size of the Universe. The temperature was of the order of $10^{32}$ K, corresponding to a particle energy of $10^{19}$ GeV.

\item {\it Grand Unified Theory (GUT) era -- up to $10^{-38}$ s}:  During this time the force of gravity ``freezes out" and becomes distinct from the other forces. The temperatures ran around $10^{27}$ K, or $10^{14}$ GeV.

\item {\it Electroweak era -- $10^{-38}$ to $10^{-10}$ s}:  The strong force ``freezes out" and becomes distinct from the other forces. In this era, Higgs bosons collide to create W and Z bosons that carry the electroweak force and quarks. The temperatures are of the order of $10^{15}$ K, or 100 GeV.

\item {\it Particle era -- $10^{-10}$ to $10^{-3}$ s}:     During this era, the Universe expanded and cooled down so that at its end protons and neutrons begin to form out of free quarks. Temperatures were of the order of $10^{13}$ K, or 1 GeV. At the end of this era, W and Z bosons were no more created and they started decaying. The electroweak force then separated from the electromagnetic force and became the short-range weak force as we know at present.   

\item {\it Nucleosynthesis era -- $10^{-3}$ s to 3 m}:   Even at 10 s, the temperature was high enough so that the energy was passed from electrons and positrons to photons. But as the temperature cooled down in this era, protons and neutrons started fusing into the deuteron, and nuclear fusion started creating helium, and tiny amounts of heavier elements. This is the era of the Big Bang that we are interested in this review.

\item {\it Era of nuclei -- 3 minutes to 400,000 yrs}:   During this era the temperature was still too hot ($\gtrsim 3,000$ K) so that electrons were not  bound to nuclei. 

\item {\it Era of atoms -- after 400,000 yrs } --   Electrons and nuclei combined into neutral atoms. The first stars were then born. We are also interested in this epoch of the Universe and what comes later, as they are related to stellar nucleosynthesis and stellar evolution. 

\item {\it Era of galaxies -- after 200,000 million yrs} --   During this time, galaxies begin to form, but at a much higher rate than at present.

\end{enumerate}

When electrons recombined with protons and helium nuclei to form neutral atoms,  the photons and baryons decoupled and the Universe became transparent to radiation at $\sim 400,000$ yrs. Oscillations in the photon and baryon densities at this time would lead to tiny variations of 1 part in $10^5$ in the temperature observed in the  microwave background.  The Wilkinson Microwave Anisotropy Probe (WMAP) has observed these anisotropies and decomposed them in terms of spherical harmonics, with  the expansion coefficients describing the magnitude of the anisotropy at a certain angle and closely correlated with the cosmological parameters.  These studies lead to the extraction of the values $h = 0.705 \pm 0.013$ for the dimensionless Hubble constant, $h=H/(100$ km.s$^{-1}$), and $\Omega_m h^2 = 0.0227\pm 0.0006$.  One also gets from these observations,  $t_0 = 13.72 \pm 0.12$ Gyr for the Universe age, $\rho_{crit} = (9.3\pm 0.2) \times 10^{-30}$ g.cm$^{-3}$ for the critical density, and $\Omega_b = 0.044\pm 0.003$ for the baryon density parameter \cite{Kom11,Per07}. The Planck mission, with a higher resolution and sensitivity than WMAP,  obtained  $h = 0.674\pm 0.014$, $\Omega_m h^2 = 0.02207 \pm 0.00033$, and $t_0 = 13.813 \pm 0.058$ Gyr \cite{Ade14}

\subsection{Proton to neutron ratio and helium abundance}

We will assume that the laws of physics were the same now and during the Big Bang so that we can predict its behavior.  During the early stages ($t\sim 0.01$ s) only $\gamma$, $e^\pm$ and three neutrino families were present in the Universe. The only existing baryons were neutrons, n, and protons, p. To agree with present observations, calculations based on known physics need to assume a value for  the {\it baryon-to-photon ratio} was $\eta \equiv n_B/n_\gamma $. We do not know why, but as the Universe cooled,  a net baryon number remained.  Right before nucleosynthesis there were protons and neutrons, not many antineutrons and antiprotons.   Sakharov \cite{Sak67} proposed that in order to generate more baryons than anti-baryons, a set of physics conditions (now dubbed as {\it Sakharov conditions}) had to apply. These are: (a) there must be a baryon number violation, as well as (b) a violation of charge (C) and charge-parity (CP) symmerty, and (c) the interactions between all particles before the baryon-anti-baryon asymmetry set in must be out of equilibrium. The combined charge-parity-time (CPT) symmetry has to be respected because antiparticles have the same mass as their corresponding particles, with opposite charges. The Sakharov condition (a) is necessary if baryons are more numerous than anti-baryons, whereas the condition (b) is also needed, otherwise the overproduction of baryons would also be  annulled by  anti-baryons being produced by the charge-symmetric equivalent reactions/decays.  Additionally, CP-symmetry needs to be violated to avoid that  left-handed baryons and right-handed anti-baryons (and vice-versa) be produced in equal amounts.  Finally,  CPT symmetry under equilibrium conditions would lead to a balance in loss and gain of baryons and anti-baryons, thus demanding out of equilibrium conditions. One has observed experimentally that CP violation exists \cite{Chr64}. But there is no evidence  that baryon number is not conserved, although the Standard Model (SM) of particle physics (and beyond it) does not require baryon number conservation \cite{Wein05}. The reasons why the Universe is made predominantly of matter, with extremely small traces of anti-matter, constitutes some of the most fascinating problems in particle cosmology. For a review on baryogenesis and the possible mechanisms for baryon production, see, e.g., Ref. \cite{Cli06}.

The baryonic density of the Universe has been inferred from the observed anisotropy of the CMB radiation which reflects the number of baryons per photon, $\eta$. The number density of photons remains constant after the epoch of electron-positron annihilation, which happened during $4-200$ s. The $\eta$ ratio remains constant during the expansion of the Universe because baryon number is also conserved. Big Bang models are consistent with the observations from WMAP yielding the value $\eta = (6.16 \pm 0.15) \times 10^{-10}$ \cite{Kom11}. The CMB observations yield $T = 2.725 \pm 0.001$ K \cite{Mat99}, and is given by $n_\gamma = 410.5$ cm$^{-3}$. At $t\sim 0.01$ s and $T \sim 10^{11}$ K  the thermal energies were  $kT\sim 10$ MeV $ \gg 2m_ec^2$. Therefore, electrons, positrons and neutrinos were in chemical equilibrium by means of charged- and neutral-current interactions, that is,
\begin{eqnarray}
&p + e^- \longleftrightarrow n + \nu_e, \ \ \ \ \ 
n + e^+ \longleftrightarrow p + \bar{\nu}_e \nonumber  \\
&p \longleftrightarrow n + e^+ + \nu_e, \ \ \ \ \
n \longleftrightarrow  p + e^- + \bar{\nu}_e . \label{wdr}
\end{eqnarray}
According to the Boltzmann-Gibbs statistics,  the neutron-to-proton ratio at this time should be
\begin{equation} {\mathrm{n} \over \mathrm{p}} = {e^{-m_n/kT} \over e^{-m_p/kT}} = e^{-\Delta m/kT} ,\end{equation}
where  $\Delta m = 1.294$ MeV.  At $T = 10^{11}$ K, $kT=8.62$ MeV and  $ {\mathrm{n} / \mathrm{p}} = 0.86 $, i.e., there were more  protons than neutrons.  Fermi's theory of the weak interaction predicts that the weak rates in Eqs. \eqref{wdr}  drop rapidly with $T^5$.  At some temperature the weak rates were so slow that they could not stay in pace with other changes in the Universe. A {\it decoupling temperature} existed when the neutron lifetime and the Hubble expansion rate, $H = \dot a /a$,  were similar. The neutrinos start to behave as free particles, and for  $T<10^{10}$ K they are ineffective,  i.e., matter became transparent to them. This temperature is equivalent to thermal energies of $\sim 1$ MeV, comparable to the energy for creation of electron-positron pairs. Electrons and positrons start to annihilate, and for some unclear reason a small excess of electrons remained. Neutrons and protons could be captured at this temperature, but the thermal energies were still too high and would dissociate the nuclei. 

\begin{figure}[t]
\begin{center}
\includegraphics[
width=3.4in
]
{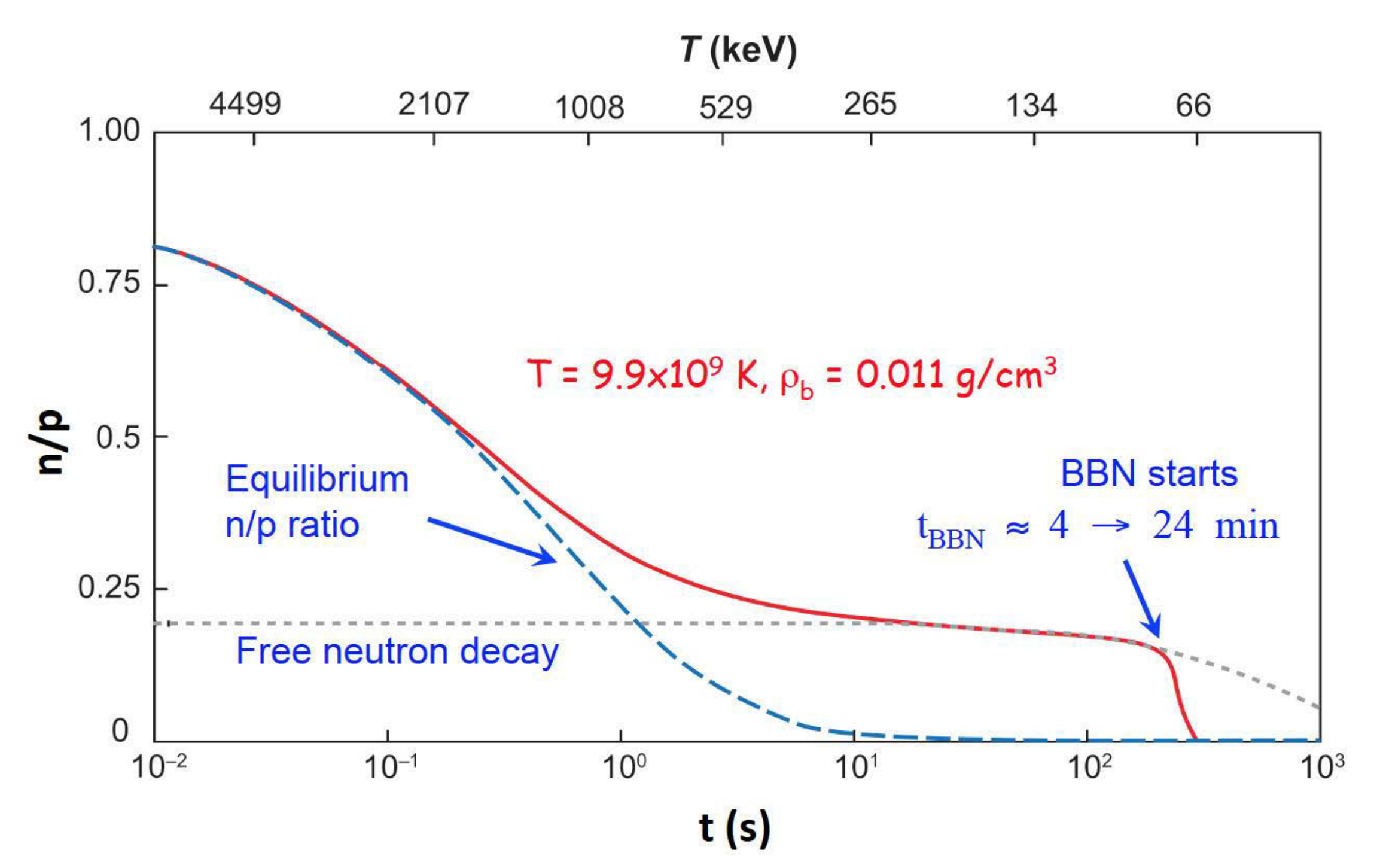}
\includegraphics[
width=3.0in,
]%
{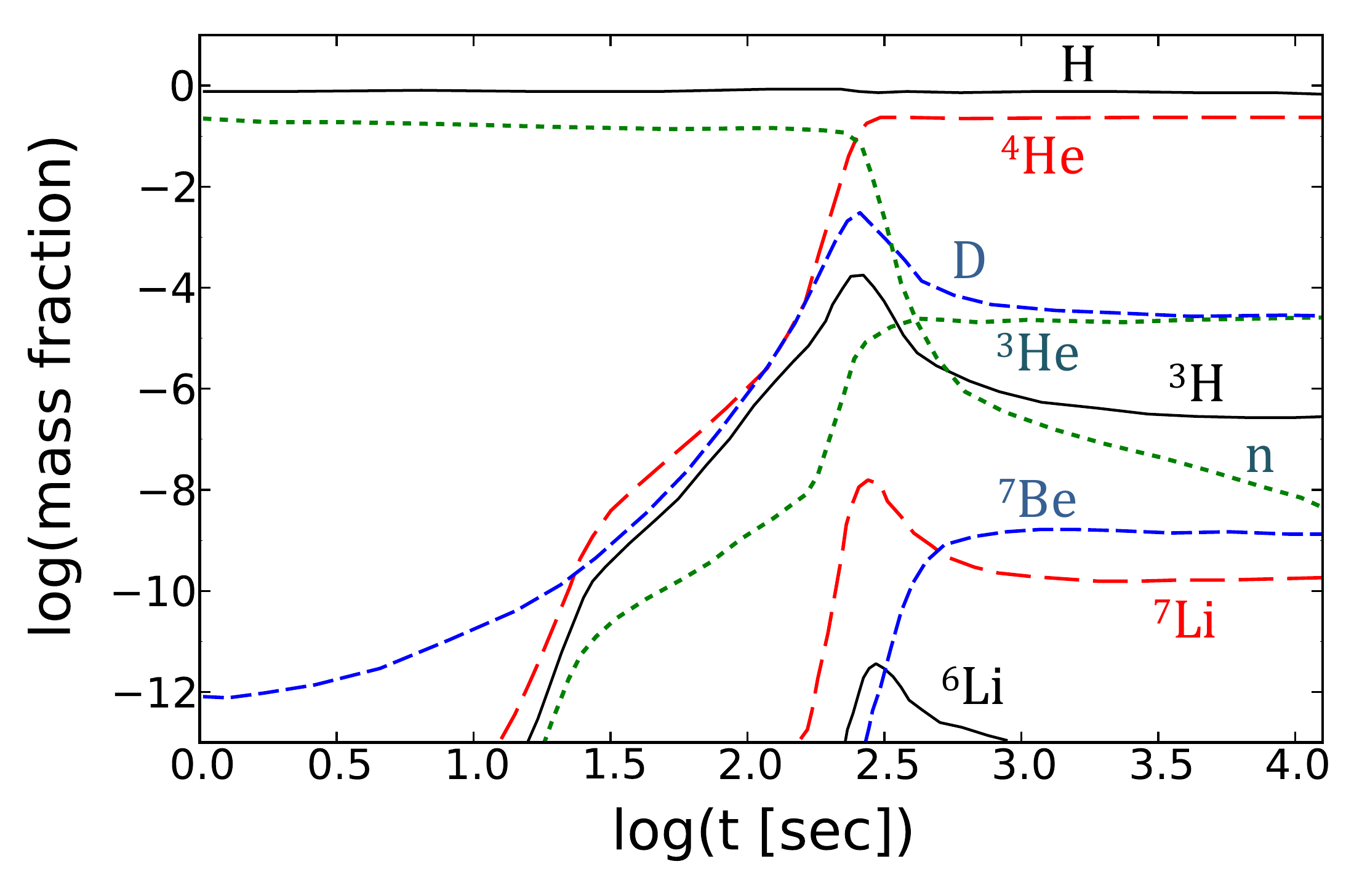}
\caption{\small {\it Left:} The neutron-to-proton ratio is shown as a function of time (lower scale) and temperature (upper scale). The  n/p  ratio at equilibrium, given by $\exp(-\Delta m/kT)$, is displayed by the dashed curve. The time-decay of free neutrons, $\exp(-t/\tau_n)$, is given by the dotted curve. The solid curve is the total n/p time dependence. The sudden fall-off at a few hundred seconds is due to the onset of BBN (Adapted from \cite{Stei07}). {\it Right:} Mass and number fractions of nuclei during the BBN as a function of temperature and time. $^4$He is presented as a {\it mass fraction} and the other elements are given as {\it number fractions}.}
\label{fig:bbndecay}%
\end{center}
\end{figure}

Because neutrinos were ineffective (i.e., decoupled),  $e^\pm$ annihilation has heated up the photon background relative  to them. A numerical calculation using the available degrees of freedom finds that the ratio of the two temperatures is \cite{Wei08}  $ {T_\nu / T_\gamma} = \left({4 /11}\right)^{1/3} $. The weak decay reaction n $\leftrightarrow$ p breaks out of equilibrium, and the  n/p ratio becomes $ {n / p} = \exp\left[-{\Delta m / kT}\right] = 0.25$, which means a large drop of the n/p ratio. This is still not 1/7, the value required to explain the observed helium abundance (see Figure \ref{hehratio}).  But   the $\nu$-induced reactions still favored the proton-rich side.  We are now at 1 second (1 MeV) after the Big Bang and in the following 10 s the n/p ratio will decrease to about 0.17 $\sim$ 1/6 \cite{Wei08}. In a timescale of about 10 minutes the neutron $\beta$-decay will drive the n/p ratio to its magic 1/7 value. This sequence of physics processes is summarized in Figure \ref{fig:bbndecay}, left, based on real calculations \cite{Stei07}. 

We also conclude that the first nuclei were formed within less than 10 min, as  $\beta$ decay cannot continue and keep the n/p ratio equal to the  1/7 steady value. The nucleon gas in the early Universe was dilute and nuclei are therefore created via two-body reactions.   The lightest two-nucleon system, the deuteron,  has only one bound-state, with a small binding energy of  2.24 MeV.   Neutrons and protons  form the deuteron by means of the radiative capture reaction 
\begin{equation}  n + p \leftrightarrow d + \gamma .\end{equation}
This reaction is very fast because it is electromagnetic and  it is therefore in equilibrium during the early Universe. Hence, the ratio of (n+p) pairs to the number of deuterons is obtained using the Boltzmann-Gibbs statistics,
\begin{equation} {N_{\rm n+p \rightarrow d +\gamma} \over N_{\rm d+\gamma}} =
{n_d n_\gamma  \exp\left[-{\Delta m/ kT} \right] \over n_p n_n }. \end{equation}
We can assume that at the time the deuteron was formed $n_p/n_\gamma \sim \eta$ and  $n_n \sim n_d$, i.e., half of the neutrons are inside deuterium nuclei.  Thus,   $ \exp\left[{-2.24 \ \mathrm{MeV}/ kT}\right] \sim \eta \sim 10^{-9}$,  and we get the temperature when deuterium is formed as $T_{\mathrm{deuterium}} \simeq 1.25 \times 10^9$ K  or   $t_{\mathrm{deuterium}} \sim 100$ s. Therefore, nucleosynthesis begins at about 100 seconds after the Big Bang.  A numerical calculation of the ratio n/p from the time of the weak freeze-out, or decoupling ($\sim$ 1 s), to 100 s yields
\begin{equation} {\rm n \over p} \sim 0.15 \sim {1\over 7}. \end{equation}
Hence, the n/p ratio is a ``fingerprint" of the thermal history of the primordial Universe.  

At the start of  nucleosynthesis, $T \sim 100$ keV, whereas we would have expected it to be about 2 MeV, the  deuterium binding energy. This is due to the very small value of $\eta$.  Based on  cross section estimates, the reaction rate for deuteron formation at $T\sim 100$ keV is
\begin{equation} 
\sigma v ({\rm p + n \rightarrow d + \gamma}) \simeq 5 \times 10^{-20}  \
{\rm cm}^3/{\rm s}. \label{ratedeut}
\end{equation}
This implies a density $\rho \sim {1 / \sigma v } \sim 10^{17} \ {\rm cm}^{-3}$ for $t\sim 100$ s. The density in baryons today is obtained from the observations of the density of visible matter, ${\rho_0} \sim 10^{-7}$ cm$^{-3}$. As the baryon density $n$ is proportional to  $R^{-3} \sim T^3$,  the temperature today should be $T_0 = ({\rho}_0/\rho)^{1/3} T_{\rm BBN} \sim 10$ K, which is  in fact close to the observed 2.725 K  temperature of the  CMB \cite{Not11}.   Numerical calculations show that  3.5 minutes after the Big Bang the temperature decreases to $3 \times10^8$ K and the density is $10^{-4}$ kg m$^{-3}$ \cite{Stei07,Wag67,Sar96,ST98,Ioc09,Jed09,Pos10,Fie11}. At this time the Universe consists of 70\% photons, 30\% neutrinos, and only $10^{-7}$\% of other particles. From those about 70 to 80\% is hydrogen and 20 to 30\% is helium, and the associated number of electrons. 

The ${\rm p} + {\rm n} \rightarrow {\rm d} + \gamma$ cross section in Eq. \eqref{ratedeut} at the BBN energies, $0.02 < E < 0.2$ MeV,  is not well known experimentally, but it can be deduced by detailed balance from its inverse reaction $\gamma +  {\rm d} \rightarrow {\rm p} + {\rm n} $, as realized in the 1940's \cite{BJ49,Bet49,BL50,Sal51,Sch50,Mar50}. The deuteron photodisintegration from the threshold to 20 MeV photon energy is largely correlated with the deuterium binding energy and with the triplet effective-range of the neutron-proton interaction and  the dependence of the cross section on the $\gamma$-energy is not influenced appreciably by the details of the interaction \cite{BJ49,Bet49,BL50,Sal51,Sch50,Mar50}. Since then, there has been many attempts to decrease the  uncertainty between theory and experiment based on the improvements in our knowledge of the nucleon-nucleon interaction \cite{MHE87}.  Perhaps worth mentioning are the efforts carried out by using effective field theoretical (EFT) methods in achieve a high accuracy, of the order 3\% for the region of interest for BBN \cite{Sav99,CRS99,PMR00,Rup00}. These results, as well as others based on the potential model \cite{SWA91}, or meson exchange currents \cite{Nag97}, are in excellent agreement with experimental measurements \cite{Tor03,Ahm08}. A much higher precision than 3\% might not be necessary, as the np fusion plays a small role in the uncertainty of the baryon density in the early universe \cite{BNT99}.

\subsection{Reaction networks}

The {\it reaction network} with the main nuclei involved in BBN is illustrated in Figure \ref{hehratio}, right. Each nucleus may be formed  through  reactions indicated by the arrows.   The number of reactions per unit volume and time is given by $r=\sigma vn_{j}n_{k}$, or, more generally \cite{Clay83,Ill07},%
\begin{equation}
{ r}_{j,k}{  =}\int{  \sigma |{\bf v}}_{j}{  -{\bf v}}_{k}%
{  |d}^{3}{  n}_{j}{  d}^{3}{  n}_{k}{  ,}%
\label{astrophys2}%
\end{equation}
where the target number density is given by $n_{j}$, the projectile number density is given by $n_{k}$, and $v=|{\bf v}_j-{\bf v}_k|$ is the relative velocity between target and projectile nuclei.  It is reasonable to assume that the velocity, or kinetic energy, distribution of nuclei  in an astrophysical plasma obey the {\it Maxwell-Boltzmann distribution} (MB). The MB distribution is based on the following assumptions: (a) the collision times are much smaller than the average time between collisions, (b) the interactions are local, (c) the particle velocities at a point are not correlated, and (d) the energy is conserved in the collisions. Alternative theories have been used to describe systems with long-range interactions and memory effects, such as those proposed in  proposed in Refs. \cite{Ren60,Tsa88}, also known as {\it non-extensive statistics}. The deviation from the Maxwellian distribution and its implications  for nuclear burning in stars have  been studied in Refs. \cite{LL05,HK08,Deg98}. For the BBN it has been shown that no appreciable changes in the abundance of light elements are obtained unless large changes are made in the MB distribution, through the modification of the non-extensive parameter $q$ \cite{BFH13}. Using the MB distribution, the reaction rates in plasmas in thermodynamical equilibrium are given by
\begin{equation}
d^{3}n_{j}=n_{j}\left({\frac{{m_{j}}}{{2\pi kT}}}\right)^{3/2}\exp\left(-{\frac
{{m_{j}^{2}v}_{j}}{{2kT}}}\right)d_{j}^{3}{v}_{j}\label{astrophys3}.
\end{equation}
The reaction rate $\left<\sigma v\right>$\ is the average of $\sigma v$\ over the velocity distribution in \eqref{astrophys3},
\begin{equation}
r_{j,k}   =\left<\sigma\mathrm{v}\right>_{j,k}{n_{j}n_{k}\over 1 +\delta_{jk}}, \ \  {\rm where} \ \ 
 \left<\sigma\mathrm{v}\right>_{j,k}=\left({\frac{8}{{m_{jk}\pi}}}\right)^{1/2}
(kT)^{-3/2}\int_{0}^{\infty}E\sigma(E)\exp\left(-{E\over kT}\right)dE.\label{astrophys5}%
\end{equation}
In this equation, $m_{jk}$\ is the reduced mass of the two nuclei and the factor $\delta_{jk}$ is introduced to account for double-counting when nuclei $j$ and $k$ are of the same species. For charged particles the cross section $\sigma(E)$ strongly (exponentially) decreases as the relative energy $E$ of the particles  decrease due to the increasing difficulty the particles have to overcome the Coulomb repulsion, namely, to tunnel through the {\it Coulomb barrier}. But as the energy $E$ increases, the Boltzmann factor also decreases exponentially. Therefore the integrand in Eq. \eqref{astrophys5} is peaked in an energy window often called by {\it Gamow window}. The peak, or centroid, energy in this window is known as the {\it Gamow energy}. For light charged particles in most astrophysical scenarios, this energy is of the order of a few tens, or at most a few hundreds, of keV.

If one of the particles described in Eq. \eqref{astrophys2}  is a photon, the relative velocity is always $c$ and the equation is not useful. But one can use time-reversal symmetry  to relate the photodisintegration cross section to the photon capture, or radiative capture cross section, leading to the photodisintegration rate given by $r_{j}=\lambda_{j,\gamma}n_{j}$, where $\lambda_{j,\gamma}$\ is averaged over a Planck distribution of photons at temperature $T$,
\begin{equation}
\lambda_{j,\gamma}(T)=\left({\frac{{Z_{l}Z_{m}}}{Z_{j}}}\right)\left({\frac{{m_{l}m_{m}}%
}{m_{j}}}\right)^{3/2}\left({\frac{{m_{u}kT}}{{2\pi\hbar^{2}}}}\right)^{3/2}
 \left<\sigma\mathrm{v}\right>_{l,m}
\mathrm{exp}%
\left(-{Q_{lm}\over kT}\right),\label{astrophys8}
\end{equation}
where $m_u=m_{^{12}C}/12$ is the {\it atomic mass unit}, $Q_{lm}$ is the {\it reaction $Q$-value}, i.e., the energy gained in the reaction, $ Q_{lm}=(m_l+m_m-m_j)c^2.$ The average $ \left<\sigma\mathrm{v}\right>_{l,m}$ is now associated to the inverse reaction rate for radiative capture,  $Z(T)=\sum_{i}(2J_{i}+1)\exp(-E_{i}/kT)$ are statistical weights, and $m_i$ are the mass numbers  of the participating nuclei. In the case of electron capture,   the velocity of the nucleus $j$\ is negligible comparison to the electron velocity ($|{\bf v}_{j}-{\bf v}_{e}|\approx v_{e}$). Depending on astrophysical conditions, the electron capture cross section has to be integrated over  either a Boltzmann, a partially degenerate, or a fully degenerate  Fermi distribution of electron energies. The electron  number density is given in terms of the {\it electron fraction} $Y_e$, i.e., $n_{e}=Y_{e}\rho N_{A}$, where $\rho$ is the matter density and $N_A$ is Avogadro's number\footnote{In astrophysics, one often uses cgs units, as is the case here.}. For a completely ionized and neutral plasma, the electron abundance is the same as the total proton abundance in nuclei, i.e.,  $Y_{e}=\sum_{i}Z_{i}Y_{i}$. The electron capture rate as a function of $T$ is then given by $ {  r}_{j}{  =\lambda}_{j,e}{  (T,\rho Y}_{e}{  )n}_{j}$.

The time evolution of number density of particles, $n_j$, is governed by  the number of reactions creating or destroying $j$ per volume and time. Defining  the number of nuclear species $i$  created or destroyed  in the reaction $j+k+l+\cdots \leftrightarrow i$ by $N^i_{j,k,l,...}$, the following balance equation holds,
\begin{equation}
\left(\frac{{\partial n_{i}}}{{\partial t}}\right)_{\rho=const}%
{  =}\sum_{j}{  N}_{j}^{i}{  r}_{j}{  +}\sum
_{j,k}{  N}_{j,k}^{i}{  r}_{j,k}{  +}\sum_{j,k,l}%
{  N}_{j,k,l}^{i}{  r}_{j,k,l}{  .}\label{astrophys11}%
\end{equation}
The first term is due to the destruction or decay of the nuclear species $i\rightarrow j$ due to either photodisintegration, electron and positron capture or neutrino induced reactions ($r_{j}=\lambda_{j}n_{j}$). The second term is due to two-particle reactions ($r_{j,k}=\left< \sigma v \right>_{j,k} n_{j}n_{k}$), and the last term is due to three-particle reactions ($r_{j,k,l}=\left< \sigma v\right>_{j,k,l} n_{j}n_{k}n_{l}$) such as the triple-alpha process $ \alpha+\alpha+\alpha
\rightarrow ^{12}{\rm C}+\gamma$.  The $N^{i}s$\ can be negative or positive numbers.  Instead of number densities, $\dot{n}_{i}$, one often works with {\it nuclear abundances}, $Y_{i}=n_{i}/(\rho N_{A})$, to exclude changes due to volume expansion or contraction of the plasma. For a nucleus with mass number $A_{i}$, $A_{i}Y_{i}$\ denotes its mass fraction, so that  $\sum A_{i}Y_{i}=1$. In terms of nuclear abundances $Y_{i}$, Eq. \eqref{astrophys11} reads
\begin{eqnarray}
\dot{Y}_{i}=\sum_{j}N_{j}^{i}\lambda_{j}Y_{j}+\sum_{j,k}N_{j,k}^{i}\rho
N_{A}<j,k>Y_{j}Y_{k}
+\sum_{j,k,l}N_{j,k,l}^{i}\rho^{2}N_{A}^{2}<j,k,l>Y_{j}%
Y_{k}Y_{l}.\label{astrophys12}%
\end{eqnarray}

The important ingredients for the calculation of nucleosynthesis, and energy generation by nuclear reactions are (a) nuclear decay half-lives, (b) electron and positron capture rates, (c) photodisintegration rates, (d) neutrino induced reaction rates, and (e) strong interaction cross sections. Additional information such as the initial electron, photon, and nuclide number densities are necessary to obtain the relative abundances of elements by means of the above equations. Figure \ref{fig:bbndecay}, right, displays an example of predicted mass fractions as a function of temperature and time during the Big Bang. These results have been obtained by a numerical solution of the coupled equations involving the evolution of the Universe temperature and density according to  Friedmann equation, energy and chemical composition changes due to Eqs. (\ref{astrophys11},\ref{astrophys12}) and the available theoretical and experimental data on nuclear cross sections and decay half-lives.  We have used a modified version of the  BBN code developed by Wagoner, Fowler, and Hoyle \cite{Wag57} and Kawano \cite{Kaw92,Wang11} and assumed the presently accepted values for the baryon/photon ratio, $\eta$, and  for the neutron decay lifetime \cite{MKS05}. At the end of the BBN almost all neutrons end up in $^4$He, and the element abundance  depends  on the neutron lifetime $\tau_n$ and on the number of neutrino families. The neutron lifetime $\tau_n$ has an influence on the weak reaction rates due to the dependence of $\tau_n$ on the weak coupling constant. For a shorter  $\tau_n$ the reaction rates remain greater  than the Universe expansion rate until a lower  freezeout temperature sets in, having therefore a strong impact on the neutron-to-proton ratio at the freezeout. The 2014 PDG  recommended value for the neutron lifetime is $\tau_n = 880.3 \pm 1.1$ s \cite{Oli14}. Another input of the  Big Bang model calculations is the number of light neutrino families,  $N_\nu$. The model can accommodate values between $N_\nu = 1.8 - 3.9$ (see, e.g., Ref. \cite{Kei02}). The measurement of the $Z_0$ width in LEP experiments at CERN imply that $N_\nu = 2.9840 \pm 0.0082$ \cite{LEP06}. A value of $N_\nu=3$ has been used in the present calculations. 

It is customary to use the concept of {\it astrophysical S-factors} by  factoring out the approximate value of tunneling through the Coulomb barrier for reactions between charged nuclei. For a reaction involving charges $Z_1$ and $Z_2$ this value is approximately given by $\exp(-2\pi\eta)$, called the {\it Gamow factor}. The Sommerfeld parameter is given by $\eta = Z_1Z_2e^2/\hbar v$, where $v$ is the relative velocity of the particles. In addition, cross-sections are essentially quantum ``areas" proportional to $\pi\lambda^2 \propto 1/E$ multiplied by reaction probabilities (here, the Gamow factor). Therefore, we can factorize the fusion cross sections as
\begin{equation}
\sigma(E) = {\exp(-2\pi\eta)\over E} S(E) ,\label{astrophys13a}
\end{equation}
where the factor $S(E)$ is the astrophysical S-factor and has a much smoother dependence with the center of mass energy $E$ than $\sigma(E)$. For neutron induced reactions there is no Coulomb barrier and the transmission probability of a neutron through a nuclear potential surface is proportional to its velocity  $v$ incase one assumes a sharp potential surface. Hence, it is more appropriate to write neutron induced cross sections  as 
\begin{equation}
\sigma(E) =  {R(E) \over v} ,\label{astrophys13n}
\end{equation}
where $R(E)$ has also a smoother dependence on $E$. Since S-factors are rather smooth at the low astrophysical energies, it is common to express them in terms of an expansion about $E=0$ \cite{Bah89}, i.e.
\begin{equation}
S = S(0)\left(1+{5kT\over 36E_0}\right) + E_0{dS(0)\over dE } \left(1+ {15kT\over 36E_0}\right) + \cdots \ ,
\end{equation}
where $E_0/kT = (\pi Z_1Z_2 \alpha/\sqrt{2})^{2/3}(\mu/kT)^{1/3}$ is the Gamow energy and  $\mu$  the reduced mass of ions with charges $Z_1$ and $Z_2$. Non-resonant reactions induced by the weak interaction, such $({\rm p},e^+\nu)$, neutrino scattering or electron capture reactions, are the smallest of all cross sections. Radiative capture reactions, such as $(p,\gamma)$ or $(\alpha,\gamma)$ reactions, have also small cross sections because they involve the electromagnetic interaction. The largest cross sections are for  reactions induced by the strong interaction such as $({\rm p},\alpha)$, (d,p), or $(\alpha,{\rm n})$ reactions. For example the $^6$Li(p,$\alpha$)$^7$Be reaction is 4 orders of magnitude larger than the $^6$Li(p,$\gamma$)$^3$He reaction.

\subsection{BBN  of elements up to beryllium}

Once deuterium is formed, reactions proceed quickly to produce helium by means of 
\begin{equation}
{\rm d(n,}\gamma)^3{\rm H}, \ \  \ \ \ {\rm d(d,p)}^3{\rm H},\ \ \ \ \
{\rm d(p,}\gamma)^3{\rm He}, \ \ \ \ \ {\rm d(d,n)}^3{\rm He},\ \ \ \ \ ^3{\rm He(n,p)}^3{\rm H} . \label{dn3H}
\end{equation}
At a longer time,  $^3$H  will $\beta$-decay to $^3$He.  At $E_n=30.5$ keV, relevant for BBN, the cross section for the reaction  ${\rm d(n,}\gamma)^3{\rm H}$ is reported as $2.23\pm 0.34$ $\mu$b \cite{Nag06} which is very small as expected for an electromagnetic process. It is much smaller than the other reactions and it is not considered in the BBN reaction network shown in Figure \ref{hehratio}, right. The existing set of data at this energy differ by a factor of 2 and so is the uncertainty in the cross section  in the range $10^7-10^9$ K. The ${\rm d(d,p)}^3{\rm H}$ reaction has been extensively measured  experimentally \cite{Grei95,Krau87,MK51,Sch72,JB90,Gan57,Arn54,Rai02,Boo56,Dav53,VEn61,Coo53,Mof52,Tie07,Leo06,Tum11,Tum14}.  A recent measurement of the  S-factor for this reaction yields $S(0)=55.3$ keV.b within a 5\% error \cite{Tum11,Tum14}. The reaction ${\rm d(p,}\gamma)^3{\rm He}$ has been measured in Refs. \cite{Gri62,Gri63,Bai70,Sch97,Ma97,Cas02}. A study performed in Refs. \cite{NB00, NH11} argue for preferring theoretical calculations \cite{Viv00,Mar05,Mar15,Kie09} over experimental results, because they might be affected by normalization errors. At $E=10$ keV the S-factor for this reaction is $S=0.286$ keV.b, according to Ref. \cite{Mar15}. The ${\rm d(d,n)}^3{\rm He}$ reaction has also been measured in several experiments \cite{Grei95,Krau87,MK51,Sch72,JB90,Gan57,Arn54,Rai02,Boo56,Leo06,Tum11,Tum14,Dav57,Hof01,Pre54,Bel90,Yin73,Bys10} and a recent measurement reports the value $S(0)=58.6$ keV.b within a 5\% error \cite{Tum11}. Data for the reaction $^3{\rm He(n,p)}^3{\rm H} $ have been gathered in several experiments \cite{BAS55,GM59,SJW61,MG65a,MG65b,CFL70,Bor82a,Bor82b} and R-matrix fits \cite{Bru99,AD03}. The cross section for this reaction near the threshold is strongly influenced by a $0^+$ s-wave resonance in $^4$He. This leads to a deviation of the cross section from the typical $1/v$ law for neutron induced reactions, where $v$ is the relative velocity of n+$^3$He. But the function $R(E)$ in Eq. \eqref{astrophys13n} is relatively constant at the  very thermal low energies $E$. The value of $\sigma E^{1/2}$ at $E=0$ is about 0.7 MeV$^{1/2}$.b. The cross section for this reaction is very large, of the order of kilobarns for thermal neutrons ($\sim 2000$ m/s).

When temperatures cool down to $\sim$100 keV,  the reactions
\begin{equation}
^3{\rm H(p,}\gamma)^4{\rm He},\ \ \ \ \  ^3{\rm H(d,n)}^4{\rm He},\ \ \ \ \
^3{\rm He(n,}\gamma)^4{\rm He},\ \ \ \ \ ^3{\rm He(d,p)}^4{\rm He},\ \ \ \ \ ^3{\rm He(}^3{\rm He,2p)}^4{\rm He}, 
\end{equation}
will follow, despite being  strongly suppressed by Coulomb repulsion. The repulsive Coulomb barrier  is given by $ e^2 Z_1 Z_2 / r $, which for  $Z_2=Z_2=2$ and $r \sim 2$ fm, yields about 3 MeV.  Hence, kinetic energies of $\sim$100 keV lead to very small tunneling probabilities through the Coulomb barriers.  The reaction $^3{\rm H(p,}\gamma)^4{\rm He}$ has a very small cross section because it forms a $0^+$ state in the entrance channel and the photon transition is suppressed because the $^4$He ground state is also a $0^+$ state.  At 40 keV the cross section is about $0.015 \pm 0.004$ $\mu$b \cite{Can00}, in contrast to the much larger cross section for $^3{\rm He(n,p)}^3{\rm H} $, as discussed above. That is why this reaction was left out in the diagram of Figure \ref{hehratio}, right. The reaction $^3{\rm H(d,n)}^4{\rm He}$ at  energies below 70 keV has been measured in several experiments \cite{BJH87,JBH84,BP57,AP51,Bal58,KST66}. The cross section is dominated by a resonance at about 50 keV and at 20 keV the S-factor is 13.4 MeV.b with a 2.4\% error \cite{JBH84}. The $^3{\rm He(n,}\gamma)^4{\rm He}$ cross section is also very small at the energies relevant for BBN, of the order of a few $\mu$b \cite{Kom93}.This reaction is also left out the BBN chain reaction  Figure \ref{hehratio}, right. The reaction $^3{\rm He(d,p)}^4{\rm He}$ has been measured in Refs.  \cite{Krau87,MB80,Arn54,Sch89}. The cross sections from different measurements at low energies is rather scattered, dominated by a resonance at about 0.2 MeV. The S-factor at 20 keV is about 7 MeV.b with a 20\% uncertainty. The cross section for the reaction $^3{\rm He(}^3{\rm He,2p)}^4{\rm He}$ is about $10^{-4}$ $\mu$b \cite{Ita03} and does not play a relevant role in BBN.  Recently,  quark masses have been linked to binding energies and reaction rates using lattice Quantum Chromo-Dynamics (QCD) and  Effective Field Theories (EFT). The measured $^4$He abundance is thought to put a constraint of $ \Delta m_q/m_q \sim 1\%$ on  variations of the quark masses \cite{BLP11} (see also, Ref. \cite{Che11}).

Once $^3$He and $^4$He are formed, only very tiny amounts of $^7$Li and $^7$Be will be produced through
\begin{equation}
^4{\rm He(}^3{\rm H,}\gamma)^7{\rm Li},\ \ \ \ \ \ \ ^4{\rm He(}^3{\rm He,}\gamma)^7{\rm Be}.
\end{equation}
Data for the first reaction has been published in Refs. \cite{HJ59,Sch87a,Gri61,Bur87,BKR94}. The accepted value for the S-factor is $S(0)=0.107$ keV.b \cite{BKR94}, guided by a comparison with theoretical calculations.  The second reaction has been measured in experiments reported in Refs. \cite{PK63,NDA69,Krae82,Rob83,Vol83,Ale84,Osb84,Hil88,Boe14}. The cross sections are very scattered at the measured energies, with an estimated S-factor of 0.4 keV.b at $E_{c.m.}=50$ keV, based on best fits to the experimental data. The scattering of the data at the lowest energies amount to a factor of 2.  When $^7$Be becomes an atom, i.e., after electrons are captured, it will be able to capture a bound electron and to decay into $^7$Li. That is when BBN should stop as the Universe expands, cools down, and it becomes increasingly more difficult for the nuclear fusion reactions to proceed. Only extremely small amounts of elements heavier than Li isotopes are formed compared to those produced in stars during the later evolution of the Universe.

If one increases the value of the baryon/photon ratio, $ \eta $, then one gets a higher BBN temperature $T_{BBN}$ and also a larger n/p ratio \cite{Stei07}.  Thus, larger $ \eta $ yields larger $^4$He abundance. Deuterium and $^3$He act as ``catalysts" because as soon as they are produced, they are immediately consumed, reaching an equilibrium value that depends on the rate of their production and consumption.   Production starts with deuteron formation, which has no Coulomb barrier.  Destruction proceeds via numerous reactions, e.g., d+d, d+p, etc, which are suppressed by Coulomb repulsion.  A larger $T_{BBN}$ enhances more destruction than production and yields less d, and $^3$He. Numerical calculations show that $^3$He is produced at the $^3$He/H $\sim$ 10$^{-5}$ level  in low mass stars and slowly will become larger over the history of the Universe \cite{Stei07}. For small $ \eta $ (low $T_{BBN}$) $^7$Li is created and destroyed by
\begin{equation} 
^4{\rm He(^3H,}\gamma)^7{\rm Li} \ \ \ \ \ {\rm and} \ \ \ \ \ ^7{\rm Li(p,}\alpha)^4{\rm He}, 
\end{equation}
respectively. The second reaction has a smaller Coulomb barrier and thus a lower $T_{BBN}$ will increase more the creation of $^7$Li than its destruction yielding more $^7$Li. On the other hand, for large $ \eta $, $^7$Li production changes because more $^3$He is produced, and $ ^4{\rm He(}^3{\rm He,}\gamma)^7{\rm Be}  \label{he4he3be7}$ becomes important, benefited from high temperatures because of the Coulomb barrier.  Also, there are very few neutrons around to deplete $^7$Be by (n,$\alpha$) reactions. Therefore, $^7$Li produced via the  decay of $^7$Be is also occurring at high T and large $ \eta $. The experimental situation for the first reaction has been unclear for a long time, due to conflicting experimental results, as discussed above. The $^7$Li(p,$\alpha)^4$He reaction was extensively studied experimentally\cite{Cas62,Man64,FK67,Lee69,STW71,Cir76,RK86,Eng88,Har89,Sch89,Eng92a,Spr00,Lat01,Pizz03,Cru09,Lam12c}
Recent data for the S-factor for this reaction yields $S(E) = 0.055  + 0.21E -0.31E^ 2$ MeV.b \cite{Cru05,Cru09}.

\begin{figure}[t]
\begin{center}
\includegraphics[
width=2.9in,
]%
{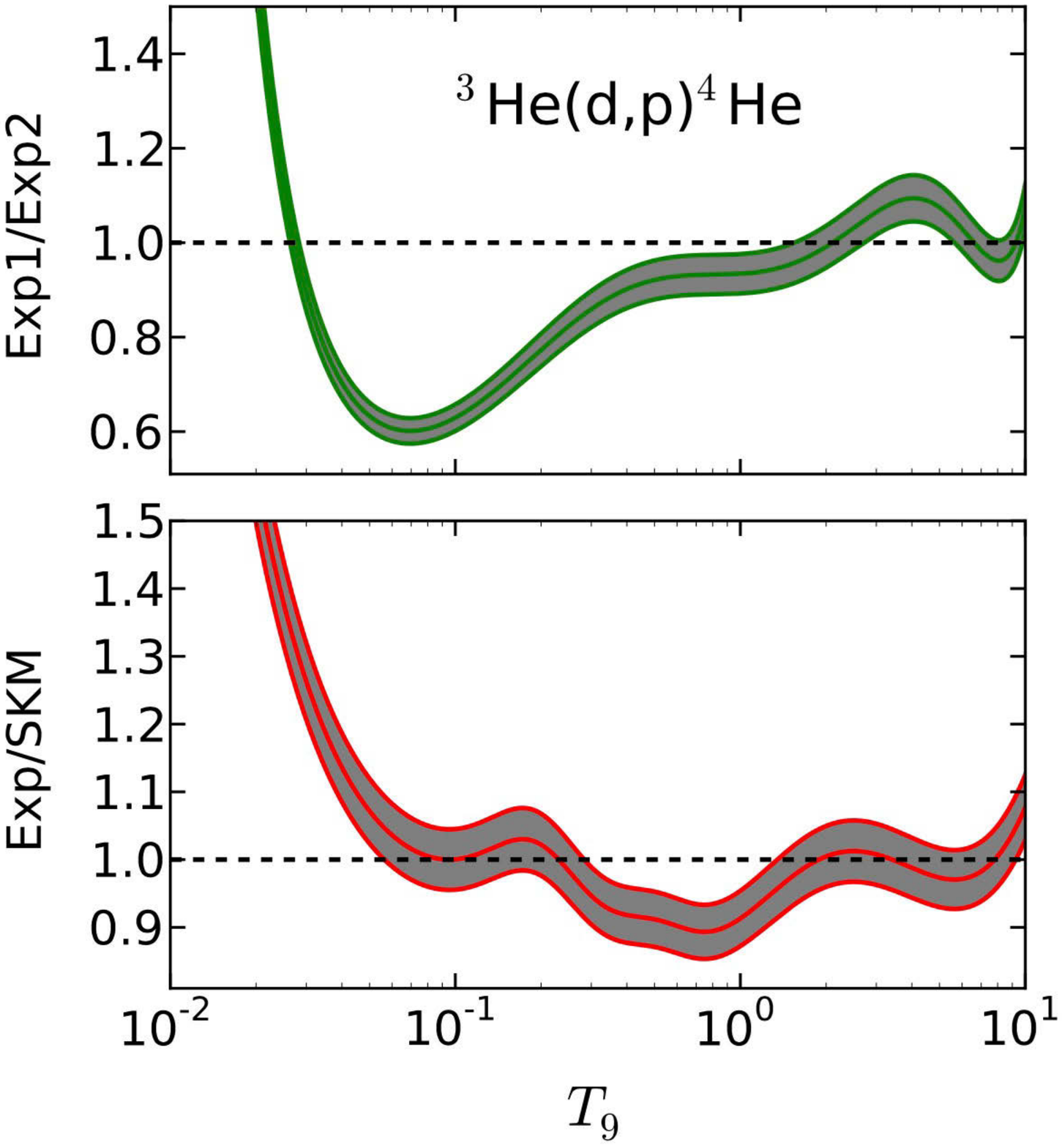}
\includegraphics[
width=3.in,
]%
{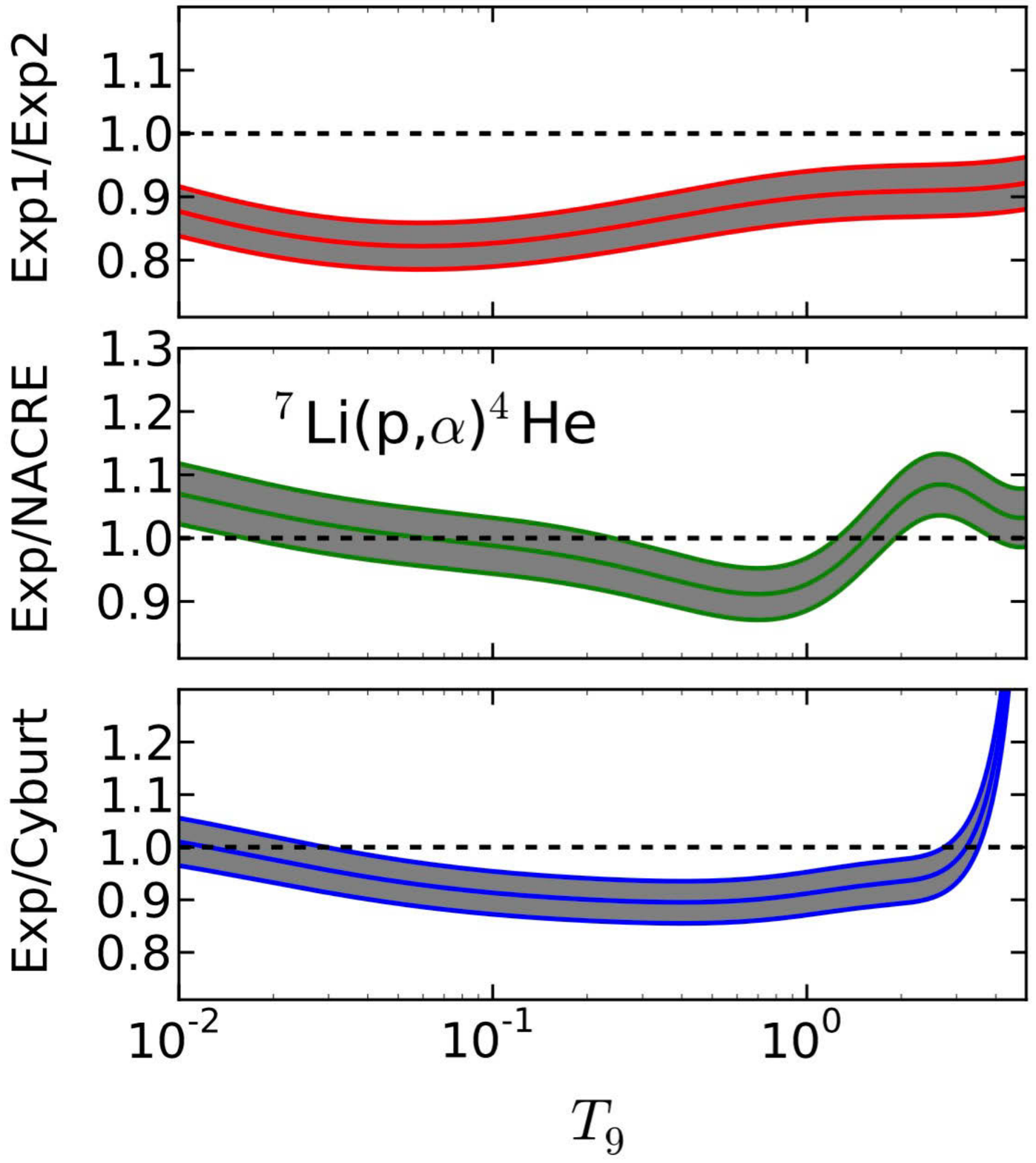}
\caption{\small {\it Left:} Ratios of the reaction rates for $^3$He(d,p)$^4$He  calculated using fits to indirect experimental data (Exp1) from Ref. \cite{Piz14} with fits to a global set of experimental data from direct measurements (Exp2) (upper panel).   The  lower panel is a similar ratio using rates published in Ref. \cite{SKM93} (SKM). {\it Right:} (upper panel) Same as upper panel of left figure, but for  the reaction  $^7$Li(p,$\alpha$)$^4$He. The ratio calculated with fits  to indirect experimental data  \cite{Piz14} and with the fits obtained by the NACRE compilation \cite{Ang99} is shown in the middle panel. The lower panel is the same but the ratio is done with the fit of Ref. \cite{SKM93}.}
\label{bbn_unc}
\end{center}
\end{figure}

The final reaction in the list of the most important ones for BBN is the  $^7{\rm Be}({\rm n,p})^7{\rm Li}$ reaction. It is strongly enhanced because of a $2^-$ resonance close to the threshold. The cross section at thermal energies is about $\sigma = 3.8 \times 10^4$ b \cite{Ajz88}. This is the largest thermal cross section known for reactions with light nuclei \cite{Ajz88,New57,Bar77,Mac58,BP63,Pop76,Sek76,Koe88}. The resonance has very different proton and neutron widths, due to a strong isospin mixing \cite{New57,Bar77}. Despite its large cross section, experimental measurements  are very hard because $^7$Be is radioactive  and the inverse reaction, which could be used together with detailed balance, can lead to excited states in $^7$Be. A R-matrix analysis of the existing experimental data using four partial waves yields the nearly constant, $\sigma E^{1/2} \simeq 5.8$ MeV$^{1/2}$.b, at thermal energies  \cite{AD03}.

There has been a large number of recent works \cite{Bru99,Ang99,Gei99,NB00,Cyb04,Des04,Leo06,Bro07,Cos08,Cru08,Pis08,Dub09,Boy10,Kir11,Mall11,Cha11,Coc12,Cyb12,Bro12,Cyb13,Coc14,Coo14,Kus15} dealing with experimental measurements, theoretical calculations, and the determination of uncertainties of the cross sections listed in Eqs. (\ref{dn3H}-\ref{he4he3be7}). The sensitivity of the BBN abundances of light elements to the variations of the fusion cross sections around the valid ranges set by the experimental data has been the focus of some of these works and one can find a detailed study in Ref. \cite{Cyb15}. It is thought that the present observation data on D/H together with new data on CMB allows one to constrain the number of neutrino families to $N_\nu < 3.2$, closer to the three neutrino families determined from the LEP experiments data at CERN.   Almost all of the reactions listed  in Eqs. (\ref{dn3H}-\ref{he4he3be7}) carry large uncertainties, at the level of 20\% and more (see, e.g., Refs. \cite{NB00,Cyb15}. Some reactions such as $^3$He($\alpha,\gamma$)$^7$Be, t($\alpha,\gamma$)$^7$Li and $^7$Li(p,$\alpha$)$^4$He have noticeably large experimental uncertainties within the Gamow window for temperatures of relevance for BBN.  As an example we show in Figure \ref{bbn_unc} with ratios obtained with fits to different sets of data which have been published. The error band is associated with the error bars in one of the chosen data sets, the smallest of the two. One clearly sees that each reaction rate is  influenced by the compilation methods and by the data set under consideration in each compilation. For certain temperatures, differences between data as large as 40\% are seen. For the $^3$He($\alpha,\gamma$ )$^7$Be and t($\alpha,\gamma$)$^7$Li the differences between different data sets can be as large as a factor of three. There has an effort to compile, fit and correlate different sets of data, also including other information than only reaction cross sections. Recent compilations based on such correlation methods have helped to narrow down the uncertainty bands in the reaction rates for BBN \cite{Cyb15}.

In Table \ref{tabbbn} we show predictions of BBN calculations using most recent values of reaction cross sections for the network in Figure \eqref{hehratio} \cite{Piz14,Cyb15}. These are compared to observations. We note that: (a) The mass fraction for $^4$He, $Y_p=0.2565 \pm 0.006$ (0.001 statistical and 0.005 systematic), is from Ref. \cite{YT10}.  (b) The mean deuterium abundance is the mean average $\left< ({\rm D/H})\right > = (2.82 \pm 0.26) \times 10^{-5}$, which is equivalent to $\Omega_m h^2 \ ({\rm BBN}) = 0.0213 \pm 0.0013$ \cite{Mea06}.  (c) The $^3$He abundances are adopted from Ref. \cite{BRB02} as  a lower limit to the primordial abundance. (d) The lithium abundance arises from observations of stars forming  the ``lithium plateau" \cite{Sbo10}. D/H is in units of 10$^{-5}$, $^3$He/H in 10$^{-5}$ and Li/H in 10$^{-10}$.

\begin{table}[htbp]
\vspace{0.0cm}
\centering
\begin{tabular}{|l|c|c|c|c|c|c|c|}
\hline
\hline
\small
{\small Yields} &Predictions \cite{Cyb15} & {\small Observed}\\  \hline
{\small  $Y_p$ }&{\small $0.24709 \pm 0.00025$}&{\small $0.256\pm 0.006^{(a)}$}  \\ \hline
{\small D/H } ($10^{-5}$)
& {\small $2.58 \pm 0.13$} &{\small $2.82\pm 0.26^{(b)}$} \\ \hline
{\small ${^3}$He/H } ($10^{-5}$)
 &$10.039 \pm 0.090$&$\geq 11.\pm 2.^{(c)}$\\ \hline
${^7}$Li/H ($10^{-10}$)
&$4.68 \pm 0.67$&$1.58\pm 0.31^{(d)} $ \\ \hline
\hline
\end{tabular} 
\vspace{0.0cm}
\caption{\label{tab:Table1} Predictions of BBN calculations \cite{Cyb15} using most recent values of reaction cross sections for the network in Figure \eqref{hehratio}. These are compared to observations \cite{Piz14}.
}
\label{tabbbn}
\end{table}        

The lithium abundance has been inferred from the absorption spectra originated from old  metal-poor stars in the halo of our galaxy or from metal-poor stars in globular clusters. Such stars are ideal to probe the primordial abundance of lithium \cite{Rya01,Rya01b}.  It has been shown that dwarf and subgiant halo main sequence  stars with  $T_{eff} \gtrsim 6,000$ K have a nearly constant Li abundance \cite{SS82}. In contrast, it was found that  the Li abundance is much smaller in cooler stars. The mechanism thought to explain this difference is that hotter stars have a thin convection zone so that Li does not reach the hot regions that can destroy it. Hot halo stars with $[{\rm Fe/H}] \equiv \log[({\rm Fe/H)/(Fe/H})_\odot] \lesssim -1.5$ have a much lower amount of Li than solar metallicity stars, and  the Li content varies very little with the metallicity. This is known as the  {\it Spite plateau} and is thought to directly give the primordial value of the Li/H abundance if these hot stars have not destroyed any of their lithium \cite{SS82,SS82b}. Stellar metallicity is usually defined in terms of  the iron content of the star ``[Fe/H]". Iron is the most abundant heavy element, and is also easy to observe in the visible spectrum. The abundance ratio is sometimes defined as in terms of that of iron as
\begin{equation}
[{\rm Fe} /{\rm H} ]=\log _{10}\left(\frac{n_{\rm Fe}}{n_{\rm H}}\right)_{\rm star} -\log _{10}\left(\frac{n_{\rm Fe}}{n_{\rm H} }\right)_{\rm Sun} ,
\end{equation}
where $n_i$ is the number density of species $i$. Population III, or primordial, stars have a metallicity of less than $-6.0$,  which is about a millionth of the abundance of iron in the Sun. Such stars are the major resources for information about elements produced in BBN. According to BBN calculations as shown in Table \ref{tabbbn}, the $^7$Li/H abundance is predicted to be $\simeq 4.5\times 10^{-10}$. But observations in low metallicity halo stars reveal an abundance of $1.5\times 10^{-10}$. One could try to explain the difference by  modeling of stellar destruction of lithium \cite{How12}. But observational data suggest an overall trend  that the lithium galactic abundance has increased with time. Several possible explanations for this discrepancy have been proposed, which range from stellar phenomena to non-standard BBN models. Among them, some authors have proposed modifications of the nuclear reaction rates \cite{Coc12,CFO04,Boy10}, new nuclear resonances \cite{Cha11,Cyb12,Bro12}, non-standard model particles in the BBN \cite{Pou15,Mat14}; axion cooling \cite{Kus13},  hybrid axion dark matter model \cite{Kus13}, or variations in the fundamental constants \cite{Ber10,Coc12,Cheoun11}, and many more \cite{Mot08,Mot09,Mot09b,Kus10,Mot13,Mat14}. Such investigations seem to show that the so-called {\it lithium problem} cannot be due to nuclear reaction rate uncertainties  \cite{Stei07,Piz14}. But the excess in the lithium abundance might be due to  an over-production of $^7$Be isotope, that radiatively decays to $^7$Li  after BBN ceases. $^7$Li is produced mainly by the electron capture process $^7{\rm Be}+e^- \rightarrow \ ^7{\rm Li}+\nu_e$. The destruction of $^7$Be via many channels could  become a possible solution for the lithium puzzle \cite{Brog12}. Other reactions such as $^6$Li($\alpha,\gamma$)$^{10}$B, $^7$Li($\alpha,\gamma$)$^{11}$B and $^7$Be($\alpha,\gamma$)$^{11}$C could also lead to a possible depletion of lithium and modify the lithium abundances appreciably due to their large experimental uncertainties \cite{He13,Har84,Xu13}.  The other lithium isotope, $^6$Li, is predicted to be formed at the $^6$Li/H = $10^{-14}$ level in BBN.  However, the primordial $^6$Li abundance has been revised  in recent spectroscopic observations of metal-poor halo stars \cite{Lind13},  and there is now only an upper limit of the same order as $\approx 10^{-14}$. Most elements are produced through stellar nucleosynthesis but $^6$Li is mainly destroyed by proton induced reactions in stellar interiors at  temperatures higher than $2 \times 10^6$ K. The radiative capture reaction $^{6}$\textrm{Li}$($\textrm{p}$,\gamma)^{7}\text{Be}$ constitutes one of the main consumers of $^6$Li and of formation of $^7$Be  \cite{He13}. The major source of $^6$Li is thought to be the spallation of galactic cosmic rays by the interstellar medium. For some time, the observed  $^6$Li/$^7$Li isotopic ratio is assumed to be $^6$Li/$^7$Li $\sim 5 \times 10^{-2}$ \cite{Asp06}, in contradiction with BBN predictions $^6{\rm Li}/^7{\rm Li} \sim 10^{-5}$ \cite{CUV14}. This discrepancy cannot be explained by means of spallation in galactic cosmic rays and is known as the {\it second lithium problem}. But the analysis of a recent experiment of the LUNA collaboration has  led to a new value of $^6{\rm Li}/^7{\rm Li} = (1.5 \pm 0.3) \times 10^{-5}$ \cite{And14}, more in line with BBN predictions. A recent theoretical analysis of the reaction $^2$H($\alpha,\gamma)^6$Li  reinforces the LUNA experimental findings, which seems to suggest that the second lithium problem might be solved without invoking non-standard astrophysics scenarios \cite{Akr16}. More work in this direction, both theoretical and experimental, is certainly desirable.

We conclude that detailed calculations of standard BBN (SBBN) use a nuclear reaction network keeping trace of the temperature and density dependence of the Big Bang. The most important nuclear reactions in the network are displayed in Fig. \ref{hehratio}, right. Some of them will be discussed later for hydrogen burning of stars (e.g., the pp chain in the Sun), because they are identical in both cases. However, BBN happens at higher temperatures than inside stars. The extension of the nuclear network calculations to include reactions pertaining to the CNO cycle (discussed below) does not change in a noticeable way the abundances of the elements which are considered primordial, i.e, up to $^7$Li. As we have seen, only p, d, $^3$He, $^4$He and $^7$Li are observed with some confidence as primordial and the evolution of their abundances through the decreasing temperatures in the Big Bang epoch are shown in Figure \ref{fig:bbndecay}, right. The primordial concentration of these isotopes constrains the baryon density of the Universe through the Big Bang parameter $\eta$. The numerical BBN calculations are consistent with 3 neutrino families, the neutron lifetime value of $\tau_n = 880$ s, and with the baryon-to-photon ratio of $\eta = 6.1 \times 10^{-10}$, in accordance with recent astronomical observations \cite{Planck}.  The sensitivity of BBN to the neutron lifetime is of particular significance in elementary particle physics. The neutron lifetime is used as a test of the Cabibbo-Kobayashi-Maskawa matrix \cite{Serebrov05},  which should be consistent with the current G$_A$/G$_V$-value measured in the asymmetry of super-allowed beta-decay experiments \cite{Hardy09,Ade11}.

The agreement between BBN predictions  and observation is  impressive, but the dependence of BBN abundances, in particular $Y_p$ (helium mass fraction), with the neutron lifetime is not negligible \cite{Cyb15}.  The lithium problem still remains, what has led to several theoretical speculations outside the nuclear cross sections realm  (see, e.g., Refs. \cite{Pou15,Kus13,Ber10,Coc12}).  Possible cosmological extensions of the standard BBN have been proposed in Ref. \cite{Yam14}, including photon cooling, radiative decay of X particles, and the possible existence of a primordial magnetic field. Although it is apparent that a a solution of the lithium problem is not related to the uncertainties in the experimental data for nuclear fusion cross sections, this is not entirely ruled out. The uncertainties in nuclear reaction data are still one of the main reasons for the precise determination of some nuclear abundances compared to improved observational data on abundances and on $N_\nu$, $\tau_n$ and $\eta$.  Possible interaction processes with dark-matter particles  could influence the production of lithium isotopes prior to the onset of classical nucleosynthesis patterns.  Examples are, e.g., the influence of  light minutely-charged dark matter particles, which would  lead to strong overproduction of $^4$He, if their masses are much smaller than the electron mass \cite{ZDT13}. But a reasonable agreement with observations is restored with larger masses and if a non-vanishing lepton asymmetry exists. As a collateral effect,  this mechanism worsens the lithium problem and one would need more enhancement of stellar depletion than typically invoked. The additional degrees of freedom due to the presence of weakly-interacting dark matter particles are not capable of draining away energy in the standard Big Bang scenario as to induce appreciable modification in BBN predictions. As shown in Ref. \cite{BFF13}, the possible parallel universes consisting of dark matter are very cold. Further consequences of the matter-dark-matter asymmetry scenario as an outcome of X-particles (X = beyond standard model) are found in several other references, e.g., \cite{DMS10}.

\subsection{Inhomogeneous Big Bang}

It has been suggested that cosmic baryon number fluctuations may have occurred in the early Universe.  Baryons would pile-up in lumps with mass of about $10^{-21}$ M$_\odot$ and higher.  Such fluctuations could have an impact on the production of  light elements during the BBN \cite{App87,Alc87,Mat96}. These inhomogeneities would occur due to a first-order  {\it QCD phase transition} or by means of electroweak baryogenesis. The motivation to introduce the inhomogeneous  Big Bang model is that it could be consistent  with $\Omega_m = 1$. But at present, $\Omega_m = 1$ does not seem necessary for a variety of other reasons, such as the cosmological baryon content, or the success of a Universe structure formation with cold dark matter,  and so on \cite{App87,Alc87,Mat96,Lar06}. We will discuss other implications of this idea to Big Bang nucleosynthesis also called {\it inhomogeneous  BBN} models. The standard Big Bang model assumes that the Universe has always been homogeneous and isotropic. However,  QCD predicts a phase transition from a {\it quark-gluon plasma} to hadronic matter around a temperature $T_{crit}= 170$ MeV. As mentioned previously, to overcome a few difficulties of BNN (e.g., the lithium problem),  inhomogeneous Big Bang models have been suggested. In these models the density in the early Universe is not homogeneous, but  lumpy.  In the beginning, a plasma of hadronic and deconfined quarks and gluons (quark-gluon plasma, or QGP) exist in equilibrium. As the Universe expanded and  its  dropped below $T_{crit}$, the QGP regions transform into hadronic bubbles. As it happens to the phase transition in water, the energy released during this transition keeps the Universe at constant temperature $T_{crit}$. At some moment the energy released is insufficient to maintain the temperature constant due to the reduction induced by the Universe expansion. The hadronic and QGP phases are no longer in chemical potential and pressure equilibrium and the two phases decouple.

In equilibrium conditions, the baryon number density is much larger in the QGP regions than in the hadronic regions. This happens because  it is statistically easier to create low mass quarks than heavy baryons. Thus, during phase-coexistence QGP regions have high baryon densities and hadronic regions have low baryon densities. As the Universe temperature decreases and the phases decouple, the leftover of QGP regions freeze out to hadronic matter, but at higher densities.  The high density bubbles could be 1 - 10 m apart at $T_{crit}$ and when BBN started these distances expanded to $10^3-10^4$ m  at $T \sim 100$ keV. Because neutrons have no charge, they can diffuse  in the plasma with a much larger mean-free path than protons. During the BBN the neutron distribution was homogeneous in the whole Universe, while protons were mainly confined to the high-density regions. The p/n ratio is then different in high and low density regions leading to a different BBN chain for each region. Inhomogeneous BBN models are very dependent on neutron capture  on light nuclei \cite{Jed94}. These reactions also occur during the s-process nucleosynthesis in red giant stars. The neutron capture cross sections need to be known over a broad energy range. The  high primordial lithium abundance could be an optimal fingerprint for such inhomogeneous scenario, as well as a high abundance of beryllium and boron isotopes. Inhomogeneous BBN reactions involve reaction chains such as \cite{App88} $^1$H(n,$\gamma$)$^2$H(n,$\gamma$)$^3$H(d,n)$^4$He(t, $\gamma$)$^7$Li(n, $\gamma$)$^8$Li, $^8$Li($\alpha$,n)$^{11}$B (n,$\gamma$)$^{12}$B($\beta^-$)$^{12}$C(n,$\gamma$)$^{13}$C, and \cite{Boyd89,Kajino90} $^7$Li(n, $\gamma$)$^8$Li(n, $\gamma$)$^9$Li($\beta^-$)$^{9}$Be,  $^7$Li($^3$He,p)$^9$Be, $^7$Li(t,n)$^9$Be(t,n)$^{11}$B,  etc., yielding very abundant $^{9}$Be, $^{10,11}$B and heavy nuclei.  But since the abundance of heavier than lithium and beryllium isotopes is very small,  and since they are  formed in stars, it is difficult to trace their abundance back to test inhomogeneous BBN models. The production of heavy elements in inhomogeneous Big Bang scenarios has been nicely reviewed in Ref. \cite{Rau94}.

\subsection{BBN constraints on cosmological models}

The BBN is of broad interest for its significance  in constraining cosmological models of the cosmic expansion of the early Universe. Cosmological observations \cite{Garn98,Riess98,Perl98} of type Ia supernovae at intermediate redshift,  together with additional observational restrictions at low and intermediate redshift,  as well as the CMB  spectrum of temperature anisotropies  and polarization \cite{Kom11,Planck} all suggest that the Universe is accelerating due to the influence of a dominant dark energy $\Omega_{\Lambda} \approx 0.7$ with a negative pressure \cite{Wang00}.
The simplest interpretation for this dark energy is the existence of a cosmological constant in Einstein's field equations. The  resulting the equation of state from this assumption is $\omega$ = P/$\rho$ = -1. 

Another possibility arises from the so-called quintessence models where the dark energy is the outcome of a slowly evolving scalar field with an effective potential.  The equation of state is negative -1$\leq \omega$ = P/$\rho \leq$ 0, but does not need o be constant  \cite{Ratra88,Peeb88,Brax00}. The  initial conditions probably appear near the inflation era.  The energy density of the quintessence field can be constrained by the BBN epoch  at 0.01 $\leq$ T $\leq$ 1 MeV. When the BBN epoch starts, many of the possibilities for the initial conditions will have already reached the tracker solution. If the energy density is close to the background energy density, BBN will be changed by the enhanced expansion rate due to the increased total energy density. However, it is conceivable that the tracker solution was not yet achieved during the BBN. In this case the strongest constraints would arise by the time the weak reaction rates freeze out, whereas the later BBN epoch might be unaffected. In the third case, the tracker solution may be achieved after the BBN epoch. The ultimate tracker curve might have a large quintessence energy density at a later time  which could be constrained by the CMB. Quintessence models are thus severely constrained by the BBN and CMB \cite{Yahiro02}.

There is recent focus in particle cosmology to understand the Einstein gravity in higher dimensional brane world cosmology which is motivated  by the superstring theories or M-theory. In this scenario our universe is presumed to be a sub-manifold embedded in a higher-dimensional space time. 
Physical matter fields are confined to this sub-manifold, while gravity can reside in the higher-dimensional space time.  The scale of gravity is lowered to the weak scale by introducing large extra dimensions.  This eliminates the hierarchy between the weak scale and the Planck scale, but  it also generates a new hierarchy problem between the weak scale and the size of the extra dimensions.  In Ref.  \cite{RS99} one has solved this new hierarchy problem by introducing non-compact extra dimensions such that our universe is described as a three-brane embedded in a five-dimensional anti-de Sitter bulk space. 
The BBN makes a strongest constraints on the new term which appears as dark radiation in Friedmann equation. It was shown that, although the observational constraints from the BBN allow only a small contribution when this dark radiation term is positive,  a much wider range of negative values is allowed \cite{Ichiki02}.

\section{Stellar nucleosynthesis}\label{stelarsynt}

A star is formed when the primordial clustering material reaches a mass of about 0.08M$_\odot$. Protostars with masses below this  value are known as {\it brown dwarfs}, with insufficient temperatures to ignite hydrogen. But some brown dwarfs, heavier than  about 13 times Jupiter's mass can fuse deuterium during a short period. Heavier stars have higher core temperatures making it easier for particles, starting with hydrogen, to tunnel their mutual Coulomb barrier and form heavier elements. Astrophysicists call stars with the initial stage of hydrogen burning by {\it main sequence} stars. Stellar masses and nuclear reactions are related by:
\begin{enumerate}
\item[a)] $0.1 - 0.5\ M_\odot $. In such stars there will be hydrogen but no helium burning.
\item[b)]  $0.5 - 8\ M_\odot$. For these stars both hydrogen and and helium burning occur.
\item[c)] $< 1.4\ M_\odot$. Such stars end up as a white dwarf (WD).  A WD is formed from the core of stars with masses $\lesssim 8 M_\odot$ after their envelope is ejected.
\item[d)] $8  - 11 \ M_\odot$. They can house hydrogen, helium and carbon burning.
\item[e)] $> 11 \ M_\odot$. Nuclear burning in these stars encompass all stages of thermonuclear fusion.
\end{enumerate}

The basic equations of stellar structure, in the absence of convection, assuming spherical symmetry, are mass conservation and hydrostatic support,
\begin{equation}
{d m (r) \over dr} =4\pi r^2 \rho(r) ,  \ \ \ \ \ \ \ \ \ {dP(r)\over d r}=-{Gm(r) \over r^2}\rho(r)  \label{st:dPdr}
\end{equation}
energy generation and (d) radiation transport,
\begin{equation}
{dL(r)\over  dr}=4\pi^2 \rho(r) [\epsilon (r) - \epsilon_\nu (r) ], \ \ \ \ \ \ \ \ \ {dT \over dr} =-{3 \over 4ac} {{\bar \kappa} \rho \over T^3} {L(r) \over 4\pi r^2}. \label{st:dTdr}
\end{equation}

In these equations, $m(r)$ is the mass within a sphere of radius $r$ from the center of the star, $\rho(r)$ is the local mass density at $r$, $P$ the pressure, $L$ the luminosity radiated away from distance $r$, $T$ the local temperature, $\epsilon(\rho,T)$ the local energy density produced by nuclear reactions and $\epsilon_\nu(\rho,T)$ the local energy density carried away by neutrino emission. The mean opacity $\bar \kappa(\rho,T)$ which takes care of the radiation energy absorbed by atomic processes, such as ionization, Compton scattering, etc. $a$ is the Stefan-Boltzmann constant.
These equations are complemented by the {\it Equation of State} (EoS), i.e., $P\equiv P(\rho, \epsilon, T)$ which is a  sum  radiation pressure, ion pressure, and electron pressure.  For the Sun the EoS is well described by the ideal gas law of a mixture of hydrogen and helium ions. The energy densities $\epsilon$ and $\epsilon_\nu$  are obtained by solving the reaction networks as in Eq. \eqref{astrophys12},  which is also responsible for the changes in the local elemental composition. If convection and entropy changes occur, the equations above need to be modified by including local changes due to convection mixing and the time-dependence of local entropy \cite{KWA12}.   In a convective region we must solve the equations above with additional equations arising from convection, such as
\begin{equation}
{P \over T}{dT \over dP} ={\gamma -1  \over \gamma},
\end{equation}
where $\gamma =c_p/c_v$ the ratio of the specific heat under constant pressure and the specific heat under constant volume \cite{KWA12}. 

Once these coupled equations have been solved, one can calculate $L_{rad}$ and the luminosity due to convective transport $L_{conv} = L - L_{rad}$.
The equations appropriate for a convective region must be switched on when the temperature gradient becomes equal to the adiabatic value, and switched off when all energy is transported by radiation. This method of solution may break down close to the surface of a star.

\subsection{The pp chain}

\begin{figure}[t]
\begin{center}
\includegraphics[
width=3.35in,
]
{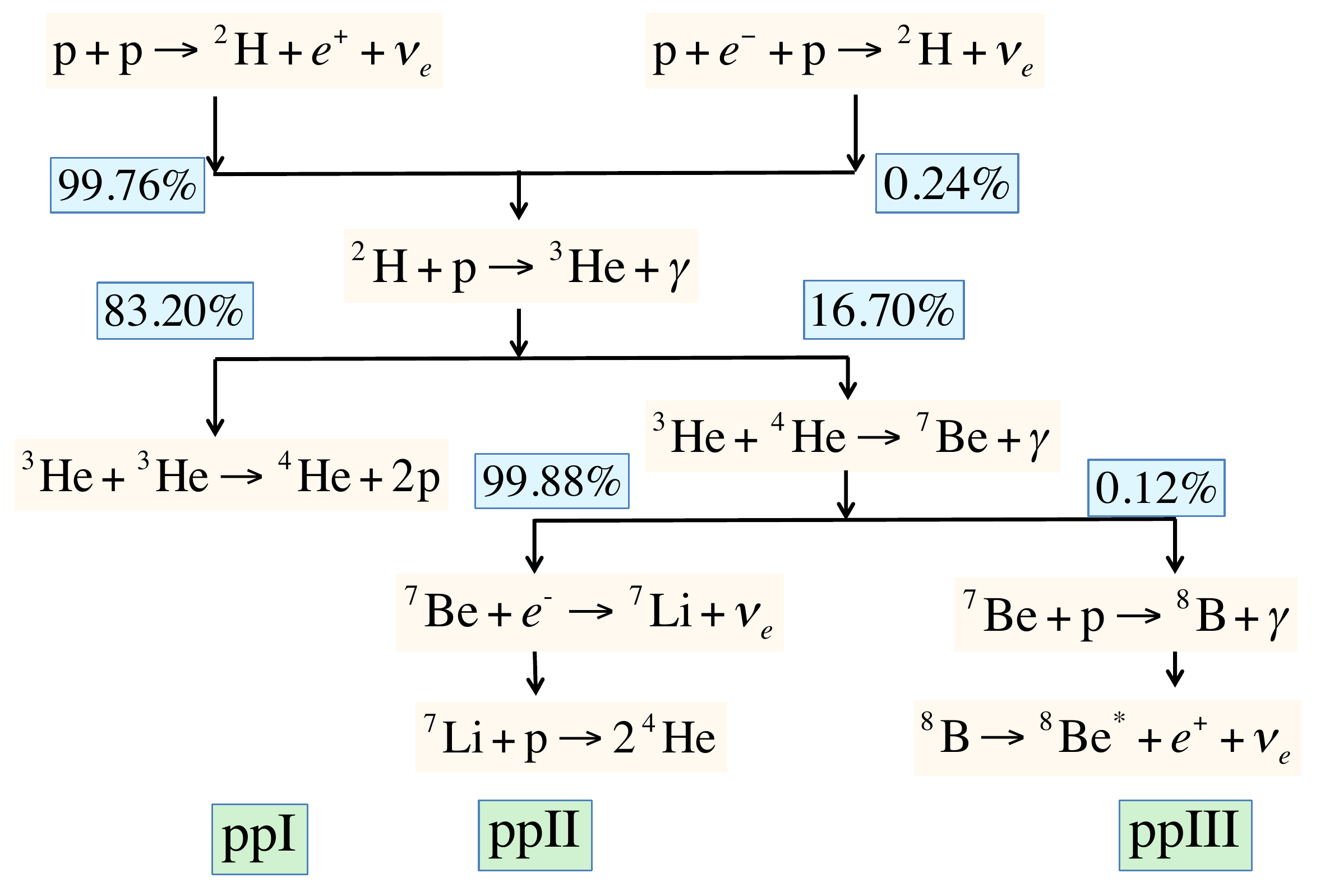}
\includegraphics[
width=3.1in,
]
{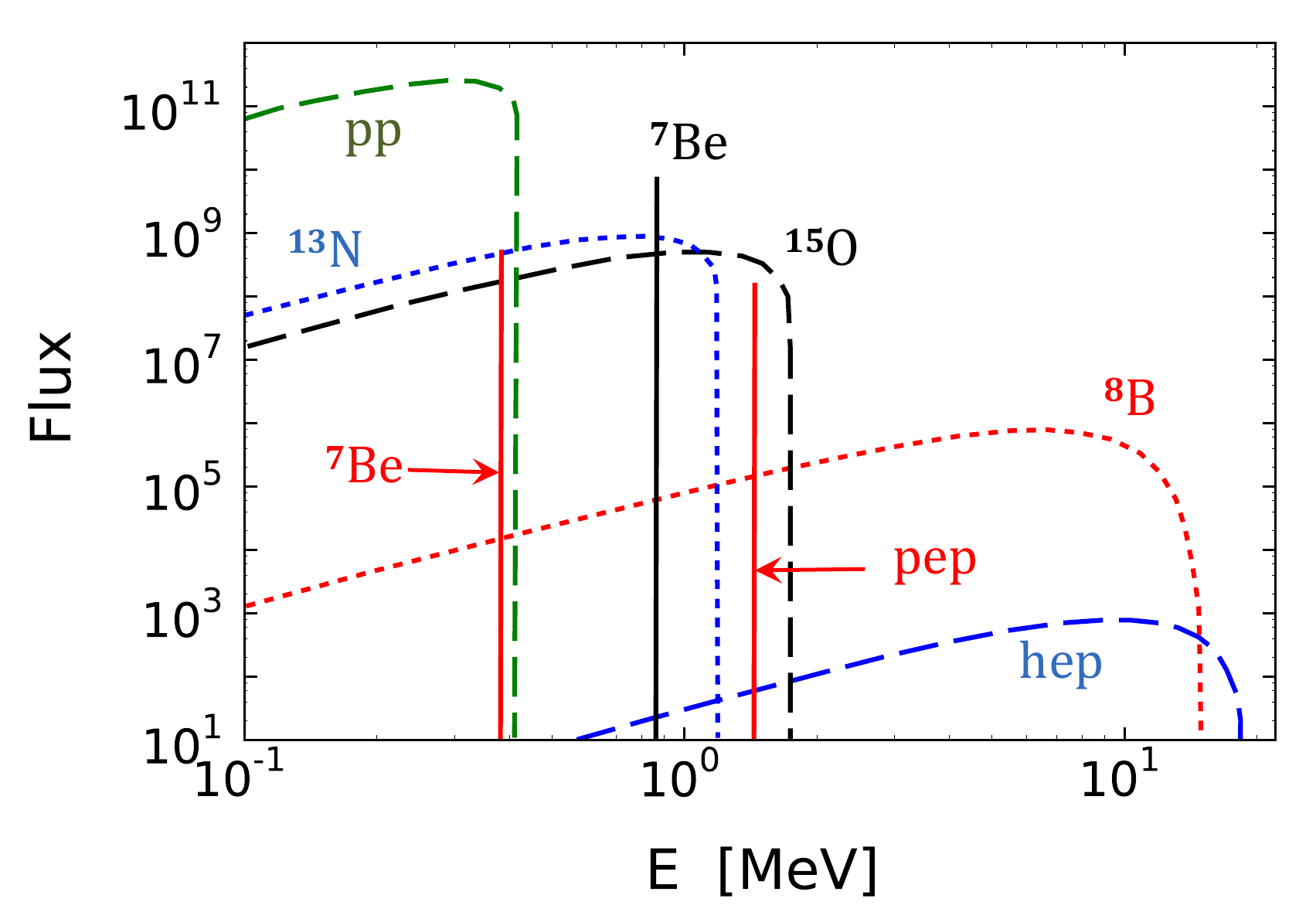}
\caption{{\it Left:} The Sun fusion scheme, or p-p chain, is composed of three branches (ppI, ppII and ppIII) reaction chains.  The percentage for each branch is shown explicitly \cite{Ade11}. {\it Right:} Flux of neutrinos emitted by the Sun, in cm$^{-2}$s$^{-2}$ according to Bahcall's standard solar model \cite{Bah89}.}%
\label{suncycle}
\end{center}
\end{figure}

The Sun is powered by the {\it pp-chain}, i.e., a chain of reactions leading to the formation of helium nuclei, as shown in Figure \ref{suncycle}. There are in fact three such chains, the ppI chain,
\begin{equation} 
{\rm p(e^-p, \nu_e)^2H},~~~~ {\rm ^2H(p,\gamma)^3}{\rm He},~~~{\rm and},~~~^3{\rm He(^3He,2p}){\rm ^4He},
\end{equation}
the ppII chain 
\begin{equation} 
{\rm p(e^-p, \nu_e)^2H},~~~ {\rm ^2H(p,\gamma)^3}{\rm He},~~~^4{\rm He(^3He,\gamma}){\rm ^7Be},~~~^7{\rm Be(e^-,\nu_e}){\rm ^7Be}~~~{\rm and}~~~
{\rm ^7Li(p,^4He)^4He},
\end{equation}
and the ppIII chain
\begin{equation} 
{\rm p(e^-p, \nu_e)^2H},~~~ {\rm ^2H(p,\gamma)^3}{\rm He},~~~^4{\rm He(^3He,\gamma}){\rm ^7Be},~~~^7{\rm Be(p,\gamma}){\rm ^8Be}~~~{\rm and}~~~
{\rm ^8B(,e^+\nu_e)^8Be},
\end{equation}
with $^8$Be promptly decaying into two $\alpha$ particles. The net product of these chain reactions is the consumption of 4 protons and the gain of 28 MeV by the formation of a helium nucleus. Energy is released away in the form of photons and neutrinos. The energy production in the Sun at $T_6=15$ is approximately given by $\epsilon \sim \epsilon_{pp} X^2 T_6^4$, where $X$ is the hydrogen mass fraction and $\epsilon_{pp} = 10^{-12}$ Jm$^3$ kg$^{-2}$ s$^{-1}$. 

\subsection{The pp reaction}

The Sun has been burning nuclear fuel for the last $4.5\times10^{9}$ years. The  primordial proton cloud got compressed and heated by gravitational contraction with the temperatures reaching  values high enough to trigger the pp fusion reaction $\mathrm{p}+\mathrm{p}\longrightarrow\mathrm{d}+\mathrm{e}^{+}+\nu_{e}$. This reaction is non-resonant and occurs in two steps. First the protons tunnel the Coulomb barrier with height $E_c=0.55$ MeV and in a second step one of the protons  $\beta$-decays by positron and neutrino emission. The Coulomb barrier height is relatively small and the Sun would burn its fuel quite quickly if this was the only impediment for the pp reaction. But, because it is a $\beta$-decay process, the final stage occurs with  a very low probability to yield a proton and a neutron system captured within a deuteron. Bethe and Critchfield \cite{Bet38} used Fermi's theory of $\beta$-decay, based on point interactions, to obtain the second part. The cross section for this reaction for protons with energy below 1 MeV is very small, about $10^{-23}\;$b. The effective energy for this reaction in the Sun is in fact much smaller than 1 MeV, i.e.,  about 20 keV. The mean lifetime of protons in the Sun due to  the p(p,d) reaction is about 10$^{10}$ y. Thus, the energy radiated by the Sun is nearly constant in time, and not an explosive process.  Using {\it Fermi's Golden rule} for this reaction, we get
$ d\sigma = {2\pi \rho(E)  \left|\left< f \left| H_\beta \right| i\right>\right|^2/ \hbar v_i} $, where $\rho (E) $ is the density of final states in the interval $dE$ and $v_i$ is the initial relative velocity. $\Psi_i$ is the initial wave function of the two protons and the final state wave function $\Psi_f$ is a product of the deuteron, the positron and the neutrino wavefunctions,  $ \Psi_f = \Psi_d \Psi_e \Psi_{\nu} $. A plane wave can be used for the electron if its energy  is large compared to $Z R_{\infty}$, where the {\it Rydberg constant} is  $R_{\infty}= 2 \pi^2 m e^4 /ch^3$. The energy release in the reaction is 0.42 MeV and therefore the kinetic energy of the electron  is $K_e \leq 0.42$ MeV. The mean energy of the neutrinos is $\left< E_{\nu}\right> = 0.26$ MeV, which is too low and both plane wave exponentials for the  electron and the neutrino wavefunctions can be approximated by the unity. The weak decay matrix element becomes
\begin{equation}\left< f \left| H_\beta \right| i\right> = g \sum_{m_{f}}\sum_j\left\vert \left< \Psi
_{d}^{\ast}\left|   t_{\pm}\sigma_{j}  \right| \Psi_{i}\right>
\,\right\vert ^{2} , \label{mwdec}
\end{equation}
where $\sum_j$ is a sum over the Pauli matrices, $\sigma_{x}$, $\sigma_{y}$ and $\sigma_{z}$, $\sum_{m_f}$ is a sum over final spins and $g$ is the weak coupling constant. For the  deuteron, $J_f^{\pi} = 1^+$, dominated by $l_f = 0$ in a triplet state $S_f = 1$. The maximum (super-allowed) transition probability has $\Delta l=0$ and the initial $\rm p+p$ wavefunction  must have $l_i = 0$. In order for their wavefunction be antisymmetric in space and spin, one must have  $S_i = 0$. Thus, the transition is $ ( S_i =0, l_i = 0 )  \rightarrow  ( S_f =1, l_f = 0 ) $, i.e., a pure Gamow-Teller transition. The full calculation yields 
\begin{equation}\sigma = {m^5 c^4 \over 2 \pi^3 \hbar^7 v_i} f(E) g^2 {M^2_{space} M^2_{spin} \over 2} \ , \label{weakdc}\end{equation}
where $M^2_{spin} ={(2J+1)/(2J_1+1)(2J_2+1)} =3$, $M_{space} = \int_0^{\infty} \chi_f(r) \chi_i(r) r^2 dr$, and the dimensionless Fermi integral $f(E)$ accounts for Coulomb distortion of the electron wavefunction. At large energies, $f(E)  \propto E^5$. The  deuteron radial wavefunction $\chi_f(r)$  and the initial two-proton wavefunction at the low stellar energies involve only the s-wave parts \cite{Fri51}.  A numerical integration of the radial integral yields a cross section of about $\sigma = 10^{-47} \rm cm^2$ at  $E_p = 1 \; \rm MeV$, that cannot be measured experimentally. 

Several theoretical works were published after  Ref. \cite{Bet38} and computed its S-factor for this reaction \cite{BM68,BM69,KB94,Sch98a,BC01,Par03}. The constant $g$ in Eq. \eqref{weakdc} becomes a combination of the Fermi and axial-vector weak coupling constants $G_V$ and $G_A$ \cite{BM69}. The most recent value for the super-allowed $0^+\rightarrow 0^+$ is $(ft)_{0^+\rightarrow \ 0^+}= 3071.4 \pm 0.8$ s  \cite{HT09}, and for For $G_A/G_V$, one can use the Particle Data Group (PDG) value $G_A/G_V = 1.2695 \pm 0.0029$ \cite{pdg}.   Radiative corrections are also  important \cite{BM69} for an accurate calculation of this reaction rate. The largest uncertainty come from the calculation of the matrix element in Eq. \ref{mwdec}. Standard calculations in Refs. \cite{KB94,BP92} were followed by those in Ref. \cite{Sch98a} based on potential models. More recently, EFT calculations up to next-to-next-to-next-to-leading order (N$^3$LO) expansion terms have been performed (see, e.g., Refs. \cite{KR01,And08,BC01,BCV02,CHR03,Par03}). Based on the latest theoretical developments the S-factor for this reaction has been calculated with an accuracy of at least 3\% and there are plans to reduce it to an accuracy of the order of 1\%, based on a combination of theory and experimental efforts \cite{Ade11}.  A conservative value for the S-factor for this reaction is about $S(0)=(4.01\pm 0.04) \times 10^{-22}$ keV b and $dS(0)/dE = S(0)(11.2\pm 0.1)$ MeV$^{-1}$ \cite{Ade11}. In the core of the Sun, $T_6 =15$, and this yields $\left<\sigma v\right>_{pp} = 1.2\times10^{-43} \rm\; cm^3 \; s^{-1}$ \cite{Ade11}. The density in the Sun's core is about $\rho = 100 \; \rm gm\; cm^{-3}$ and if one assumes an equal mixture of hydrogen and helium, then $X_H = X_{He} =0.5$, and the average life of a hydrogen nucleus to be converted to deuterium is $\tau_H(H) = \left<\sigma v\right>_{pp}/N_H \sim 10^{10}$ y, comparable to the age of old stars.  

\subsection{$^3${\rm He} formation and destruction}

Once deuterium is produced in the Sun, it is promptly consumed by a nonresonant direct capture reaction to the ground state of $^3$He,
${\rm d} + {\rm p}\  \rightarrow \ ^3{\rm He} + \gamma $, with a Q-value of 5.5 MeV and the S-factor at zero energy is $S(0) = 0.214 \left({+ 0.017\atop - 0.016}\right)$ eV b according to Ref. \cite{Ade11} based on a combination of fits to experimental data \cite{Gri62,Sch96,Ma97,Cas02}  and theory \cite{Mar05}. Notice that the deuteron induced reactions 
\begin{equation}
\rm d(d,p)t, \ \ d(d,n)^3He, \ \ d(^3He,p)^4He, \ \ \ \ and \ \ \
 d(^3He, \gamma)^5Li\end{equation} 
 have a larger cross section than that for d(p,$\gamma)^3$He. But because of the much larger number of protons in stars, the process of Eq. d(p,$\gamma)^3$He dominates. The rate of energy production in the pp-chain is still determined by $\mathrm{p}+\mathrm{p}\longrightarrow\mathrm{d}+\mathrm{e}^{+}+\nu_{e}$  which is much slower than d(p,$\gamma)^3$He.

The equation for the evolution of  deuterium (D) abundance is given by its production through the pp-reaction and its destruction though the pd-reaction, i.e.,
\begin{equation} {d {\rm Y_D} \over dt}
= {{\rm Y_H}^2\over2} \rho N_A \left<\sigma v\right>_{pp} - {\rm Y_HY_D} \rho N_A \left<\sigma v\right>_{pd} \ .\end{equation}
In thermodynamical equilibrium, this yields
\begin{equation}{\rm Y_D\over Y_H} = {\left<\sigma v\right>_{pp} \over 2\left<\sigma v\right>_{pd}}.\label{sundeq}
 \end{equation}
Using the values of the S-factors mentioned above yields a ratio of $5.6\times 10^{-18}$ at $T_6 =5$ and $1.7\times 10^{-18}$ at $T_6 = 40$. However,  the observed $({\rm Y_D/Y_H})$ ratio in the Cosmos is $\sim 10^{-5}$, which is much larger due to BBN, before stars were formed.  In stellar interiors deuterium is destroyed by means of the reaction d(p,$\gamma)^3$He. The equilibrium condition in Eq. \eqref{sundeq} is reached in about $\tau_d  = 2 \ {\rm s}$, which is the lifetime for the consumption of deuterons. That is, deuterons are burned almost instantaneously in stellar environments.

After  $^3$He is formed, it can be consumed via the non-resonant capture reaction
$\rm ^3He \; + \; ^3He \rightarrow \;p \; + \; p \; + \; ^4He,\label{he3he3}$
with a $Q = 12.9$ MeV, and an S-factor $S(0) =5.11`\pm 0.22$ MeV b. The laboratory data  at low energies is contaminated by electron screening effects and the data from different groups are quite spread out at $E\le 350$ keV \cite{Kra87,Jun98,Bon99,Kud04}. The quoted S-factor is based on phenomenologically guided fit \cite{Ade11}.  Another non-resonant capture reaction consuming $^3$He is
$\rm ^3He \; + \; D \rightarrow \; ^4He \; + p $,  which has an S-factor, $S(0) \sim 6000$ keV b \cite{Kra87}. Both reactions have comparable $S(0)$ factors, but because the deuterium concentration is very small, the first reaction dominates. $^4$He  synthesized in the BBN can react with $^3$He and lead to the ppII and ppIII parts of the chain in Figure \ref{suncycle}. The reactions leading to the formation of $^7$Be and $^8$B, via $^3{\rm He}(\alpha, \gamma)^7{\rm Be}$ and  $^7{\rm Be}({\rm p}, \gamma)^8{\rm B}$, are responsible for the production of high energy neutrinos in the Sun  \cite{Bah89}. The first reaction occurs 14\% compared to the one in Eq. \eqref{he3he3}. In this route, $^7$Be can be consumed in two alternate ways to complete the fusion process of four protons transforming into a helium nucleus, $4 {\rm H} \rightarrow  \; ^4{\rm He}$, in the ppII and ppIII chains.

\subsection{Electron capture on $^7${\rm Be}}

After $^7$Be is formed, the first step of the ppII chain is the capture of an atomic electron by means of  $^7$Be$ \; + \; {\rm e}^- \rightarrow \; ^7{\rm Li} \; + \; \nu_e$. The capture can leave $^7$Li in its  ground state (90\%) or in its first excited state (10\%) at $E_x = 0.48$ keV, and $J^{\pi}={1\over2}^-$. The  Q-value for the capture to the ground state is 380 keV and to the excited state is $0.86 \; \rm keV$, carried away by monoenergetic neutrinos, with mean life for the capture given by $\tau = 77$ d. The  initial and final wavefunctions of the nuclei vanish rapidly outside the nuclear radius and the electron wavefunction within the nucleus can be approximated by its value at the origin. The neutrino wavefunction can taken as a plane wave normalized to the nuclear volume V, so that $H_{if} = \Psi_e(0) g  \int \Psi^*_{^7\rm Li} \Psi^{}_{^7\rm Be} d^3 r  $. The capture rate is then given by 
\begin{equation} \lambda_{EC} = {1\over \tau_{EC}} = \left({g^2 M_{if}^2 \over \pi c^3
\hbar^4}\right) E_{\nu}^2 |\Psi_e(0)|^2 \ ,\end{equation}
where $M_{if}=\int \Psi^*_{^7\rm Li} \Psi^{}_{^7\rm Be} d^3 r  $ is the nuclear matrix element for the transition.

The electrons from the atomic K-shell give the dominant contribution to the capture. In the core of the Sun,  $T_6 = 15$, and the $^7$Be are mostly ionized. But they are immersed in a sea of free electrons  and electron capture from continuum states can occur.  Since the factors in the calculations with continuum wavefunctions are approximately the same as for atomic electron capture, except for the corresponding electron densities, the $^7$Be lifetime in the Sun, $\tau_s$, and  the terrestrial lifetime, $\tau_t$, are related by ${\tau_{s} / \tau_t} \sim 2 |\Psi_t(0)|^2 / |\Psi_{s}(0)|^2$, where $|\Psi_{s}(0)|^2$ can be identified as the density of free electrons in the Sun, $n_e = \rho/ m_H$. The factor of 2 accounts for the two spin states in calculation of  $\lambda_t$, while  $\lambda_{s}$ is calculated by averaging over these two orientations. The electron wavefunctions are distorted due to  Coulomb interaction with  hydrogen (of mass fraction $X_H$) and heavier nuclei in the plasma, and one has instead 
\begin{equation}\tau_{s} = { 2|\Psi_t(0)|^2 \tau_t \over
(\rho/M_H)[(1+X_H)/2] 2\pi Z \alpha (m_e c^2 /3 kT)^{1/2}},\end{equation}
where we used $|\Psi_e(0)|^2 \sim (Z/a_0)^3/ \pi$. This equation leads to the  lifetime  $^7$Be  in the Sun \cite{Bah89,GB97}, 
\begin{equation}\tau_{s} (^7{\rm Be}) = 4.72 \times 10^8
{T^{1/2}_6 \over \rho(1+X_H)} \ \rm s.\end{equation}
The thermally averaged Coulomb corrections on the electron wavefunction yields the temperature dependence in this formula. One gets the continuum capture rate of $\tau_{s}(^7{\rm Be}) = 140$ d whereas on Earth  $\tau_t = 77$ d  \cite{Bah69b}. Considering also partially ionized $^7$Be atoms under solar conditions, one gets  another 21\% increase in the decay rate, leading to  a  lifetime of $^7$Be in the Sun as $\tau_{\odot} (^7{\rm Be}) = 120$ d. This result seems to be robust and recent claims that it might be influenced by additional physics have been found to be unsubstantiated \cite{BGJ08}.

\subsection{$^8$B, solar neutrinos, and neutrino masses}\label{solarneut}

$^7$Be can also be consumed $0.12 \%$ of the time by means of $^7{\rm Be} ({\rm p}, \alpha) ^8{\rm B}$, completing the ppIII part of the pp-chain in the Sun. This reaction occurs at energies much smaller than the 640 keV resonance in $^8$B. The weighted average for its S-factor is $S(0) = 21.0 \rm \; eV b$  \cite{Ade11} but the experimental situation is not completely satisfactory. This has led to the need of support from theory, with the development of better microscopic reaction theories, sometimes denoted as {\it ``ab-initio" calculations} \cite{Nav11}.  Within a lifetime of $\tau = 1.1$ s,  $^8$B($J^{\pi} = 2^+$) decays by means of  $^8{\rm B}\  \rightarrow \  ^8{\rm Be} + {\rm e}^+ + \nu_e$, mainly to the broad ($\Gamma = 1.6$ MeV) resonance in $^8$Be at  $E_r= 2.94$ MeV ($J^{\pi} = 2^+$), which promptly decays into two $\alpha$-particles. The  neutrinos acquire an average energy in the decay of $\bar{E_{\nu}}(^8{\rm B}) = 7.3 $ MeV. 

The  {\it solar neutrino problem} was a long-standing (from mid 1960s to 2002) problem first associated with a pioneering experiment on neutrino capture on $^{37}$Cl  carried out for many years in the Homestake gold mine in the U.S. The experiment was proposed by Bahcall and Davis to measure the neutrino fluxes based on a Standard Solar Model (SSM) developed by Bahcall \cite{Bah99} based on the solution of Eqs. (\ref{st:dPdr}-\ref{st:dTdr}) and the network for the pp-chain and CNO chain reactions. Davis experiment was particularly sensitive to the high energy neutrinos coming from $^8$B decay. They observed a neutrino flux that was substantially smaller than predicted  \cite{Bah89,BP92,BKS98,TL93}. The observed event rate since 1970 was $2.56\pm 0.23$ SNU (1 SNU = $10^{-36}$ interactions per target atom per second). Bahcall's Standard Solar Model (SSM) prediction was $8.1$ SNU. The puzzle led to the development of new experiments with more sensitiveness to the lower energy neutrinos.  The predictions of the SSM are shown in Figure \ref{suncycle}. The puzzle was solved by the discovery of  {\it neutrino oscillations}, i.e. oscillations between different neutrinos types \cite{BKS98,Fu98}. Some of the electron-neutrinos reaching the chlorine detector become muon-neutrinos on their way from the center of the Sun to the Earth  explaining the smaller number of electron-neutrinos observed on the Earth. The fluxes of $^{8}$B and $^{7}$Be neutrinos  are given by \cite{BKS98}
\begin{equation}
\Phi({\rm B})\propto S_{17}\frac{S_{34}}{\sqrt{S_{33}}}T_{\odot}^{20} ,\;\;\;\;\;\;\;\;\;\;\;\;\Phi({\rm Be})\propto \frac{S_{34}}{\sqrt{S_{33}}}T_{\odot
}^{10},\label{sunneutr1}
\end{equation}
where $S_{ij}$\ are the  astrophysical factors for reactions between nuclei $i$ and $j$. The fluxes are strongly dependent on  $T_{\odot}$ which means that  $^{7}$Be and $^{8}$B neutrinos are good thermometers for the Sun's core temperature. 

If neutrinos have masses,  they can change from one type to another \cite{Pon57,Gri69}. For neutrinos, their mass and weak interaction eigenstates span the same two-neutrino space but are not the same: their mass eigenstates $|\nu_{1}\rangle$ and $|\nu_{2}\rangle$, with masses $m_{1}$ and $m_{2}$, are related to the weak interaction eigenstates through $ |\nu_{e}\rangle  =\cos\theta_{v}|\nu_{1}\rangle+\sin\theta_{v}|\nu_{2} \rangle$ and  $|\nu_{\mu}\rangle  =-\sin\theta_{v}|\nu_{1}\rangle+\cos\theta_{v}|\nu_{2}\rangle$, where $\theta_{v}$ is the vacuum mixing angle \cite{Fis01}. Therefore, a state produced as  $|\nu\rangle=|\nu_{e}\rangle$ or  $|\nu\rangle=|\nu_{\mu}\rangle$  does not remain a pure flavor eigenstate as it propagates. The different mass eigenstates  will accumulate different phases as they evolve in time. This is know as \textit{vacuum oscillations}. The probability that a neutrino state remains as a $|\nu_{e}\rangle$ at time $t$, after a distance $x$ from the source is found to be
\begin{equation}
P_{\nu_{e}}(t)   =|\langle\nu_{e}|\nu(t)\rangle|^{2}=1-\sin^{2}2\theta_{v}\sin^{2}\left(  {x\over L_\nu}\right) . \label{Pnuosc}
\end{equation}
where $\delta m^{2}=m_{2}^{2}-m_{1}^{2}$, and $L_\nu$ is the {\it oscillation length}, 
\begin{equation}
L_\nu={\frac{4\pi\hbar cE}{\delta m^{2}c^{4}}}. \label{Pnuosc2}
\end{equation}
If this length is shorter or comparable to one astronomical unit, one expects a decrease in the solar $\nu_{e}$ flux on Earth. Neutrino oscillations change appreciably when they propagate in  dense environments because of changes in the neutrino effective mass. This  greatly enhances oscillation probabilities as a  $\nu_{e}$ can be transformed into a $\nu_{\mu}$ as it passes by a critical density in the Sun \cite{Wol78,MS85}. The resulting adiabatic electron neutrino survival probability at a position $x$,  is \cite{Bet86} 
\begin{equation}
P_{\nu_{e}}^{\mathrm{adiab}}={\frac{1}{2}}+{\frac{1}{2}}\cos2\theta_{v}
\cos2\theta_{i}
\end{equation}
where $\theta_{i}=\theta(x_{i})$ is the mixing angle at the point where the neutrino is created, given by
\begin{equation}
\sin2\theta(x)   ={\frac{\sin2\theta_{v}}{\sqrt{\Theta^{2}(x)+\sin^{2}2\theta_{v}}}},\ \ \ \ \ \ \ \ 
\cos2\theta(x)   ={\frac{-\Theta(x)}{\sqrt{\Theta^{2}(x)+\sin^{2}2\theta_{v}}}},
\end{equation}
where $\Theta(x)=2\sqrt{2}G_{F}\rho(x)E/\delta m^{2}-\cos2\theta_{v}$. One sees that $\theta(x)$ varies from $\theta_{v}$ to $\pi/2$ as the density $\rho(x)$ varies from 0 to $\infty$.  If the density is not constant, the mass eigenstates evolve in time as the density changes. This phenomenon is known as the {\it Mikheyev-Smirnov-Wolfenstein (MSW) effect} \cite{Wol78,MS85}.  It provides an explanation for the observed reduction in the solar neutrino flux.  The 2002  Nobel Prize in Physics was given to Ray Davis and Masatoshi Koshiba for their experimental work on the ``missing" solar neutrinos.

Cosmic rays collide with nuclei at the top of the atmosphere and produce secondary showers of hadrons, leptons, and neutrinos, such as the reaction
$p + p \rightarrow p + n + \pi^+ .$ The pion decays as  $\pi^+ \rightarrow e^+ + \nu_e \mathrm{~or~} \mu^+ + \nu_\mu .$ The muon then decays by $\mu^+ \rightarrow e^+ + \nu_e + \bar{\nu}_\mu$ and the net result is a neutrino flavor ratio ${ (\nu_\mu + \bar{\nu}_\mu ) / \nu_e } = 2 $. But experimentally one finds that this  ratio  is closer to one. This  puzzle is also explained in terms of  neutrino oscillations. The path-length is limited by the Earth's diameter and for a neutrino energy of the order of 1 GeV one finds that {\it atmospheric neutrinos} are sensitive to $ \delta m^2 \gtrsim 10^{-4}$ eV$^2$. The Super-Kamiokande collaboration obtained the first strong evidence for neutrino oscillations with observations that muon-neutrinos produced in the upper atmosphere consistently changed into tau-neutrinos. It was observed that fewer neutrinos were detected coming through the Earth than coming directly from above the detector.  The Sudbury Neutrino Observatory (SNO) obtained another evidence for solar neutrino oscillations by finding that only about 35\% of the arriving solar neutrinos were $\nu_e$s, with the rest being $\nu_\mu$s and $\mu_\tau$s. The observed number of neutrinos agrees quite well with the SSM model for the Sun. The 2015 Nobel Prize for Physics was awarded to  Takaaki Kajita from the Super-K Observatory and Arthur McDonald from Sudbury Neutrino Observatory for experiments on neutrino oscillations. There has been many estimates of the lower limit for neutrino masses from such type of experiments and the various neutrinos flavors have been estimated to have a total mass of $0.06$ eV. Recently, Planck data have shown a small discrepancy between CMB and lensing observations.  The neutrino masses are thought to suppress the growth of dense structures which are responsible for the formation of galaxy clusters. In addition to 3 neutrinos one also allows for the a sterile neutrino,  which does not take part in weak interactions but can appear through flavor oscillations in the same way as the other neutrinos. The neutrino mixing should include all neutrinos. For a 3-neutrino mixing $\left|\nu_l\right> =\sum_j U_{lj} \left|\nu_j\right>$, where the $3\times 3$ matrix $U_{lj}$ is known as the  Pontecorvo-Maki-Nakagawa-Sakata (PMNS) or Maki-Nakagawa-Sakata (MNS) mixing matrix.  Adding the sterile neutrino to the mixing obviously extends it to a $4\times 4$ matrix. 

Neutrino astrophysics is a thriving research field, with several nations supporting experiments which connect neutrinos to  deep questions of scientific interest. Examples are the above mentioned Sudbury Neutrino Observatory, which completed data taking in 2006, and had vital contributions to the neutrino oscillation discovery \cite{SNO01} as well as the Super-Kamiokande observatory which was designed  for proton decay searches, and the detection of solar, atmospheric and supernova  neutrinos. A follow-up observatory, SNO+ aims to measure neutrinoless double-beta decay to study lepton number violation and to measure the neutrino mass \cite{SNO+}. The Super-Kamiokande detector is still active and under new improvements \cite{Abe14}.  The Karlsruhe Tritium Neutrino Experiment (KATRIN)  is another experiment aimed at  measuring the electron neutrino mass with a precision of  0.2 eV \cite{KATR14}. The ANTARES detector \cite{Adr14} is a directional neutrino telescope aimed to observe neutrinos from the cosmos  in the  southern hemisphere,  whereas IceCube \cite{Abba09} detects neutrinos from the North.  The list of neutrino detectors is impressive and it is hard to make justice to their contribution to physics in this limited space. A list and description of the several neutrino detectors can be found online \cite{WikiNeut}.  One of the greatest scientific discoveries of the 21st century was the recent detection of gravitational waves by LIGO  \cite{Abb16}, as predicted by Eintein's theory of General Relativity about 100 years ago. Physicists from IceCube and ANTARES have already checking whether one had registered an neutrinos in time coincidence with the LIGO observations. These and other possibilities will open an new era of astronomy with gravitational waves and neutrinos, two of the weakest interacting fields in nature, becoming valuable tools together with photons and cosmic rays to probe the structure and history of our Universe.

\section{Nucleosynthesis in massive stars}\label{nuclsynt}
\begin{figure}[t]
\begin{center}
\includegraphics[
width=3.5in
]
{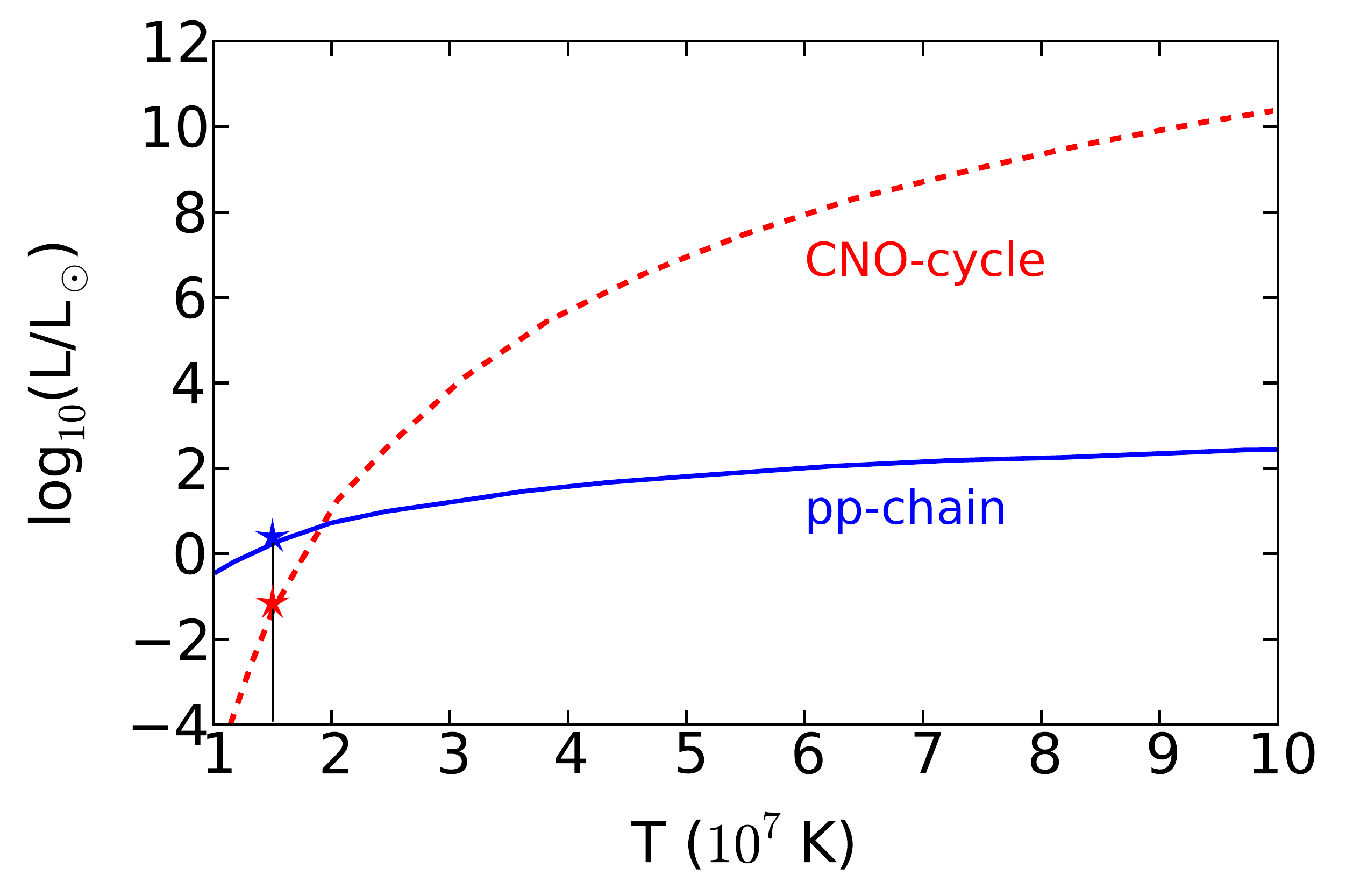}\ \ \ \ 
\includegraphics[
width=2.8in
]
{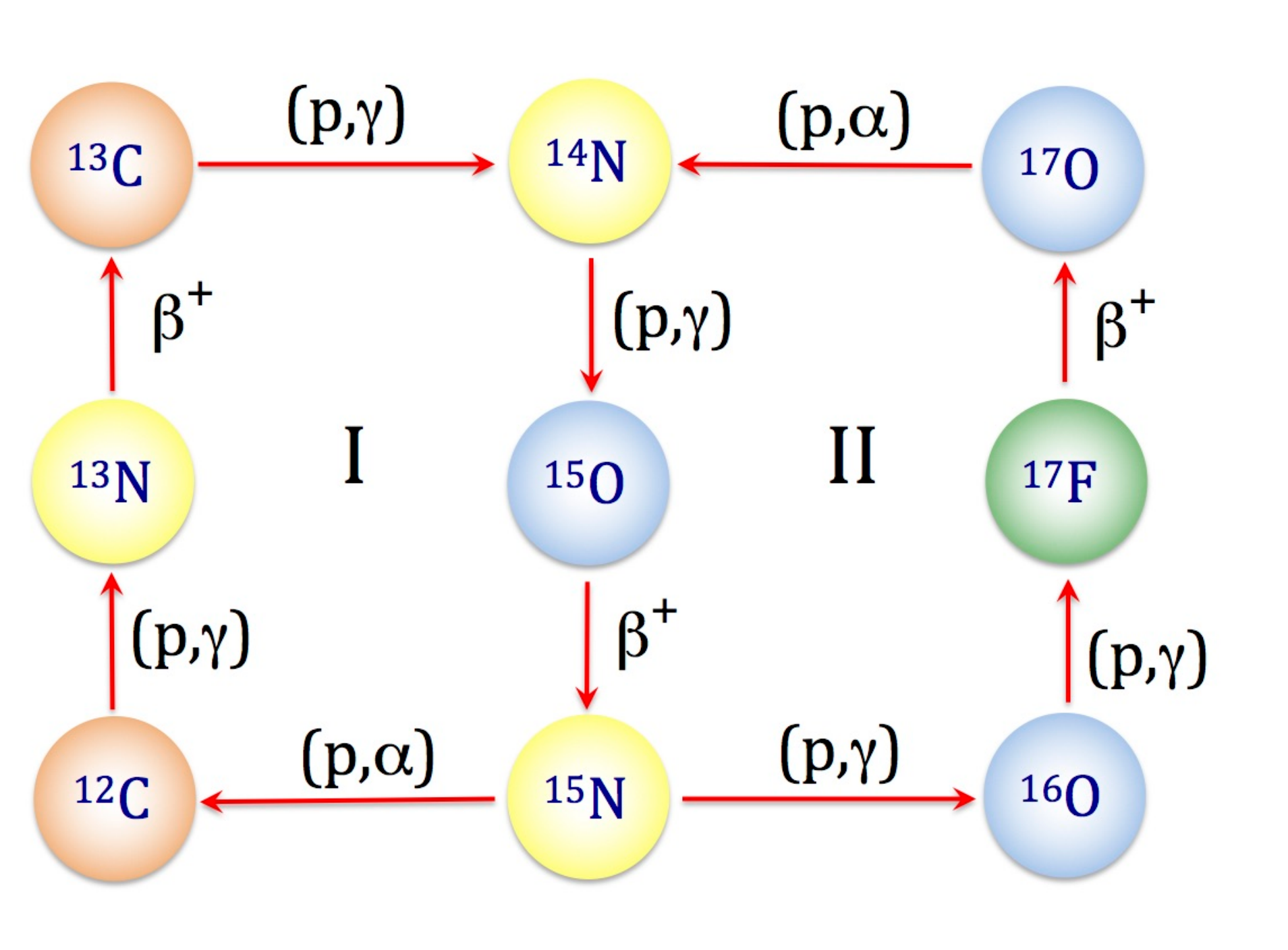}
\caption{\small {\it Left:} Temperature dependence of the energy production by the pp-chain and CNO cycles, assuming a CNO composition as that of the Sun. The passage by the Sun's core temperature is marked by ``star" symbols. {\it Right:} The (``cold") CNO cycle is composed of cycles I and II.  Cycle I is also known as the CN cycle \cite{Bet37} and produces only about 1\% of the solar energy, but leads to a significant  flux of neutrinos from the Sun. Cycle II is a breakout off the CN cycle and is also known as the ON cycle. Cycle I operates 1000 times for each chain of reactions completing cycle II.}
\label{pp_cno}%
\end{center}
\end{figure}

The pp-chain is very effective to generate energy in low temperature stars such as our Sun. But as temperatures increase for more massive stars,   the CNO cycle becomes more effective, provided the CNO-nuclei are already present (from earlier stellar generations). This is shown in Figure \ref{pp_cno} (left) where the energy output of the CNO cycle is compared to that of the pp-chain as a function of the temperature. The CNO energy output, assuming the CNO solar abundance, dominates for  for $20 \le T_6 \le 130$, corresponding to Gamow peak energies of 30-110 keV. 

The CNO-cycle and rp-processes presented in this section are fundamental for the description of cataclismic events, discussed in section \ref{cataclysm}. We will introduce them in this section first as we make reference to some of the reactions involved in the ensuing text.

\subsection{The {\rm CNO} cycle}

Population II and population III are early generation of stars which generated their energy mainly through the pp-chain. One still observes such stars in {\it globular clusters} and the center of a galaxy. They usually have masses smaller than the Sun, are older, cooler and have fewer heavy elements, i.e. low metallicity.  In contrast, population I stars are later generation stars which formed from the debris of heavier stars containing heavy elements. Later generation stars,  heavier than the Sun,  achieve higher central temperatures  due to higher gravity. In such a scenario, hydrogen is more efficient through reaction chains involving C, N, and O. Such nuclei have a non-negligible abundance of  1\% and slightly more compared to Li, Be, B which as we have seen before, have a very low abundance.

The CN cycle, shown in Fig. \ref{pp_cno} (right) as cycle I, is the chain of reactions 
\begin{equation}^{12}{\rm C(p},\gamma)^{13}{\rm N(e}^+\nu_e)^{13}
{\rm C(p},\gamma)^{14}{\rm N(p},\gamma)^{15}{\rm O(e}^+\nu)^{15}{\rm
N(p},\alpha)^{12} {\rm C},\end{equation}
catalyzing an $\alpha$-particle from four protons, $4 p \rightarrow ^4{\rm He} + 2 {\rm e}^+ + 2\nu_e$, and releasing $Q=26.7$ MeV.  The reactions in this cycle are relatively well-known experimentally. The $^{12}$C( p,$\gamma)^{13}$N reaction has been studied over a wide energy range down to  70 keV \cite{You63,BS50,HF50,LH57a,RA74,Kem74,BF80,HV80,BK91,Li06,Bur08}. The S-factor at 25 keV is found to be $(1.75 \pm 0.22)$ keV.b \cite{Bur08}. The  reaction $^{13}{\rm C(p},\gamma)^{14}{\rm N}$ has been measured over a wide energy interval down to  $E_{c.m.}\simeq 70$ keV \cite{Bur08}. For $E_{c.m.}$ below 200 keV, it is dominated by the tail of the s-wave capture to a broad $1/2^+$ resonance at $E = 420$ keV.  The $S(E)$ factor at the stellar energies of about 25 keV is found to be $S = (1.75 \pm 0.22)$ keV b \cite{Bur08}. The $^{13}{\rm C(p},\gamma)^{14}{\rm N}$ reaction  is also related to the {\it s-process}, to be discussed later,  because it depletes the seed nuclei for the reaction $^{13}$C($\alpha$, n)$^{16}$O which is a neutron source for reactions in {\it AGB stars} with solar metallicity.  Measurements for the later reaction have reached down to 100 keV, with R-matrix fits yielding $S(0) \simeq 5-8$ keV b \cite{Muk02,Muk10,Gen10}. The slowest process in the CNO cycle is the $^{14}$N(p,$\gamma)^{15}$O reaction \cite{Sch87,LH57,CF88,Ade98,Ang99,AD01,Ber01,Muk03,Yam04,For04,Imb05,Run05,Lem06} with a cross section of 0.24 picobarns at 70 keV \cite{Lem06}.  This reaction needs to be measured with a better accuracy to reduce the uncertainty in the  CNO neutrino production rate.  The  $^{15}{\rm N (p}, \gamma)^{16}$O reaction leads to a loss of catalytic nuclei from the CN cycle. But the catalytic material returns to the CN cycle  by means of the CNO cycle II,
\begin{equation}^{15}{\rm N (p}, \gamma)^{16}{\rm O(p}, \gamma) ^{17}{\rm F(e}^+\nu_e)^{17}{\rm
O(p},\alpha)^{14}{\rm N}\ .\end{equation}
Two low energy neutrinos are emitted from the beta decays of $^{13}$N ($t_{1/2} = 10 \ \rm min$) and $^{15}$O ($t_{1/2} = 122 \ \rm s)$. The slowest reaction is $^{14}{\rm N(p}, \gamma)^{15}$O in the CN cycle because it involves $Z=7$ nuclei and therefore its Coulomb barrier is largest, and also because it is governed by electromagnetic forces, whereas the other reactions involving N isotope,  $^{15}{\rm N(p}, \alpha)^{12}$C, is governed by strong forces and as a consequence is faster. Thus the rate of energy production in the CN cycle is dictated by the $^{14}{\rm N(p}, \gamma)^{15}$O reaction. Sometimes the CNO cycle II is called the ON cycle and, together with the CN cycle, it constitutes the CNO cycle. The  ON cycle is much slower than the CN cycle, by a factor of 1000, because the S-factor for $^{15}{\rm N(p}, \alpha)^{12}$C  is about a 1000 times larger than that for $^{15}{\rm N(p}, \gamma)^{16}$O \cite{Cac11}. The energy released in the CNO cycle at $T_6=20$ is approximately given by $\epsilon \sim \epsilon_{CNO} XX_{CNO} T_6^{19.9}$, where $X_{CNO}$ the mass fraction of oxygen/carbon/nitrogen (an average of the three) and $\epsilon_{CNO} = 8.24 \times 10^{-31}$ Jm$^3$ kg$^{-2}$ s$^{-1}$. Notice the much larger temperature dependence  compared to the pp chain ($\epsilon \sim T_6^4$. That is why lower mass stars, with cooler core temperatures, generate most of their energy with pp chains, whereas in more massive stars  with higher core temperatures the CNO cycle is more important. 

\begin{figure}[t]
\begin{center}
\includegraphics[
height=1.9in
]%
{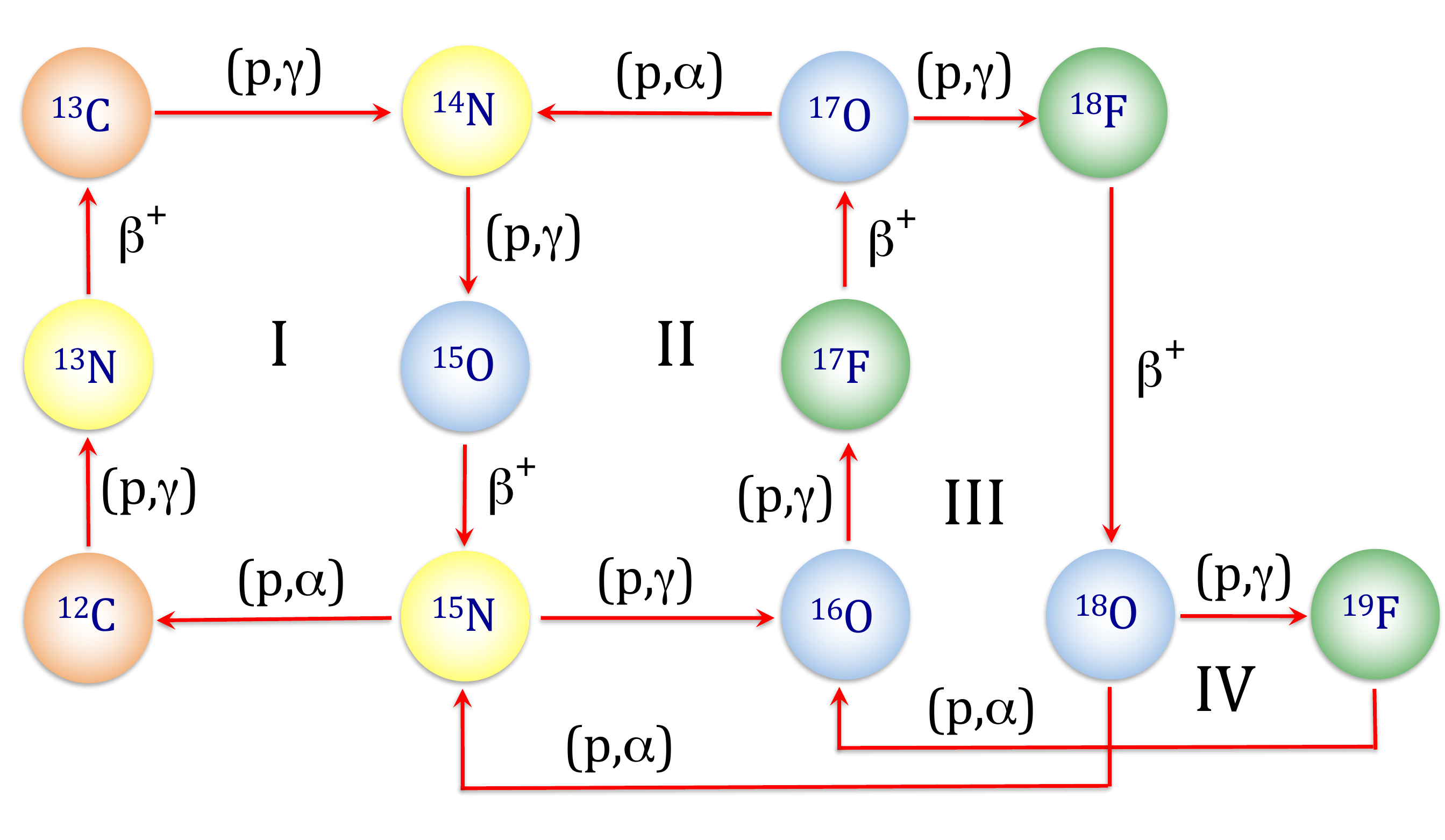} \ \ \
\includegraphics[
width=3.in
] 
{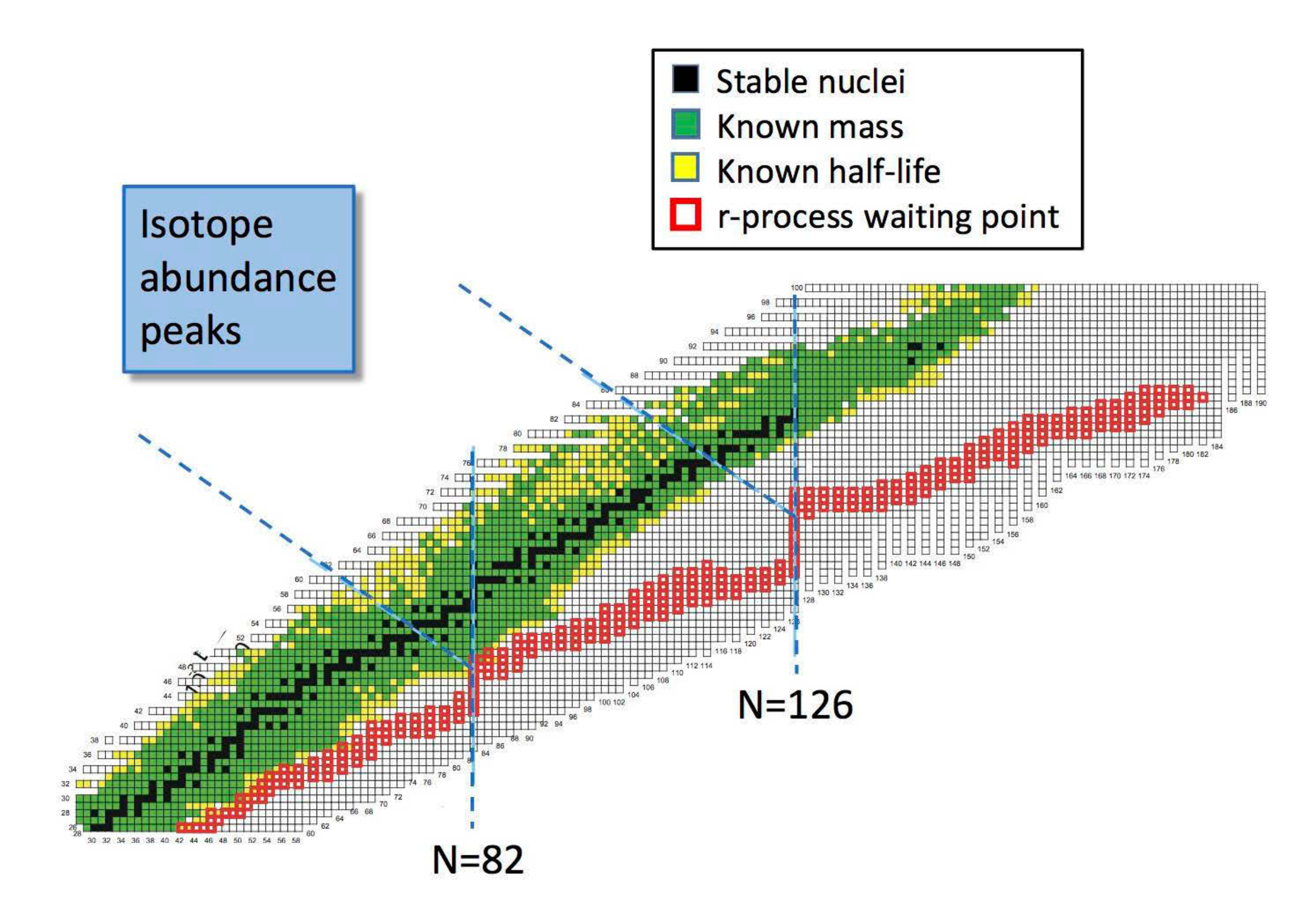}
\caption{\small {\it (Left)}:
The hot CNO cycle occurs at high temperatures because of the additional cycles leaking from the CNO cycle through the $^{17}{\rm O}({\rm p}, \gamma)^{18}$F reaction. {\it (Right)}: Schematic view of the r-process (red squares) in the nuclear chart. The process occurs close to the neutron drip  drip line. The waiting point nuclei at the neutron magic numbers are responsible for peaks in the abundance of isotopes as a function of the nuclear mass.
}
\label{fig:CNOcycle}
\end{center}
\end{figure}

\subsection{Hot CNO and rp-process}

The  CNO cycle works at  $T_6 \geq 20$ and is found in stars with the solar composition, and slightly more massive than the Sun, and burning hydrogen slowly. But  CNO cycles also work at much larger temperatures ($T \sim 10^8 - 10^9 \rm \; K$) in other stellar sites, such as (a)  the accreting surface of a neutron star and (b) explosive {\it novae}, i.e., burning on the surface of a WD, or (c)  in the exterior layers of the supernova shock heated material. Such stellar temperatures can induce the {\it hot CNO cycles} which operate on a timescale of only few seconds, as shown in Figure \ref{fig:CNOcycle} (left).  In the hot CNO hydrogen burning is constrained by the $\beta$-decay lifetimes of the participating proton-rich nuclei, e.g., $^{14}$O and $^{15}$O. At $T \geq 5 \times 10^8$ K, material from the CNO cycle can leak and lead to the formation of heavier nuclei by means of the {\it rapid proton capture}, or  rp-process.

The rp-process follows a path in the nuclear chart analogous to the {\it r-process} due to neutron capture. It converts  CNO nuclei into proton-rich isotopes near the  {\it the proton drip line}. Following a succession of fast proton captures, a mass number $A$ is reached when another proton capture must wait until  the occurrence of $\beta^+$-decay. The rp-process rate is hindered as the charge and mass numbers increase due to the increasing Coulomb barrier. Therefore, the rp-process does not really reach the proton drip line, being more effective by running near the beta-stability valley where the $\beta^+$-decay rates are of similar magnitude as the proton capture rates. Fig. \ref{fig:CNOcycle} (right) shows schematically the paths for occurrence of the rp- and the r-processes in the nuclear chart.

\subsection{Triple-$\alpha$ capture}

When the core temperature and density in a star reach $T_6 = 100-200$ and $\rho_c = 10^2 - 10^5$ gm  cm$^{-3}$, it starts burning $^4$He steadily.  The most probable path to heavier elements is the fusion of three $\alpha$ particles into $^{12}$C \cite{Opi51,Sal52,Sal57}. The reaction occurs in two steps: (a)  first by the fusion of two $\alpha$ particles into $^8$Be, which is unstable to decay into two $\alpha$ particles, (b) followed by the fusion of $^8$Be with another $\alpha$-particle into $^{12}$C as displayed in Figure \ref{R_matrix}, left. $^8$Be has a very short lifetime of $10^{-16}$ s (with a decay width, $\Gamma = 6.8$ eV), nonetheless long compared to  $10^{-19}$ s, the time  that two $\alpha$-particles stay close to each other. The Q-value of the reaction is $Q= -92.1$  keV. The reaction $\alpha +\ ^8{\rm Be} \rightarrow \  ^{12}{\rm C}$ has therefore time to occur before $^8$Be decays, for a sufficiently large density of $^8$Be. At lower temperatures $T_6=10$, the $\alpha$-particle energies are not high enough to produce the $^8$Be($0^+_1$ ) resonance and the process is non-resonant. The {\it Saha Equation} for particle concentrations in the reaction $1 + 2 \rightarrow 3$ 
\begin{equation} n_{3}={n_1 n_2 \over 2} \left({2\pi
\over \mu kT}\right)^{3/2} \hbar^3 {(2J_3 +1) \over (2J_1+1) (2J_2 +1)} \exp \left(-{E \over k T}\right) \ , \label{chpt8:saha}\end{equation}
can be used to calculate the equilibrium concentration of $^8$Be at  $T_6 =100$ and density $\rho = 10^5$ gm  cm$^{-3}$. One obtains, using $n_{3}=n_{^{8}\rm Be}$, $n_1=n_2=n_\alpha$  one gets ${{\rm n}(^8{\rm Be})/ {\rm n}(^4{\rm He})} =  10^{-9}$.
At such concentrations, the amount of $^{12}$C produced  is not enough to explain the observed abundance of $^{12}$C, unless the reaction passes by a resonance \cite{Hoy53,Hoy54}. If one assumes an s-wave ($l=0$) resonance in $^{12}$C slightly above the threshold, the $^8{\rm Be} + \alpha$ reaction is greatly enhanced. Both $^8$Be and $^4$He  have $J^{\pi} = 0^+$, and an s-wave resonance would imply a $0^+$ resonance state in $^{12}$C. Fred Hoyle proposed that the excitation energy in $^{12}$C had to be $E_r \sim 7.68$ MeV \cite{Hoy53,Hoy54}. In fact, this state was discovered experimentally only a few years later \cite{Coo57}.  The state became known as the {\it Hoyle state}. It has a total width of $\Gamma = 9 $ eV \cite{RR88}, mostly due to $\alpha$-decay.  The $\gamma$-decay to the ground state of $^{12}$C is not allowed by angular momentum conservations, as all states have $J^{\pi} = 0^+$. The  $\gamma$-decay width is several thousand times smaller than the $\alpha$-decay width, i.e, $\Gamma = \Gamma_{\alpha} + \Gamma_{rad} \sim \Gamma_{\alpha}$. Experimentally, it was determined that  $\Gamma_{rad} = \Gamma_{\gamma} + \Gamma_{e^+e^-} = 3.7$ meV, dominated by the $\gamma$-decay width, $\Gamma_{\gamma} = 3.6$ meV. 

The triple-$\alpha$ reaction rate is given by
\begin{equation}r_{3\alpha} = \rho^2 N_A^2 Y_{^8{\rm Be}} Y_{\alpha} \left< \sigma v \right>_{^8{\rm Be} + \alpha}\ .\end{equation}
The cross section can be parametrized by a Breit-Wigner form for a resonance reaction. For a narrow resonance with width $\Gamma\ll E_{r}$, the Maxwellian exponent $\exp\left(  -E/kT\right)  $  in  Eq. \eqref{astrophys5} can be taken outside the integral, yielding
\begin{equation}
\left\langle \sigma v\right\rangle =\left(  \frac{2\pi}{\mu kT}\right)^{3/2}\hbar^{2}\left(  \omega\gamma\right)_{r}\exp\left(  -\frac{E_{r}} {kT}\right)  \ ,\label{svres}
\end{equation}
where $\left(  \omega\gamma\right)_{r}$ is the\textit{\ resonance strength}, given by
\begin{equation}
\left(  \omega\gamma\right)_{r}=\frac{2J_{r}+1}{(2J_{j}+1)(2J_{k} +1)}\ \left(  1+\delta_{jk}\right)  \ \frac{\Gamma_{\mathrm{r}}\ \Gamma
_{\gamma}}{\Gamma}.\label{svres2}%
\end{equation}
In the equation above, $\Gamma_r=\Gamma_{\alpha}$ dominates over $\Gamma_{\gamma}$ so that $(\Gamma_\alpha \Gamma_\gamma / \Gamma) \sim \Gamma_\gamma$. 

Using Saha's equilibrium equation \eqref{chpt8:saha} one finds the equilibrium concentration of $^8$Be,
\begin{equation}n(^8{\rm Be}) = n_{\alpha}^2 \left(  \omega\gamma\right)_{r} f {h^3 \over \left(2\pi \mu_{\alpha \alpha} kT\right)^{3/2}} \exp\left(-{E\over kT}\right)\ ,\end{equation}
where a factor $f$ was introduced to account for the enhancement of the cross section due to electron screening in  stellar environments \cite{BeGa10}.
The triple-alpha reaction rate is obtained by multiplying the equilibrium concentration of the resonant state in $^{12}$C  by the gamma-decay rate
$\Gamma_{\gamma}/\hbar$ leading to the ground state of $^{12}$C. That is,
\begin{equation}r_{3\alpha} = \rho^2 N_A^2 Y_{^8{\rm Be}} Y_{\alpha} \hbar^2
\bigg({2\pi \over \mu_{\alpha ^8{\rm Be}} kT}\bigg)^{3/2} \left(  \omega\gamma\right)_{r} f  \exp \left(-{E^{'}\over kT}\right) \ .
\end{equation}
Combining all equations in one, 
\begin{equation}r_{3\alpha \ \rightarrow \ ^{12}C} = \rho^3N_A^3{Y_{\alpha}^3 \over 2} 3^{3/2} \bigg({2\pi \hbar^2 \over M_{\alpha} kT}\bigg)^3 f  \omega{\Gamma_{\alpha} \Gamma_{\gamma} \over \Gamma \hbar} \exp \left(-{Q\over kT}\right) \ ,\end{equation}
where $\omega$ now denotes only the spin-related part of the Eq. \eqref{svres2}. The Q-value is the sum of $E'(^8{\rm Be} + \alpha)=E_r = 287 \rm \; keV$ and $E(\alpha +\alpha) = |Q| = 92 \; \rm keV$, i.e., $Q_{3\alpha} = (M_{^{12}{\rm C}^*} - 3 M_{\alpha})c^2 = 380$ keV.  The energy production rate is given by
\begin{equation}\epsilon_{3\alpha} = {r_{3\alpha} Q_{3\alpha} \over \rho} = 3.17 \times 10^{14}\ {\rho^2 X_{\alpha}^3 \over T_9^3} f \; \exp\left( - {4.4 \over T_9}\right) \rm \; MeV \; g^{-1} \; s^{-1}\ ,\end{equation}
leading to an estimate of energy production in red giants of 100 ergs/g/s at $T_8\sim 1$.
The triple-$\alpha$ reaction runs more efficiently a temperature value of $T_8 = 1$. Making a a Taylor series expansion around $T_8=1$ leads to
$ \epsilon_{3\alpha} \sim T _8^{41}$.
Hence, a small temperature increase leads to a much larger reaction rate and energy production. AGB stars experience thermal pulses when the He burning shell is suddenly activated at high temperatures and densities and the helium burning becomes explosive. When this occurs, for a stellar core under degenerate conditions, an explosive condition known as  {\it helium flash} develops.  The luminosity from this process can reach values much larger than $ 10^{11}L_\odot$. It is like the luminosity of a supernovae. However, this energy does not reach the surface as it is  absorbed by the expansion of the outer layers. As the flash continues, the core also looses its degeneracy and expands. Helium flash models are complicated and not very accurate \cite{JI11}. Maybe the ignition really occurs due to neutrino losses in the dense regions of the star. In dense regions at temperatures of $T > 10^8$ K, electron-electron collisions may produce neutrino anti-neutrino pairs, instead of photons. This process can cool the regions close to the center of the star. Modeling of helium flashes may also include  chemical mixing in the core  and asymmetries.

Due to the impossibility to measure this reaction directly, the triple-$\alpha$ reaction has been revisited in many theoretical works in recent years. In Ref. \cite{Oga09}  a very large reaction rate was obtained  compared with previous rate  estimates \cite{Nom95,Ang99,DP09}. Using these reaction rates in stellar evolution models at typical helium burning temperatures,  these results are incompatible with the observation \cite{DP09,Fyn05}. The abnormally large reaction rates have been questioned in subsequent calculations \cite{Gar11,Ngu12,Ngu13,Ishi13,Suz13,Hir15} which reported a much smaller rate, compatible with previously results \cite{Ang99}. Also recently,  nuclear lattice simulations, have allowed  first steps towards ab-initio calculations of the Hoyle state \cite{Epe11}. Using lattice QCD calculations the dependence of the triple-alpha process, the position of the $^8$Be ground state and the Hoyle state relative to the threshold upon the light quark masses and the electromagnetic fine structure constant  appears strongly correlated with the binding energy of the $\alpha$-particle. This puts a constraint on  $2\%$ in the adopted values of the quark masses and $ 2\%$ in $\alpha_{em}$. Such a fine-tuning has been linked to the so-called {\it Anthropic principle}. The ``principle" states that the values of all physical and cosmological quantities are restricted by the requirement that life exists in the Universe" \cite{Mei15}.

\subsection{Red giants and AGB stars}

After the hydrogen fuel in the core of a star is exhausted, the core contracts and its temperature rises. The outer layers of the star expand and cool down. The star luminosity greatly increases turning it to a  {\it red giant}. When its core temperature reaches about $3\times 10^8$ K, helium burning starts. The star's cooling stops and its luminosity is further increased. When helium burning completes the star has followed a path in the {\it Hertzprung-Russell} (HR) diagram first to the right, then to the left and then to the right again and up. Therefore, based on its path on the HR diagram, it has been named {\it asymptotic giant branch} (AGB) star. Stars with initial masses  $< 9 M_\odot$ reach the AGB phase in the final phases of their evolution. He burning turns from convective to radiative, but eventually switches off and the convective envelope penetrates the inner He inter-shell and brings to the surface the  He burning products, a phenomenon known as the {\it dredge up}. This process can lead to carbon and elements heavier than iron generated by the s-process.  Dredge up can occur a few hundreds of times depending on the stellar mass and the mass-loss rate as the star gravitationally contracts and heats up again. H burning resumes again until another flash occurs. AGB stars emits newly synthesized material into the interstellar medium in strong stellar winds,  eroding its envelope within a million years \cite{Her05}. With the formation of $^{12}$C via the triple-alpha capture, the following  $\alpha$-capture reaction can occur,
\begin{equation} ^{12}{\rm C} + \alpha \rightarrow ^{16}{\rm O} +
\gamma \ .\label{Careact}\end{equation}
For large reaction rates for this process, all the carbon will end up into oxygen. But, after hydrogen, helium and oxygen, carbon is the most abundant element in the Universe, and even the cosmic C/O ratio is of the order of 0.6 \cite{RR88}. In fact,  the main outcome of He burning in red giants are C and O. 
The reaction in Eq. \eqref{Careact} is very complicated one because of capture to resonances, non-resonant capture, and interfering sub-threshold resonances.  One believes that the S-factor is about 0.3 MeV b \cite{LK83,Str08}.   Using the same procedure as described in the triple-alpha reaction case, based on Saha's equation, one gets the reaction rate 
\begin{equation} \omega_{12C} = \left({n_\alpha \over 7.5 \times 10^{26}/\mathrm{cm}^3}\right)
\left(2.2 \times 10^{13}/\mathrm{s}\right) e^{-69/T_8^{1/3}} T_8^{-2/3}, \end{equation}
which has a very strong temperature dependence. 
For a red giant density of 10$^4$ g/cm$^3$ and an $\alpha$ fraction of 0.5, we get  lifetimes for  $^{12}$C  of $1.8 \times 10^9$ y for $T_8=1$ and $1.8 \times 10^3$ y for $T_8=2$. 

\begin{figure}[tb]
\begin{center}
\includegraphics[width=2.8in]{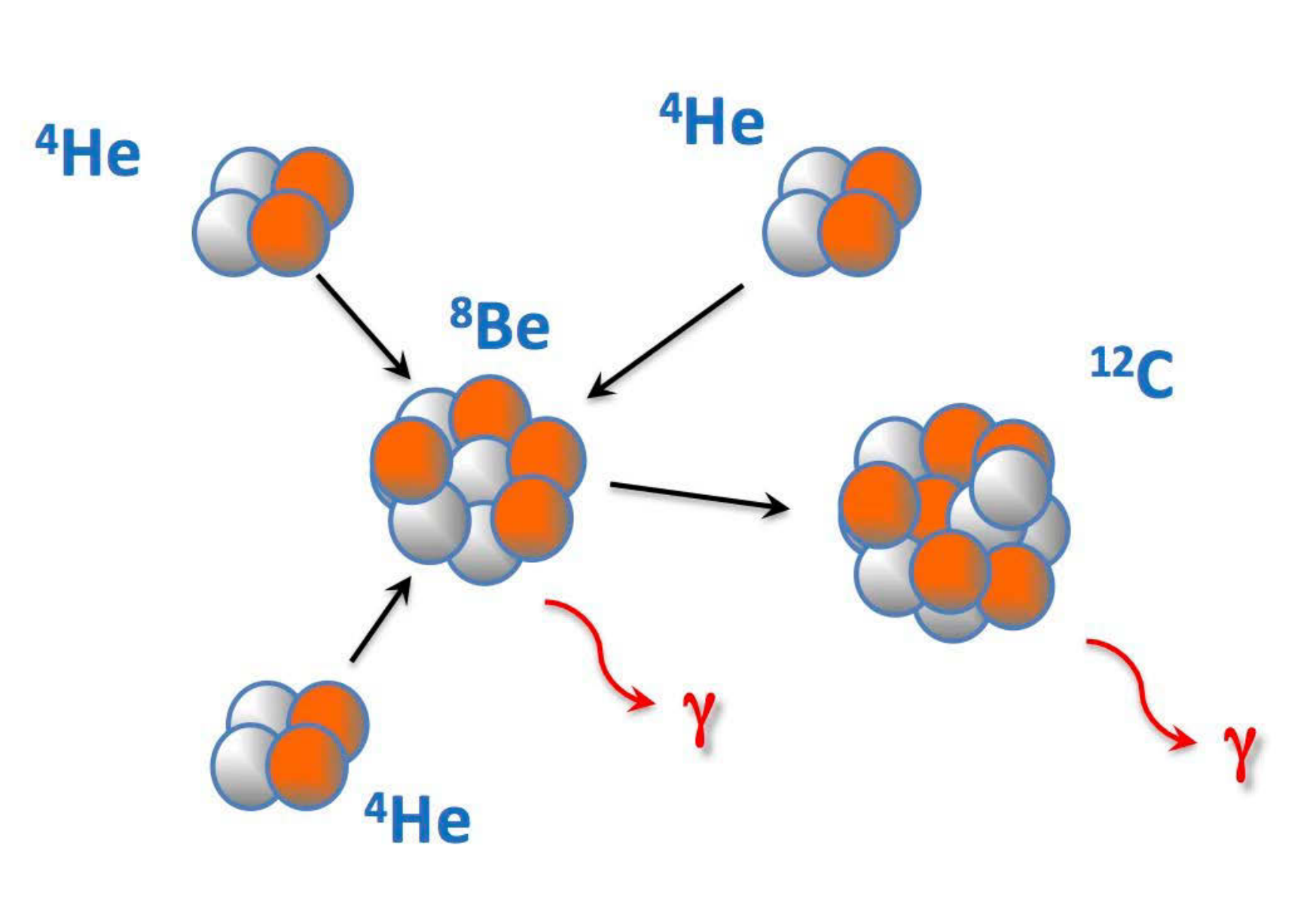}\ \
\includegraphics[width=3.5in]{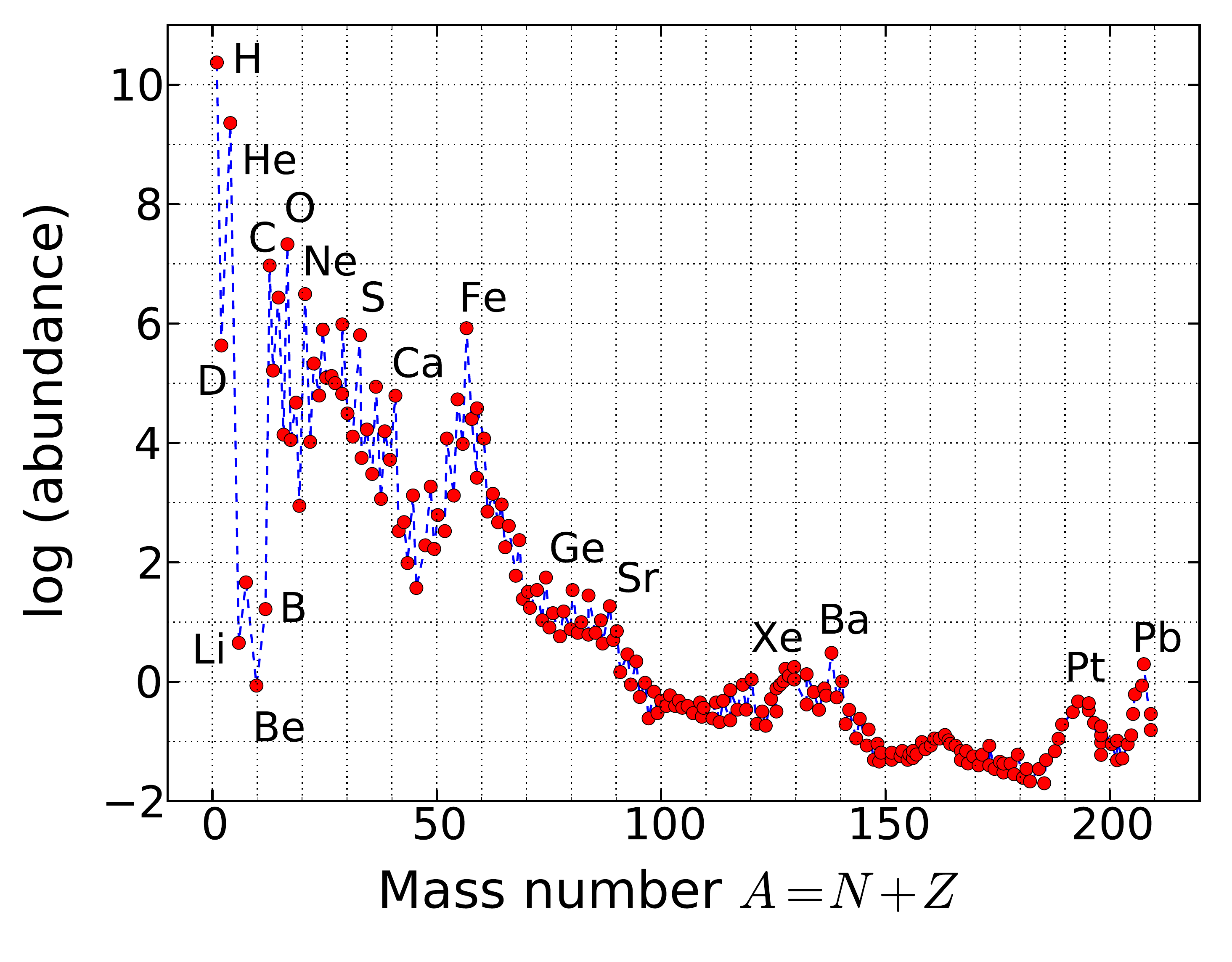}
\caption{\small {\it (Left):} Schematic view of the sequential triple-alpha process. {\it (Right):} Isotopic mass dependence of the relative abundances of elements in the solar system. The abundances have been normalized so that the abundance of silicon is $10^6$. The data have been collected from numerous sources (see, e.g., \cite{LPG09}).}
\label{R_matrix}
\end{center}
\end{figure}

At temperatures $T_9 = 0.1$ and above, the Gamow window is about $E_0 =0.3$ MeV. This energy region is on the  low energy tail of a broad resonance at $E_{c.m.} = 2.4$ MeV above the $\alpha+{\rm C}$ threshold. This is a $J^{\pi} = 1^-$ resonance, with a width of  400 keV, 9.6 MeV above the ground state of $^{16}$O. Two sub-threshold resonances exist in $^{16}$O at $E_x = 7.1$ and  6.9 MeV. These are $-45$ keV ($J^{\pi} =1^-$ ) and $-245$ keV ($J^{\pi} =2^+$) below the $\alpha$-particle threshold. Both of these sub-threshold resonances contribute to the reaction rate as their  tails extend to the Gamow window. Isospin selection rules suppress the electric dipole (E1) $\gamma$-emission from the 7.12 MeV state. Taking into account  the resonances and sub-threshold states leads to an S-factor of about $0.3 \; \rm MeV \; barn$. This is  not enough  to burn  $^{12}$C completely to $^{16}$O, and one obtains the ratio ${\rm C/O} \sim 0.1 $. Microscopically this is very hard to explain and one has to resorts to phenomenological descriptions such as the R-matrix formalism \cite{Ber07} to fit  the data and extrapolate them  to $E_0=300$ keV. This is shown in Figure \ref{R_matrix}. Direct measurements of this reaction are very difficult at the stellar energy of 300 keV region and  experiments show that the uncertainty of the rate of the $^{12}$C($\alpha,\gamma$)$^{16}$O reaction is  within a factor of two \cite{Cau85,Cau88,PD74,Kre88,Buch93,Buch93b,Buch93c,Zha93,Zha94,Bar95,Buch96}. An indirect  method has been used to determine this cross section, using the decay $^{16}$N($\beta^-$)$^{16}$O*$\longrightarrow \ ^{12}$C$+ \alpha$ \cite{Buch93,Buch93b,Buch93c,Zha93,Zha94}. It yields  gives $S(300) = 146 $ keV.b. It is often mentioned that if the cross section for this reaction is twice larger than the presently accepted value, a 25M$_\odot$ star will not produce enough $^{20}$Ne since carbon burning would cease. The impact of this result is enormous as an oxygen-rich star probably collapses into a black hole whereas carbon-rich stars would collapse into neutron stars \cite{Woo86}. More recently, efforts are being made to study the $^{12}$C($\alpha,\gamma$)$^{16}$O by means of the inverse process of  photo-dissociation \cite{Gai14} using new laser facilities such as the HI$\gamma$S facility  at Duke University or the proposed ELI \cite{ELI} and IZEST \cite{IZEST}. The burning of $^{16}$O via $^{16}{\rm O}(\alpha, \gamma)^{20}$Ne reaction is very slow during helium burning in red giant stars because of its very small cross section. The major ashes in red giants are therefore carbon and oxygen. The   $^{16}{\rm O}(\alpha, \gamma)^{20}$Ne reaction is  identified as the ``end-point" of the  reaction chain $^4$He(2$\alpha,\gamma)^{12}$C($\alpha,\gamma)^{16}$O($\alpha,\gamma)^{20}$Ne. The reaction rate is very small because there are no resonances in the Gamow window, around $E_0= 300$ keV \cite{Ale72,Kar83,Hah87,Moh05}. The reaction is dominated by non-resonant direct radiative capture, but  at higher energies resonant capture through several low energy resonances may occur. The contribution of the direct capture is not known \cite{Ale72,Kar83}.

\section{Carbon, neon, oxygen, and silicon burning}\label{COSburn}

When helium burning ceases the core of the stars becomes a mixture so C and O.  Since H and He are the most abundant elements,  C and O are formed  in appreciable large amounts \cite{Woo86}. Helium burning still continues in a layer surrounding the C and O rich core and hydrogen burning in another layer surrounding the helium burning shell. When helium burning ceases, there is not enough pressure to keep the star gravitationally stable and it begins to contract again. The temperature of the helium-poor core rises again \cite{Cla84} but the contraction continues until the next burning fuel becomes effective  or until electron degeneracy pressure stops the contraction.

\subsection{Carbon burning}

After helium burning,  stars with masses $M \geq 8-10M_{\odot}$ have their CO-rich core contract  until temperatures and densities reach $T \sim 5 \times 10^8 K$ and $\rho = 3 \times 10^6 \; \rm g cm^{-3}$ and carbon starts burning. Energy is released and the contraction stops, and a quiescent  carbon-burning continues. Two $^{12}$C nuclei can fuse, leading to a  compound nucleus of $^{24}$Mg nucleus with an excitation energy of 14 MeV. The Gamow window at such temperatures is about 1 MeV, a region with many excited levels in $^{24}$Mg \cite{Cla84}. The reaction rate for  $^{12}$C +  $^{12}$C is found to be  \cite{Ree59}
\begin{equation}\log (r) = \log (f) + 4.3 -
{36.55\left(1 + 0.1T_9\right)^{1/3}\over T_9^{1/3} } -{2\over3} \log (T_9) \ ,\end{equation}
where $f$ is the electron screening factor. The protons and alpha particles produced in the decay channels are quickly
burned by means of the reaction chain 
\begin{equation}^{12}{\rm C(p}, \gamma)^{13}{\rm N(e}^+ \nu_e)^{13}{\rm C} (\alpha, {\rm n})^{16}{\rm O}.
\end{equation}
Thus, the free proton is converted into a neutron  and the $\alpha$-particle fuse with $^{12}$C into $^{16}$O. We have discussed the reaction$^{12}{\rm C(p}, \gamma)^{13}{\rm N}$ when we referred to the CNO chain.  In AGB stars some protons diffuse from the convective envelope into the He inter-shell at the end of each dredge up and create $^{13}$C explaining the known abundances of the s-process elements at the surface of AGB stars \cite{Gal98,Bus99,GN00,Lug03,Cri09}.  A thin layer known as the $^{13}$C pocket is produced by means of the  $^{12}{\rm C(p}, \gamma)^{13}{\rm N(e}^+ \nu_e)^{13}{\rm C}$. When the temperature reaches $T_6\sim 100$  the $^{13}{\rm C} (\alpha, {\rm n})^{16}{\rm O}$ reaction starts generating the neutrons triggering the s-process \cite{Gal88,HI88,Kae90,Kae11}.  The cross-section is strongly influenced by a  subthreshold resonant tail contribution \cite{Hei08}. Experimental  measurements of this reaction  reached down to 270 keV, outside the Gamow window at $190 \pm 40$ keV for $T_6=100$. The kinetic energy is much below the Coulomb barrier, leading to an extremely small cross section \cite{Hei08,Sek67,Dav68,Bai73,Kel89,Dro93,BLK93,Har05}.

Carbon burning produces nuclei such as $^{16}$O, $^{20}$Ne, $^{24}$Mg and $^{28}$Si with  each pair of $^{12}$C  releasing on average about 13 MeV \cite{Ree59} of energy. After the completion of the carbon burning phase, reactions such as $^{12}{\rm C} + ^{16}$O and $^{12}{\rm C} + ^{20}$Ne will occur with reaction rates smaller than for $^{12}{\rm C}+^{12}$C  due to the larger Coulomb barriers. The star will lose most of tis energy during the carbon-burning and later stages, by neutrino emission, a process that is very sensitive to the core temperature. The neutrino luminosity soon becomes larger than the photon luminosity at the surface of the star. The timescale for energy production in the star by neutrino emission is very short compared to the timescale for gravitational cooling, given by $t_{th} = GM_s^2/R_s L_{s}$, where $M_s$, $R_s$ and $L_s$ are the mass, radii, and the star surface luminosity, respectively.  Because of the slow photon diffusion, there is not enough time for this energy to propagate to he surface,  and its temperature does not change considerably during and beyond the carbon burning phase.

\subsection{Neon and oxygen burning}

During the carbon burning stage mainly neon, sodium and magnesium are synthesized, with small traces of .aluminum and silicon due to the capture of $\alpha$, p and n released during that stage. After carbon is exhausted, the core contracts and its temperature raises again. When $T_9 \sim 1$, the high energy photons present in the tail of the Planck distribution begin to disintegrate $^{20}$Ne  through the reaction  $^{20}$Ne($\gamma$,$\alpha$)$^{16}$O. Since $\alpha$-decay processes release $\alpha$-particles  at nearly the same energy as of an emitted nucleon,  by time-inversion, one concludes that the most common photodisintegration reactions are $(\gamma, {\rm n}), (\gamma, {\rm p})$ and $(\gamma, \alpha)$.  The photonuclear rate passing by  an excited state $E_x$ is 
\begin{equation}r(\gamma, \alpha) = \left[ \exp \left(-{E_x \over kT}\right) {2J_R +1 \over 2J_0 + 1} {\Gamma_{\gamma} \over \Gamma}
\right] {\Gamma_{\alpha} \over \hbar} \ ,\end{equation}
where the term in brackets is the probability of finding the nucleus in the excited state $E_x$  with spin $J_R$ ($J_0$ is the ground state spin) and the 
 decay rate to alpha particle emission is given by $\Gamma_{\alpha} / \hbar$. Using $E_x = E_R + Q$, 
\begin{equation}r(\gamma, \alpha) = {\exp(- Q/kT) \over
\hbar (2J_0 +1)} (2J_R +1) {\Gamma_{\alpha} \Gamma_{\gamma} \over \Gamma} \exp(-E_R / kT).
\end{equation}
The 5.63 MeV level in $^{20}$Ne  dominates the photodissociation process for $T_9 \geq 1$. When $T_9 \sim 1.5$, the photodissociation rate is greater than the alpha capture rate on $^{16}$O leading to $^{20}$Ne, which is the reverse reaction \cite{Moh00,Vog01,Vog02,Arn04,Moh06}.  The released $\alpha$-particle then also begins to react with $^{20}$Ne by means of  $^{20}$Ne($\alpha$,$\gamma$)$^{24}$Mg.
The net result is the conversion of  two $^{20}$Ne nuclei in the form  $^{20}$Ne($^{20}$Ne,$^{16}$O)$^{24}$Mg  with $Q= 4.58$ MeV. The neon burning stage is very quick and yields a core with a mixture of $^4$He, $^{16}$O and $^{24}$Mg nuclei. At the end of Ne burning, the core contracts and the temperature reaches $T_9 \sim 2$. Then $^{16}$O begins to react by means of $^{16}$O($^{16}$O,$\alpha$)$^{28}$Si    
and $^{16}$O($^{16}$O,$\gamma$)$^{32}$Si   each occurring approximately $50 \%$  of the time. The oxygen burning phase also produces, Ar, Ca and traces  of Cl, K, $\cdots$, up to Sc. 

\subsection{Silicon burning}

At $T_9 \sim 3$, the produced $^{28}$Si produced during the oxygen burning phase begins to burn in  the Si burning phase. During the neon burning phase, not only heavier nuclei are produced, but  photons are sufficiently energetic to dissociate neon, before the temperature is high enough to
 ignite reactions involving oxygen. Thus $\alpha$-particles are produced in the neon burning phase, a trend which continues in the silicon burning phase. In general,  photodissociation of the nuclear products of neon and oxygen burning continues when the temperature surpasses $T_9 \geq 3$ because of the high energy photons in the tail of Planck's distribution. In fact, photodissociation destroys nuclei with small binding energies  and leads to many nuclear reactions involving protons, neutrons and $\alpha$-particles with  nuclei in the mass range $A= 28-65$ such as
\begin{equation}
\gamma+\ \ _{14}^{28}\mathrm{Si}\longrightarrow\ \ _{12}^{24}\mathrm{Mg}%
+\ _{2}^{4}\mathrm{He},\ \ \ \ \ _{2}^{4}\mathrm{He}+\ \ _{14}^{28}%
\mathrm{Si}\longrightarrow\ \ _{16}^{32}\mathrm{S}+\ \gamma,\ \mathrm{etc.}%
\label{astrophys26}%
\end{equation}
The large number of free neutrons leads to several (n,$\gamma$)-reactions, i.e., to {\it radiative neutron capture}. Elements with the largest binding energies in the iron mass region are produced abundantly. $^{56}$Fe has the maximum value for binding energy per nucleon, leading to a full stop of the fusion reactions around the iron-group.

The compound nuclear levels in the nuclei with $A= 28-65$ formed during silicon burning are dense and overlapping. At the high temperatures of $T_9 = 3-5$, a quasi-equilibrium is reached of reactions occurring backward and forward between the nuclei in this mass region. Among these there are a few slow reactions known as {\it bottlenecks}. As the the nuclear fuel is consumed, the thermal energy decreases by neutrinos escaping from the star, and many nuclear reactions stop occurring leading to a {\it freeze-out} of the nuclear burning process.

\section{Slow and rapid nucleon capture}\label{slowrapid}

Due to the large Coulomb barriers, the abundance of heavy elements above the iron peak would drop very much if they were to be synthesized by reactions during the silicon burning phase \cite{Sue56} and would require larger temperatures. Higher temperatures also increase photon energies in the Planck distribution, and eventually  the photodisintegrations of nuclei will dominate. Therefore, a statistical chemical equilibrium will be reached during silicon burning and a maximum abundance of nuclei will be reached at the highest binding energies.  This effect can only be circumvented with high densities, like in the rp-process on accreting neutron stars, where the high densities still favor capture reactions, acting against
photodisintegrations. However, the study of reaction chains used in stellar evolution show that large neutron fluxes can be produced at the stellar core.   Electron capture and beta decay reactions influence the reaction flow by changing the value of  $Y_e$  (the number of electrons per nucleon) changing the stellar core density and entropy not only in the silicon burning phase, but also in earlier oxygen burning phases. Fusion reactions involving charged particles cannot explain the formation of nuclei with $A>100$. Such heavy nuclei are formed by the successive capture of slow neutrons followed by $\beta^{-}$-decay \cite{Bu57}. This process can explain the maxima of the observed elemental abundance for $N=50,$ $82,$ 126 which are due to the small capture cross sections for nuclei with magic numbers, leading to  a peak of nuclear isotopes with these neutron values.

The speed of the process to produce heavy elements is determined by the timescales of the nuclear transformations encountered along the process path. In case of an only small neutron density available neutron-captures are slower than beta-decays, in the other extreme of very high neutron densities beta-decays are slower than neutron captures. Therefore, the  s-process nucleosynthesis runs very close to  the valley of $\beta$-stability. On the other hand, rapid neutron capture, or {\it r-processes} can only occur when $\tau_n \ll \tau_{\beta}$. It requires extremely neutron-rich environments, because the capture timescale is  inversely proportional to the  neutron density. The r-process path includes very neutron rich and unstable nuclei, far from the valley of stability. Nuclear half-lives are modified in the stellar environment and  excited states can also be thermally populated.

\subsection{The s-process}

The s-process path occurs along the valley of stability and avoids some nuclei because, when $N$ and $Z$ are even, ($N,Z$) and ($N+2,Z-2$)  are stable to $\beta$ decay, whereas the neighboring odd-odd nuclei ($N+1,Z-1$) are unstable.  These odd-odd nuclei have an unpaired proton and also an unpaired neutron, and have therefore a rather large ground state energy.  The production of the ($N+2,Z-2$) nucleus is therefore suppressed in the s-process. The observation of such isotopes with non-negligible abundances is a hint that other processes than the  s-process might occur.  The reaction chain along the s-process process is of the form
\begin{equation} {dn_A(t) \over dt} = n_n(t)n_{A-1}(t) \langle \sigma v \rangle_{A-1}
- n_n(t)n_A(t) \langle \sigma v \rangle_A
-\lambda_\beta(t)n_A(t) ,
\end{equation}
where $n_n(t)$ is the neutron density and $\lambda_\beta$ is the $\beta$ decay rate.  The first term accounts for neutron capture to form nucleus $A$, the second and the third are related to its destruction by neutron-induced reactions and by $\beta$-decay.  A simple estimate can be applied for the s-process by assuming a slow $\beta$-decay rate, (i.e., the last term above)  and also by assuming that $ \langle \sigma v \rangle = \sigma_A \langle v \rangle $.
Then one finds
\begin{equation} {dn_A(t) \over dt} = \langle v \rangle n_n(t) \left(\sigma_{A-1}
n_{A-1} - \sigma_A n_A \right), \end{equation}
and, at equilibrium, $dn/dt=0$, yielding
\begin{equation} {n_A \over n_{A-1}} = {\sigma_{A-1} \over \sigma_A}. \end{equation}
The abundance for mass $A$ is inversely proportional to the neutron cross section to $A$. If the cross section is small, mass $A$ will be produced abundantly.  A similar argumentation also works if  $\beta$ decay is faster than neutron capture because the  small capture cross sections at the closed shells will also result in mass peaks in that region, compatible with the observations.
Equilibrium will also occur more quickly in the plateaus between the magic numbers because mass will pile up around the closed shells before they can be surpassed. Calculations suggest that a probable site for the s-process is the helium-burning shell of a red giant \cite{Arl99,Gal99},  with sufficiently high temperatures to produce neutrons by means of $^{22}$Ne($\alpha,n)^{25}$Mg, where the $^{22}$Ne nucleus is produced during the helium burning of the $^{14}$N ashes in the CNO cycle  \cite{Ang99,Jae01,LIK12}. This occurs via the reaction chain $^{14}$N($\alpha,\gamma$)$^{18}$F($\beta^+\nu$)$^{18}$O($\alpha,\gamma$)$^{22}$Ne \cite{Kae94}. Both $^{22}$Ne($\alpha$,n)$^{25}$Mg and the competing $^{22}$Ne($\alpha,\gamma$)$^{26}$Mg reactions are poorly known. The s-process is ineffective beyond $^{209}$Bi because it leads to a decay chain that ends up with $\alpha$ emission when the neutron is captured in this isotope \cite{Aok01}.  That is the end product of the s-process and  {\it transuranic elements} must be produced in some other way. 

\subsection{The r-process}

Nuclei with mass up to about $A=90$ above the iron group  are mainly produced by the s-process in  massive stars. For $A \ge 100$ the contribution of the s-process is very small in massive stars and one believes that most of the s-process contribution in this mass range  occurs in AGB stars. The competing process, i.e., the r-process, is very fast and lasts for a few seconds in dense neutron environments with neutron densities of $n_n^{\rm r\ process} \sim 10^{20} - 10^{25}$ cm$^{-3}$, much larger than those for the s-process,  $n_n^{\rm s \ process} \sim 10^8$ cm$^{-3}$.  
During compression of electrons in a core-collapse supernova, $\beta^-$ decay process in nuclei are Pauli-blocked due to the Fermi energy of the environment electrons being larger than the energy in the beta decay. However, electrons can still be captured by nuclei, yielding a higher neutronized matter and a large flux of  neutrons and higher temperatures. Neutron capture on heavy nuclei is faster than $\beta^-$ decays when the matter expands and cools down again. The r-process produces highly unstable  neutron rich  nuclei, and its path runs  close to the {\it neutron-drip line}   (see Figure \ref{fig:CNOcycle}, right). More than 90\% of elements such as europium (Eu), gold (Au), and platinum (Pt) in the solar system are believed to be synthesized in the r-process  \cite{Bur00}.

If the radiative capture reaction $ A + n \rightarrow (A+1) + \gamma $ is {\it exothermic}, and assuming a resonant reaction in a high level density nucleus, and further neglecting spins, Eq. \eqref{svres} leads to
\begin{equation} \langle \sigma v \rangle_{(n,\gamma)} = \left( {2 \pi \over \mu kT} \right)^{3/2} {\Gamma_n \Gamma_\gamma
\over \Gamma} e^{-E_R/kT} .\end{equation}
But the resonance energy is very close to zero, i.e.,  $E_R \sim 0$ and the rate is simply given by
\begin{equation} r_{(n,\gamma)} \sim n_n n_A \left( {2 \pi \over \mu kT} \right)^{3/2}
{\Gamma_n \Gamma_\gamma \over \Gamma}. \end{equation}

The high-energy tail of the Planck distribution for thermal photons (with $\hbar=c=1$)  is given by $ n(E_\gamma)\sim (E_\gamma^2/n_\gamma \pi^2) e^{-E_\gamma/kT}  $,
where $ n_\gamma \sim ({\pi / 13}) (kT)^3 $, is the approximate value of the total photon density  obtained by integrating Planck's distribution over all energies. For a resonant reaction, and using the asymptotic form of the Maxwell-Boltzmann distribution in the reaction rate integral, we get
\begin{equation} r_{(\gamma,n)} \sim 2 n_{A-1} {\Gamma_\gamma \Gamma_n
\over \Gamma} e^{-E_R/kT}. \end{equation}
In equilibrium, i.e., when the $(n,\gamma)$ and $(\gamma,n)$ rates are the same, and using $n_A \sim n_{A-1}$, leads to
\[ n_n \sim {2 \over (\hbar c)^3} \left( {\mu c^2 kT \over
2 \pi} \right)^{3/2} e^{-S_n/kT}, \]
where we assumed that $E_R$ is the same as neutron separation energy, $S_n$.  If we assume that the r-process works under the conditions of $n_n \sim 3 \times 10^{23}$/cm$^3$ and $T_9 \sim 1$, we obtain that $S_n\sim$ 2.4 MeV.  Hence,  neutrons are bound by about 30$kT$, a small value  compared to typical binding energies  of 8 MeV in a normal nucleus.  This means that contour line of constant neutron separation energy is close to the neutron drip-line.

Assuming that one knows under what conditions the r-process operates, the goal is to reproduce the relative abundance of elements observed in our Galaxy, shown in Figure \ref{R_matrix}, right. For $A<100$ it decreases exponentially with $A$, while for $A>100$ it is nearly constant, except for the peaks around the magic numbers $Z=50$  and $N=50,\ 82$, and  126. At the  shell closure gaps $N =82,  126$, the neutron number stays fixed because the nuclei cannot overcome the gap energy. They have to wait for the several $\beta$-decays to overcome the shell closures. Moreover, the $\beta$ decays are slow at the shell closures. That is why the closed neutron shells are called {\it waiting points}.  The abundance of an isotope is inversely proportional to the $\beta$ decay time, opposite to what occurs in the s-process, where it is proportional to the time scale of neutron capture (i.e., inversely proportional to the neutron-capture cross section).  Therefore, mass accumulates at the waiting points, yielding the peaks shown in Figure \ref{R_matrix}, right. The abundance peaks near  $A=130$ and $A=195$ are a signature of an environment of rapid neutron capture near closed nuclear shells. But calculations of the elemental abundances  just close to those peaks are often underproduced by one order of magnitude or more. This has been recently explained by means of fission fragments from the recycling of material in a neutron-rich environment such as that encountered in neutron star mergers \cite{Eich15,Gor15,Shi15}. 

When the r-process ends, the nuclei $\beta$-decay to the valley of stability.  Neutron spallation  moves the peaks to to a lower value in $A$. But the peaks are also shifted to lower values of $N$ because $\beta$-decay transforms neutrons into protons during the r-process. At the end of  the r-process  very heavy nuclei ($A \sim 270$) might be produced until $\beta$-delayed and neutron-induced fission occurs leading the nuclei back to $A \sim A_{max}/2$.  For the r-process to exist one needs very high densities and temperatures and reactions occurring at very short timescales,  e.g.,
$n_n \sim 10^{20}$ cm$^{-3}$, $T \sim 10^9$K, and $ t \sim 1$ s. But these are explosive conditions, which could occur (a) in the neutronized region above the proto-neutron star in Type II supernova,  (b) or in neutron-rich jets from supernovae or from neutron star mergers \cite{Nish06,Wint12,Most14,Nish15},  (c) during the  inhomogeneous Big Bang,  (d) in He/C zones in Type II supernovae, (e) in red giant He flashes, and (f) in neutrino spallation on neutrons in the He zone  \cite{Ter01,Kaj02,Ots03,Sum01}. Numerical calculations show that for the scenarios (d-f), the neutron density $n_n$ are lower than those for the scenarios (a-c). The critical role of light neutron-rich nuclei in the r-process rucleosynthesis in supernovae has been studied in Ref. \cite{Ter01,Kaj02,Sas05,Sas05b}. Most of the neutron-capture reaction rates rely on statistical model predictions  \cite{Gor08} with the need for detailed nuclear structure parameters \cite{Bea14}. Because direct measurements with neutrons are very difficult, experimentally one has resorted to the inverse photodisintegration reaction \cite{BBR86,Ueb14} or (d, p) neutron transfer \cite{Esc07}.

One possible scenario for the occurrence of the r-process are neutron stars mergers.   In fact, recent calculations suggest that heavy elements with $A >110$ are difficult to be produced in core-collapse supernovae due to a not sufficiently neutron-rich environment \cite{Wan11,Wan13}. Neutron star mergers take a long time to merge (about 100 Myr) at a low rate. They occur in  neutron star  binary systems that lose their energy very slowly due to gravitational radiation.  It has been shown in several calculations that neutron star mergers might be responsible for the full range of nuclei produced by the r-process 
\cite{LS74,LS76,Lat77,Sym82,Eic89,Mey89,FRT99,GBJ11,Kor12,BGJ13,Ros14,Wan14,Yut15}. There is a strong debate going on in the literature to define if the site of the r-process elements is located in supernovae or in  binary neutron stars mergers.  Recent progress on understanding the r-process has been reported in Refs. \cite{Yut15,Lor15,Eich15,Gor15,Shi15,Voo15,She15,Ces15,Wie15}.

Long-lived radioactive nuclei are often utilized as {\it cosmo-chronometers}.  They are useful for investigating the nucleosynthesis process history of the cosmic evolution before the  formation of the solar system, an idea proposed by Rutherford about 70 years ago \cite{Rut29}. Only a handful of cosmo-chronometers with half-live values useful for dating the cosmic age, $1-100$ Gyr, are known, e.g., $^{40}$K, $^{176}$Lu, $^{187}$Re, $^{232}$Th and $^{238}$U. The pair of nuclei $^{187}$Re-$^{187}$Os, was also proposed as cosmo-chronometer to investigate the galactic history of the r-process \cite{Clay64}. $^{187}$Re predominantly decays by $\beta^-$ after the freezeout of the r-process. The s-process nuclei $^{186,187}$Os are not  produced in the r-process but $^{186}$Os is produced in the s-process and its observed abundance normalizes the overall $^{186}$Os production.  The half-life for $^{187}$Re to decay into $^{187}$Os is  $4.35\times 10^{10}$ yr \cite{Lin86}, longer than the age of the Universe. After subtraction of the contributions of s-process, the epoch from an r-process nucleosynthesis event  can be determined by the presently observed abundances of $^{187}$Re  and $^{187}$Os.  The robustness of the $^{187}$Re-$^{187}$Os chronometer has confirmed in several publications (see, e.g., Ref. \cite{Hay05,Mos07}).

\section{Electron screening and pycnonuclear reactions}\label{elctscr}

Astrophysical environments are full with free, ionized, electrons due to the very  high densities and temperatures. The electron screening is very important as it increases the  reaction cross sections by lowering the Coulomb barrier. The reaction rates are thus accelerated by the electron screening effect in astrophysical plasmas. The reaction rate integral  \eqref{astrophys5} is then modified in the presence of electron screening, yielding approximately
\begin{equation}
 \left<\sigma\mathrm{v}\right>_{j,k}^{\ast}=f(Z_{j},Z_{k},\rho,T,Y_{i})
 \left<\sigma\mathrm{v}\right>_{j,k},  \label{sscreen}
\end{equation}
there the screening factor $f$ depends on charges of the reacting nuclei, density, temperature, and nuclear abundances, $Y_i$.

The screening effect can be calculated by assuming that locally the energy of the charges  in the presence of a Coulomb field $V(r)$ of an ion are obtained from the statistical Boltzmann distribution. One finds
\begin{equation} V(r)={Z_ie\over r}
\exp\left(-{r\over R_D}\right), \label{nucsys:deby}
\end{equation}
where $R_D$ is the {\it Debye radius}, given by 
\begin{equation}
    R_D^2={kT\over 4\pi e^2\sum_i Z_i^2c_{i0}} ,
\end{equation}
where $c_{i0}$ is the spatially uniform concentration of  either positive of negative charges in the plasma, if the screening effect would not be present. The {\it weak screening} corresponds to  $ Z_ieV(r)/ kT \ll 1$, so that the local charge density at a distance $r$ from then plasma is $ \rho(r)=-(e^2V(r)/kT) \sum_iZ_i^2 c_{i0}$. Therefore, the Coulomb potential is modified between the strong interaction radius $R$ and the classical turning point $R_0$, leading to a modification of the barrier penetrability. In the weak screening limit, $R_D \gg R, R_0$, and one can approximate $V(r)$ around $r=0$, leading to
$
V(r)=Z_1Z_2e^2/ r +U(r)
$,
where the Debye-Hueckel {\it screening potential},
$U(r)=U(0)=const.$, is given by
$
U_0=-Z_1Z_2e^2/ R_D
$.
due to the reduction of the barrier penetrability,  the astrophysical reaction rates are enhanced and a screening factor $f=
\exp\left({U_0/ kT}\right)$ appears in Eq. \eqref{sscreen}. In the weak screening limit, it is just $f \simeq 1-U_0/kT$,
or, including charge, density, and temperature dependence,
\begin{equation}
f=1+0.188{Z_1Z_2\rho^{1/2}\xi^{1/2}\over T_6^{3/2}},  \ \ \ {\rm
where} \ \ \xi =\sum_i (Z_i^2 +Z_i)^2 Y_i .
\end{equation}

It is very important to include the screening effect in reaction rate calculations. For example, in our Sun, the reaction $^7$Be(,$\gamma$)$^8$B, leads to the formation of $^8$B. Screening effects increase this reaction rate by about 20\%, thus affecting appreciably  the emission of high energy neutrinos through the $\beta$-decay of $^8$B.  At high densities and low temperatures, electron screening can lead to enhancements of nuclear reactions by  several orders of magnitude, leading to {\it pycnonuclear ignition}
\cite{Ga38,Wil40,SvH69}. Because of the low temperatures in some stars (e.g., WD), nuclei are arranged in lattice structures. The zero-point vibrations of such nuclei is responsible for the tunneling through a wide Coulomb barrier. Because barrier penetrability  is an exponential effect with energy, screening can strongly modify its value.   Pycnonuclear reactions depend little on the temperature  and, due to zero-point motion, they can occur even at $T = 0$.  They also increase rapidly with density, e.g.,  carbon fusion into heavy elements is strongly  enhanced  at $\rho \gtrsim 10^{10}$ g.cm$^{-3}$. Pycnonuclear thermonuclear reactions  happens in compact astrophysical objects and is regulated by the {\it plasma parameter} $\Gamma = Ze^2/ak T$, where $a$ is the distance between particles. Reactions occur in such conditions when
\begin{equation}
\Gamma \ll 1\ll E_0/kT,
\end{equation}
where $E_0$ is the Gamow peak energy. Screening effects and ion-ion correlations become important when $\Gamma\sim 1$. When $\Gamma\gtrsim 1$, one needs to use advanced statistical theories for the plasma \cite{IOT94}. One  can still use the approximate electron screening given by 
Eq. \eqref{nucsys:deby} if the Debye length is replaced with the inter-particle distance $a$. For $E_0 > k T$ the reaction rate correction 
due to electron screening, is about $\exp(\Gamma_e)$, with $\Gamma_e=V_C(a_e)/k T$ being the plasma parameter at the mean inter-electron distance $a_e$ ($V_C$ is the Coulomb ion-ion potential). If one includes ion-ion correlations, the reaction rates are enhanced by a factor  $\exp (\Gamma)$.  The screening corrections are large at low temperature, as one can show that $T$ (keV) = 0.02$Z_1Z_2[\rho$(g/cm$^3$)]$^{1/3}/\Gamma$.

Electron screening also play a fundamental role in obtaining the cross sections of interest at the low  energies \cite{Ass87,Rol95,Rol01}. Screening causes a modification of the Coulomb potential leading to an enhancement in cross sections that have been observed experimentally \cite {Eng88,Eng92,Ang93,Pre94,Gre95}. This effect can be calculated \cite{Ass87} using the adiabatic approximation \cite{Sho93}, in which one assumes that the electron cloud of the reacting nuclei adjust to the ground-state of a ``molecule'' consisting of the two nuclei as they approach each other. A {\it screening potential} is obtained  given by $U_e=B(Z_1+Z_2)-B(Z_1)-B(Z_2)$ where $B(Z_1+Z_2)$ and $B(Z_i)$ are the binding energies of the two joint charges of the nuclei and of the individual nuclei, respectively. Energy conservation demands that the relative energy of the ions increase by 
\begin{equation}
\sigma \;(E+U_e)= \exp \left[ \pi \eta (E)\frac{U_e}E\right] \
\sigma (E)\;,  \label{sig1}
\end{equation}
as they react. This leads to a modification of the astrophysical S-factor by an amount $f=\exp \left[ -2\pi \eta (E)\right] $, where $\eta=Z_1Z_2e^2/\hbar v$, where $v$ is the ion-ion  initial relative velocity. For light systems, the atomic electron velocities are comparable to that  of the projectile and a dynamical calculation is necessary \cite{Sho93}.  The so-called  {\it adiabatic approximation}, based on frozen ion-ion velocity is an {\it upper limit} of the enhancement of laboratory screening.

The cross section for the reaction $^3$He(d,p)$^4$He has been studied in Ref. \cite{Ali01}, yielding  $U_e = 219\pm 15$ eV, unexpectedly larger than the adiabatic limit value of $U_{ad} = 119$ eV. Theoretical attempts to solve this discrepancy have apparently failed  \cite{Sho93,BBH97,FZ99,HB02,Fio03}).  
The amplification of small effects have been considered  in Ref. \cite{BBH97}, and a solution based on the corrections to the electronic stopping power was proposed \cite{LSBR96,BFMH96} and further studied in Refs. \cite{BD00,Ber04}. The experimentally determined values of the screening potential for several reactions lead to an observed discrepancy with theoretical calculations by about one order of magnitude larger and larger screening values have been reported for liquid and solid targets \cite{Cze01}.  The screening problem is not well understood both theoretically and experimentally. It is not ruled out that the solution might be due to improper analysis of the experimental results \cite{Ade11}. Recently it has ben advocated that the electron screening problem in fixed target experiments might be due to polarization and alignment of cluster-like structures in light nuclei \cite{Spit06}.  

In the crust of neutron stars,  neutron-rich nuclei are  located on a lattice. If the lattice  spacing  shrinks enough due to the gravitational pressure of accreting material from a companion stars, pycnonuclear reactions can set in at a critical lattice spacing. This is thought to be a possible source of energy for $\gamma$-ray bursts in neutron stars. In neutron stars $\Gamma$ can be very large, e.g., $\Gamma > 170$, in which case the ions are frozen in a crystalline lattice, with each of them oscillating with a  frequency $\omega$ about its equilibrium point. The oscillation frequency is found to be $\hbar \omega = (4\pi n Z^2e^2/3m)^{1/2}$, where $Z$, $m$ , and $n$ are,  the ion's charge, mass, and number density. When $\hbar \omega/k T \gtrsim 10$( the ion thermal energy),   the oscillation amplitude is small compared to the inter-particle distance $a\sim \rho ^{-1/3}$. Therefore, they can only interact with neighboring in the lattice.  The Coulomb penetrability is nearly independent of the temperature, being proportional $\exp(-\rho^{1/6})$, and the reaction rate is $\left< \sigma v\right>=10^7 \rho^{-0.6}_8 \exp (-260\rho^{-1/6}_8 )$ cm$^3$/s,
where $\rho _8$ is the density in units of $10^{-8}$ g/cm$^3$ \cite{IOT94,SvH69}.  Cold WD with a carbon core can suddenly release energy through such pycnonuclear reactions, when the core is compressed to densities of $10^9$ g/cm$^3$.  The reaction rate for carbon fusion increases rapidly as it it were at a very high temperature and low density, as in a normal star. Strong plasma screening in dense matter and its relation to thermonuclear nuclear burning in white dwarfs and neutron stars have been studied recently in Ref. \cite{KY14}.

The heat released by accretion in neutron stars ignites nuclear reactions in its surface and energy is radiated emitted by photons, being unable to heat up the star interior. But, due to its own weight, the accreted material  sinks into the star crust increasing its density and igniting $\beta$-capture, neutron absorption and emission, and pycnonuclear reactions. Highly exotic nuclei,  stable in dense matter ($\rho \gtrsim 10^9$ g cm$^{-3}$), are formed.  At higher densities,  $\rho \sim 10^{12} - 10^{13}$ g cm$^{-3}$,  hundreds of meter below the neutron star surface, pycnonuclear reactions develop, increasing the crust heating by thermal conductivity and warming up the whole neutron star. For iron nuclei at the crust, at $\rho\sim 10^9$ g cm$^{-3}$, reactions of the type $^{34}{\rm Ne}(^{34}{\rm Ne},\gamma)^{68}$Ca; $^{36}{\rm Ne}(^{36}{\rm Ne},\gamma)^{72}$Ca; and $^{48}{\rm Mg}(^{48}{\rm Mg},\gamma)^{96}$Cr begin to ignite and release about $ 1.5$ MeV per accreted nucleon \cite{Sch90,Kit00}.  The whole  power generated is thus  determined by the {\it mass accretion rate}, $dM/dt$. One thus estimates that the luminosity due to such process is given by
\begin{equation}
L\simeq 1.5 \ {\rm MeV} \left( {dM/dt\over m_N}\right) \approx 8.8 \times 10^{33} \left( {dM/dt\over 10^{-10}M_\odot \ {\rm y}^{-1}}\right)
 \ {\rm erg} \ {\rm s}^{-1},
\end{equation}
with $m_N$ being the nucleon mass. 

\section{Nucleosynthesis in cataclysmic events}\label{cataclysm}

\subsection{Novae}
{\it Novae} (or ``new") stars increase  their brightness by a factor of a million before their visual disappearance.  The process lasts for a few days only, reaching peak luminosities at $L = 10^4 - 10^5L_\odot$ and ejected energies of $10^{45}$ ergs,  and takes a few months to decrease. More than half of all stars are believed to be part of a binary system, consisting of a WD star  orbiting  around an orange/red dwarf or a red giant \cite{Wal54,Kra64}.  The fuel originating the outbursts by the WD are gases falling from the larger star.  The novae events leave the participating stars almost intact and the phenomenon can recur within $10^4-10^5$ yr. {\it Supernovae}, on the other hand,  are one-time events leading to the total destruction of the star.  Novae occur by a few dozen a year in our galaxy.  Observations show that nova ejecta are rich in  He, C, N, O, Ne, and Mg \cite{WT76}, but they only contribute by 1/50 as much intergalactic material  as supernovae, and only by 1/200 as much as red giant and supergiant stars. In novae,  the matter accretes from the binary companion and accumulates at the surface of the WD, building up a hydrogen layer, below which is the main components of the WD, mainly of carbon and oxygen. The accreting material leads to an increase in pressure and temperature and the WD surface grows, eventually becoming hot enough to burn hydrogen into helium. When the mass accreted  reaches $M_\odot/100,000$,  a nuclear explosion starts at the base of the accreted material. The surface layer is ejected at   $\sim 1000$ km/s or greater.  The WD is left intact underneath the explosion and the cycle can repeat again as long as the companion star can provide fresh hydrogen-rich matter.  In classical novae, the triggering reaction is $^{12}$C(p,$\gamma$)$^{13}$N leading to the reactions $^{13}$N($\beta^+$)$^{13}$C(p,$\gamma$)$^{14}$N  with is part of the CNO cycle. As  the temperature  increases and $\tau_{{\rm p},\gamma}[^{13}N] < \tau_{\beta^+}[^{13}N]$ the reactions  $^{13}$N(p,$\gamma$)$^{14}$O  and $^{16}$O(p,$\gamma$)$^{17}$F,  and $^{14}$N(p,$\gamma$)$^{15}$O occur. The reactions $^{18}$F(p,$\alpha$)$^{15}$O, $^{25}$Al(p,$\gamma$)$^{26}$Si, $^{30}$P(p,$\gamma$)$^{31}$S have been identified as major sources of nuclear uncertainties \cite{Ili02}. The endpoint for classical nova nucleosynthesis, as predicted by theory  and obtained in observations, is around the  Ca isotopes.

\subsection{X-ray bursters}
\begin{figure}[t]
\begin{center}
\includegraphics[width=3.5in]{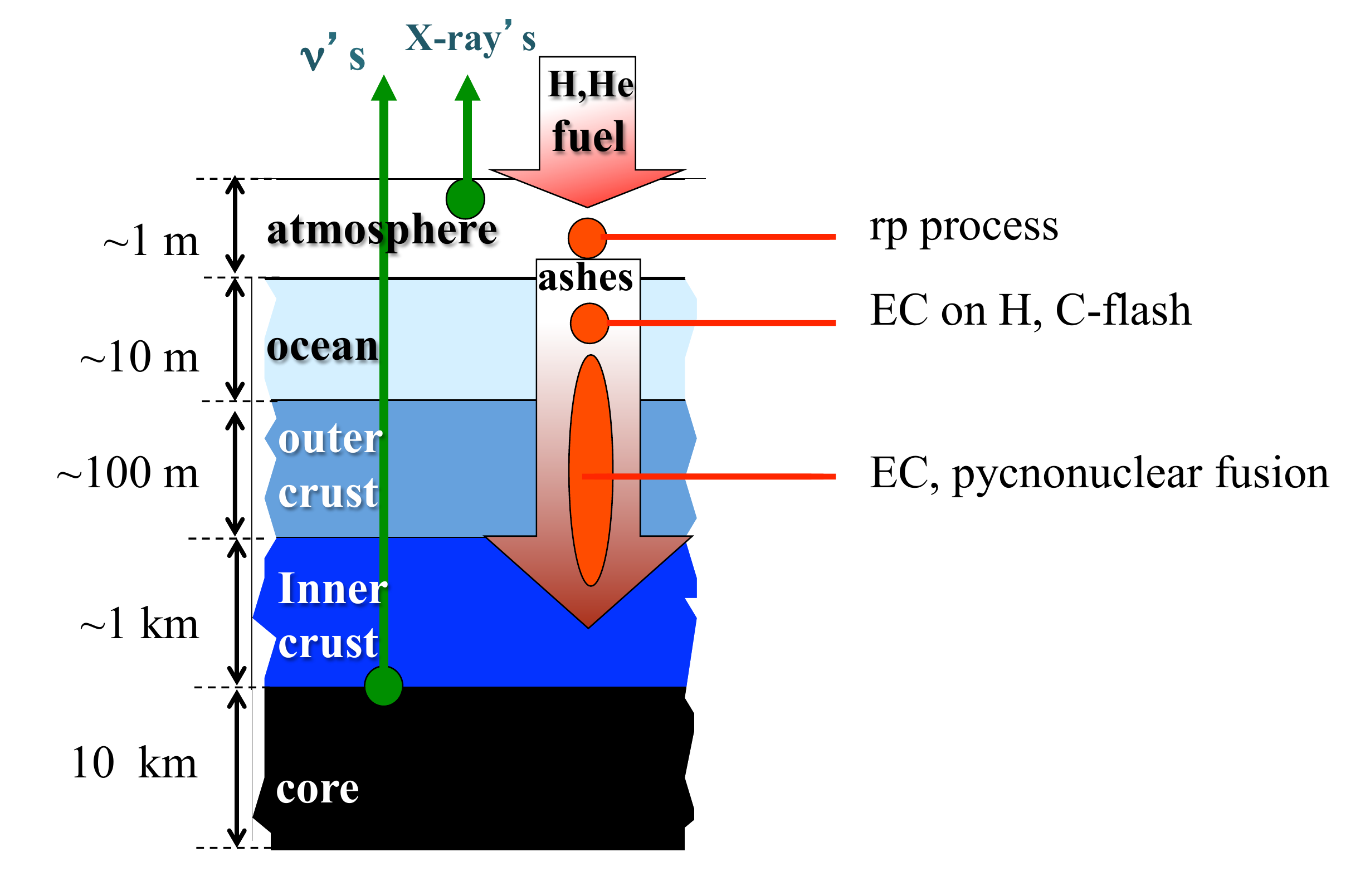}
\includegraphics[width=2.8in]
{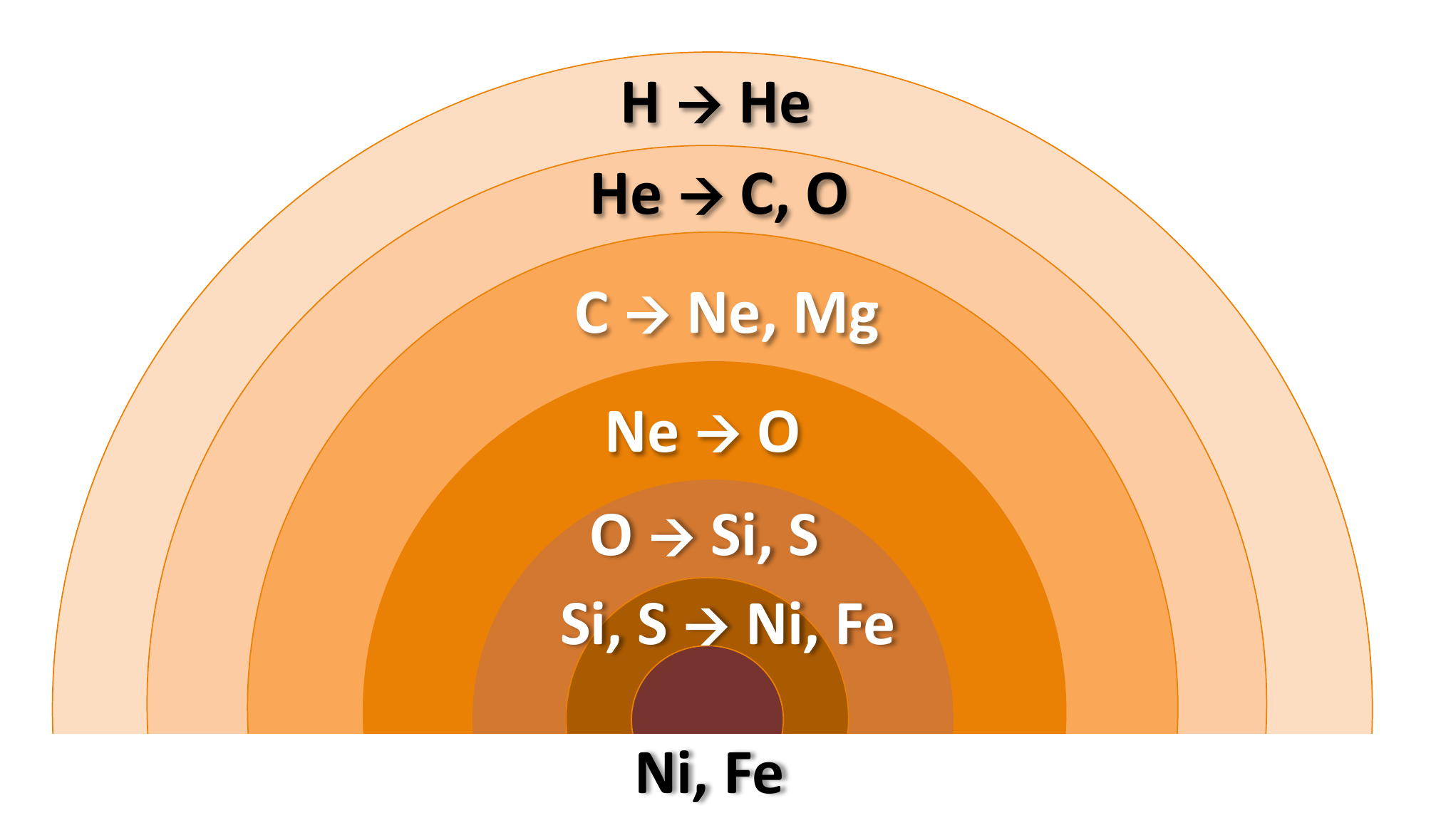}\caption{\small {\it (Left):} The accreted matter on a neutron star leads to X-ray bursts and a runaway  reaction chain involving  H, He, and C capture, electron capture (EC) and also pycnonuclear reactions in the deeper crust of the NS.  (Courtesy of H. Schatz). {\it (Right:)} Schematic view of the \textquotedblleft onion-like\textquotedblright\ structure of a pre-supernova
20M$_{\odot}$ star. }%
\label{xburst}%
\end{center}
\end{figure}

{\it X-ray bursts} are also a common phenomenon besides novae and supernovae \cite{Grin76}. In  X-ray binary systems, the compact object is not a WD but either a {\it neutron star} (NS) or a {\it black hole}. The larger gravitational field of the compact stars gives rise to large velocities of the accreting gas  rich in hydrogen and helium (Figure \ref{xburst}, left). The falling material collides with the already accreted material, leading to the formation of an accretion disk and releasing a humongous amount energy within a short time ($\sim 10^{38}$ erg s$^{-1}$ within 1 to 10 s). The energy is mostly released as X-ray photons. The accreted material forms a dense layer of degenerate electron gas at the surface of the star, due to its strong gravitational field. Changes in temperature do not yield appreciable changes in pressure in a degenerate electron gas, but after enough material accumulates on the surface of the star, thermal instabilities occur triggering  nuclear fusion reactions, further increasing its temperature, now greater than $10^9$ K, and finally enabling a runaway thermonuclear explosion. Nucleosynthesis starts with the hot CNO cycle and quickly becomes an {\it rp-process}, i.e. a succession of rapid proton captures before the nuclei are able to $\beta$-decay by $e^+$ emission.

X-ray bursts can also be recurrent (from hours to days) because they  are not powerful enough to disrupt either the binary system orbit or the stars. The recurring bursts occur at irregular periods, from few hours to several months \cite{ZKC14}. Modeling of X-ray bursts and the subproducts of the nuclear reaction networks have been published by several authors   \cite{WT76,WW81,Han83,Taa93,CW92,Taa96,Sch98,Koi99,Sch99,Sch01,Zin01,Woo04,Koi04,SR06,FST08,Elo09,Jos10,Mal11}. They involve  rp-process, the 3$\alpha$-reaction,  and $\alpha$p-process (i.e.,  ($\alpha$,p) combined with (p,$\gamma$) reactions). In some of these modeling several hundreds of nuclei are included in the reaction network. It is not easy to identify the reactions of most relevance \cite{Ili99,Thi01,Amt06,Cyb10,Par08,Par09}, although some reactions such as $^{65}$As(p,$\gamma$)$^{66}$Se, $^{61}$Ga(p,$\gamma$)$^{62}$Ge, $^{12}$C($\alpha$,$\gamma$)$^{16}$O and $^{96}$Ag(p,$\gamma$)$^{97}$Cd have been identified as key reactions along with about 50 other reactions out of 7500 included in the calculations \cite{JI11}.  These reactions as well as the masses of several nuclei would have a strong influence on the predicted yields. More recently attention has been given to new phenomena called by X-ray superbursts \cite{Cor00,CB01,Wij01,Kuu02,Kuu04,Cum05}  which possibly result from thermonuclear runaways occurring in deeper layers at densities greater than $10^9$ g cm$^{-3}$ \cite{CB01}.

\subsection{Supernovae}\label{nucs:sn}

 {\it Type I Supernovae} (SNI) are thermonuclear stellar explosions triggered by material accreted into a massive carbon-oxygen  WD in a binary system. They play an important part in astronomy as ``{\it standard candles}", compared to which luminosities of other stars are  used to infer cosmic distances, because all are thought to blaze with equal brightness at their peaks. Hundreds of nuclear isotopes and thousands of nuclear reactions participate in the process,  at densities of up to $\rho \sim 10^{10}$ g/cm$^3$  and temperatures up to $T \sim 10^{10}$ K. The explosion can lead to  turbulent instabilities and energy deflagration during the expansion of a SNI. Detailed 3D hydrodynamic modeling is necessary and accurate nuclear physics input must be provided. 

 {\it Type II Supernovae} (SNII) have hydrogen in their spectra and  their light curves  peak  around $10^{42}$ erg s$^{-1}$. In contrast, SNI do not show hydrogen in their spectra. If their spectra contain some silicon lines, they are called {\it Type Ia}. If the spectra contain helium lines they are called {\it Type Ib}. If they do not contain helium lines, they  are called {\it Type Ic}. The luminosities of  Type Ia supernovae can reach a peak value of about $2 \times 10^{43}$ erg s$^{-1}$. For more details on Type Ia supernovae, see, e.g., Ref. \cite{Seit13}. SNII  show  H-lines in their spectra and, in contrast, SNI  lack H in their ejecta.  Type II supernovae develop from stars with mass larger than  10$M_\odot$, burning  hydrogen in their core, and contracting when hydrogen is exhausted,  until densities and temperatures are so high that the 3$\alpha \rightarrow\ ^{12}$C reaction ignites. He burning follows, producing nuclear ashes which are also burned and an ensuing cycle of fuel exhaustion, contraction, followed by ignition of the  ashes from the previous burning cycle repeats, finally bringing the star to the ignite an explosive burning of $^{28}$Si up to Fe nuclei \cite{Jank12,Bur13}. This evolution is fast for heavy stars, with a 25M$_\odot$  star going through all of the cycles within 7 My, and  the final  Si explosion stage  only taking a few days.  The start will then posses a  ``{\it onion-like}" structure, as shown  in  Figure \ref{xburst}, right.  Starting from the center, there is an iron core, remnant of silicon burning, followed by successive thick layers of $^{28}$Si, $^{16}$O, $^{12}$C, $^{4}$He, and $^{1}$H. At the interfaces between these regions, nuclear burning continues. Silicon burning exhausts the nuclear fuel and  the heat from nuclear reactions cannot hold the  gravitational contraction of the iron core. A collapse occurs inevitably. But stars  with $M > 20 - 30M_\odot$ might not explode as a supernova and instead  collapse into black holes \cite{Maz08}. Supernova models together with recent observations of low metallicity stars have confirmed that the ejecta of SNII are characterized by elevated ratios of $\alpha$-elements (a convenient designation for the observation that some even-Z elements  such as O, Mg, Si, S, Ca, and Ti) relative to iron: [($\alpha$-elements)/Fe] $\simeq + 0.5$. Therefore, the contributions of SNIa to $^{56}$Fe  comprise $\simeq 2/3$ of the observed galactic iron. Since SNIa produce $\simeq 0.6 M_\odot$ per event while SNII produce only $\simeq 0.1 M_\odot$ per event, than SNII have been about 3 times more numerous when averaged over the galactic history.

The  core in a pre-supernova has densities and temperatures  around $2 \times 10^9$ g/cm$^{-3}$ and  $kT \sim$ 0.5 MeV, or $T \sim10^{10}$ K, respectively.  The core is composed of  $^{56}$Fe nuclei and electrons. The collapse of the core is accelerated by two factors.
The nuclei and photons are initially in thermal equilibrium, that is,
\begin{equation}
\gamma+\;_{26}^{56}\mathrm{Fe}\longleftrightarrow13(^{4}\mathrm{He}%
)+4\mathrm{n}-124\ \mathrm{MeV}.\label{nastro4}%
\end{equation}
For the densities and temperatures mentioned above, half of $^{56}$Fe is dissociated, taking energy away from the core and causes pressure loss. The collapse is therefore accelerated. But if the core mass  exceeds the Chandrasekhar mass $M_{Ch}\sim 1.5 M_\odot$, nuclei capture electrons so that the  Pauli principle is avoided, that is,
\begin{equation}
\mathrm{e}^{-}+(Z,A)\longrightarrow(Z-1,A)+\nu_{e}.\label{nastro5}%
\end{equation}
As the core collapse continues,  nuclear densities of the order of $\ \rho_{N}\approx10^{14}$ g/cm$^{3}$ are reached, and matter becomes a degenerate Fermi gas of nucleons. Because nuclear matter has a finite compressibility, the collapse in the inner part of the core decelerates and starts to increase again due to the increase of the nuclear matter. A  \textit{shock wave} is then developed which propagates to the external part of the star which, during collapse  continued to contract and reached supersonic velocities. An outward bound shock wave could lead to the supernova explosion, but crucial pieces of the physics are unknown, including a good equation of state of the nuclear matter under such diverse conditions \cite{Jan07}. The compressibility influences the intensity of the shock wave, approximately given by $10^{51}$ erg, needed to eject the mantle of the star.  But as the shock wave must have initially more energy because as travels outwards, it heats and melts the iron on its way, at a  loss of about 8 MeV/nucleon.  An enhancement in  electron capture rates occurs on the free protons left over by the shock wave. This together with  a sudden reduction of the neutrino opacity ($\sigma_{coherent} \sim N^2$), greatly enhances neutrino emission  (\textit{neutrino eruption}), leading to more energy loss, a process which can lead to the shock to stall.
The neutrinos will escape and take away energy from the core, it looses pressure and the collapse, is again accelerated. The chemical equilibrium condition for the reaction $e^- + p \rightarrow \nu_e + n$ is $ \mu_e + \mu_p = \mu_n + \langle E_\nu \rangle$, with $\mu_i$ being the chemical potential of particle $i$. But since the neutrinos escape and the electron Fermi energy increases as the density increases, the reaction above breaks equilibrium and leads to a rise of  neutronization of the matter.  The neutrinos also carry off energy and lepton number.  All these features lead to a very fast collapse, as attested by accurate numerical simulations as reported by several groups (see, e.g., Refs. \cite{RHN02,Gam03,GKO05,GN05,RN07,BG08,Mae10}) with first genuine 3D calculations being reported recently \cite{AMV14,Cou15,MJM15}. 

\begin{figure}[t]
\begin{center}
\includegraphics[
width=3.3in
]%
{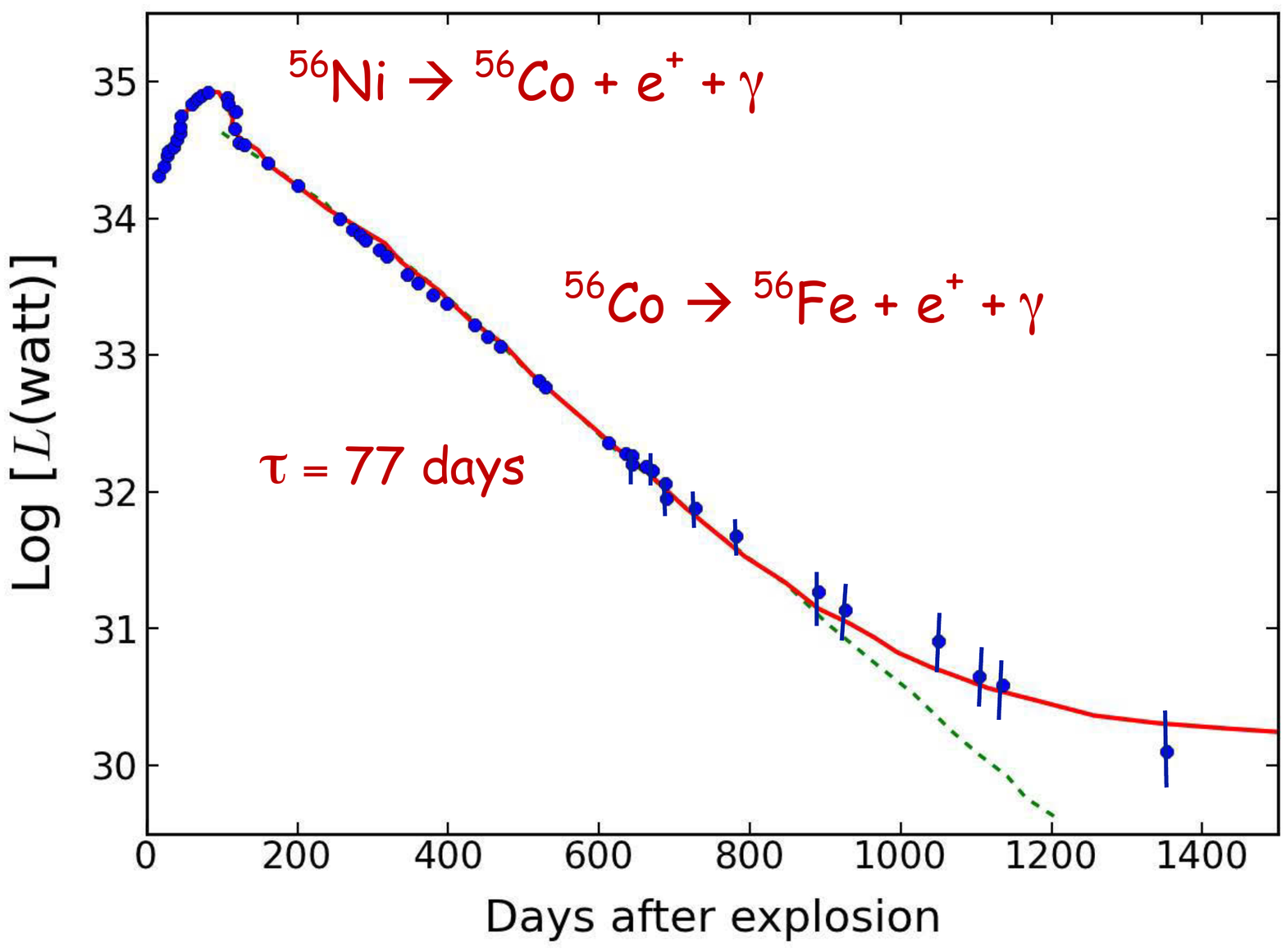}
\includegraphics[width=3.in]%
{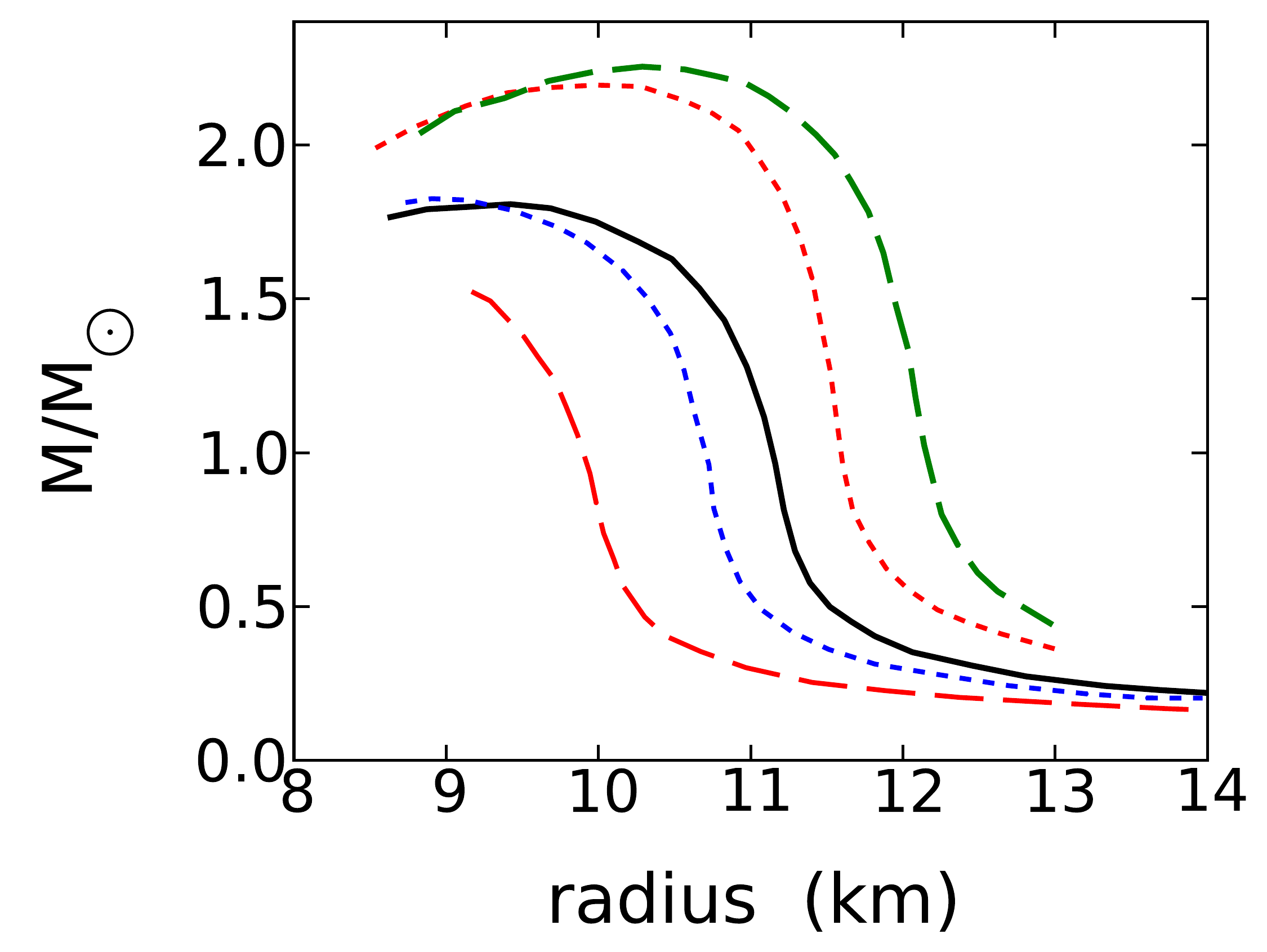}
\caption{\small {\it (Left):} Observed luminosity of supernova SN1987 A as a function of time (data from \cite{Sun92}). The dashed line is the exponential decay of $^{56}$Co. The solid curve is based on calculations. {\it (Right):} Nuclear masses  of neutron stars in units of solar masses for a few equations of state as a function of its radius in kilometers.}
\label{snlight}
\end{center}
\end{figure}

\subsection{Neutrinos}

For an iron core  of mass $\sim 1.2-1.5M_\odot$ models predict that the collapse happens at  about 0.6 of the free fall time \cite{Cla84} for matter under gravitation, given by $ t_{fall}=( {3/ 8\pi G \rho} )^{1/2}$. The iron core has a density of $\sim 10^8$ g cm$^{-3}$ at its edge  and  $\sim 3 \times 10^9$ g cm$^{-3}$ at its center when the collapse starts. Thus, $t_{fall}= 0.13 \ \rho_8^{-1/2} \ {\rm s}$, where $\rho_X$ is the density in $X$g.cm$^{-1/3}$. The neutrino mean free path for scattering on matter is $\lambda_\nu= 1/\left< n\sigma_\nu \right>$, and for a typical neutrino energy of $20$ MeV  it is only $\lambda_\nu \sim 0.5 \rho_{12}^{-1}$ km. The time it takes for the neutrinos to diffuse to a radial distance $R$ is given by $ t_{diff}= {R^2/ 3\lambda_\nu c}$. The neutrino scattering off matter occurs through both coherent neutral  current and charged current processes.  The neutral current neutrino  scattering cross section is determined by the total nuclear weak charge, proportional to $N^2$, where $N$ is the  neutron number. The neutrino random walks in the core, leading to its diffusion up to to the core surface. On its way, neutrinos can induce nucleosynthesis, which can be enhanced with neutrino oscillations \cite{Yos04,Yos06,Sas05,Sas05b}. For a core with mass $1.5 M_\odot$ and and a neutrino energy approximately equal to the approximate local Fermi energy,  $E_F \sim 35 \rho^{1/3}_{12}$ MeV, one finds $t_{diff} = 5 \times 10^{-2} \rho_{12} \ {\rm s}$. If we compare the diffusion time scale with the free fall time scale, we get ${t_{diff}/ t_{fall}} = 40  \ \rho_{12}^{3/2} \ {\rm s}$. Hence, for density larger than $\ 10^{11}$ g cm$^{-3}$ the neutrinos become fully confined in the core. More energy is released by the continued gravitational collapse and the star's still existing lepton number are trapped within the core. The fact that $t_{diff}$ much larger than $t_{fall}$  reinforces the relevance of the neutrino diffusion process.  Accurate calculations show that  neutrino diffusion occurs  on a time scale of  about 2 s. All three species of neutrinos are generated by pair production in the hot medium. A new shock wave can be reinstated by the outward push due to neutrino diffusion which carry most part of the gravitational energy liberated in the collapse of the core ($\approx10^{53}$ erg). For a new shock wave to develop one needs only  $\sim 1$\% of the energy contained the neutrinos, if this energy is converted into kinetic energy of nuclear matter by means of coherent neutrino-nucleus scattering. The stellar explosion ignited in this way is known as the \textit{retarded mechanism} for supernova explosion. The first shock wave stalls at a radius of $200-300$ km.   The shock wave revival probably occurs at about 0.5 s later. To know the exact  mechanism of a supernova explosion it is necessary to know the  electron capture rates, the nuclear compressibility, and the neutrino transport properties. The remnant of the explosion is an  iron core, and it will either become a neutron star, and maybe a \textit{pulsar} (rotating neutron star), or even a \textit{black-hole}, for more massive stars, with masses $M \geq 25-35M_{\odot}$.

The daughter nucleus in Eq. \eqref{nastro5} may beta decay and the original nucleus can be restored, with the creation of  an electron-neutrino and electron-antineutrino ($\bar\nu_e$) pair and the emission of  the neutrinos from the core. This process can repeat, draining energy from the core by the escaping neutrinos. The two-step process is given by
\begin{eqnarray}(N, Z ) + e^-  & \longrightarrow  \ \ (N+1, Z-1) +\nu_e\ \  \ \ \ {\rm (electron~capture)}\nonumber ,\\ 
(N+1, Z-1) & \longrightarrow  \ \ (N, Z)+e^-+\bar\nu_e \ \ \ \ \ \beta{\rm -decay).}
\end{eqnarray}
For a core of $M\sim 1.5M_\odot$ and a 10 km radius, an estimate of its binding energy is $ {G M^2 / 2R} \sim  10^{53} \ {\rm ergs}$, which is  roughly the trapped energy later radiated by the  neutrinos. The electron captures on nuclei occur at small momentum transfer, and they are dominated by Gamow-Teller (GT) transitions; i.e.  the response of nuclei to spin-isospin ($\sigma\tau$) operators, in which a proton is changed into a neutron. The electron chemical potential $\mu_e$  grows with density like $\rho^{1/3}$ and when it is is of the same order as the reaction $Q$-value, the electron capture rates are very sensitive to  the detailed GT distribution of the involved  nuclei. Experiments are very difficult to carry out in order to obtain such nuclear matrix elements to obtain the weak-interaction rates for $A \sim 50 - 65$, which are relevant at such densities \cite{Pin00,Heg03,Hix03,LM03}.  An alternative is to use charge-exchange reactions such as (p,n), ($^3$He,t) or heavy ions (Z,Z$\pm$1), performed in inverse kinematics \cite{Goo80,Tad87,Etch88,Bre88,Lens88,Lens89,Ost92,Ber93,Ste96,BL97,Dmit01,Bau05,Neg05,Fuj11}. 

Also worth mentioning are efforts to use charge-exchange reactions to access double-beta decay matrix elements \cite{AFR07,Cap15}.  As we discussed in Section \ref{solarneut}, the  matrix $U_{ ij}$ represents the PMNS matrix, when the standard three neutrino theory is considered. It reads \cite{Eid04}
\begin{eqnarray}
\begin{bmatrix}
U_{e 1} & U_{e 2} & U_{e 3} \\
U_{\mu 1} & U_{\mu 2} & U_{\mu 3} \\
U_{\tau 1} & U_{\tau 2} & U_{\tau 3}
\end{bmatrix} 
&=& \begin{bmatrix}
1 & 0 & 0 \\
0 & c_{23} & s_{23} \\
0 & -s_{23} & c_{23}
\end{bmatrix}
\begin{bmatrix}
c_{13} & 0 & s_{13} e^{-i\delta} \\
0 & 1 & 0 \\
-s_{13} e^{i\delta} & 0 & c_{13}
\end{bmatrix}
\begin{bmatrix}
c_{12} & s_{12} & 0 \\
-s_{12} & c_{12} & 0 \\
0 & 0 & 1
\end{bmatrix}
\begin{bmatrix}
e^{i\alpha_1 / 2} & 0 & 0 \\
0 & e^{i\alpha_2 / 2} & 0 \\
0 & 0 & 1 \\
\end{bmatrix} \\
&& {\rm Atmospheric~~~~~~~~~Reactor~~~~~~~~~~~~~Sun~~~~~double-beta~decay} \nonumber
\end{eqnarray}
where $c_{ij} = \cos \theta_{ij}$ and $s_{ij} = \sin \theta_{ij}$, and the last line gives which type of experiment that are sensitive to each sub-matrix. To a good approximation, the experimental data can be understood in terms of oscillations between just two neutrino flavors. Atmospheric $\nu_\mu$ neutrino data has determined the mixing angle $\theta_{23}$ while the solar neutrinos $\nu_e$ were used to determine $\theta_{12}$ \cite{nobel15}. The phase factors $\alpha_1$ and $\alpha_2$  are nonzero only if neutrinoless double beta decay occurs and if neutrinos are Majorana particles, i.e. if the neutrino is identical to its antineutrino. The phase factor $\delta$ is non-zero only if neutrino oscillations violate CP symmetry.  Double beta decay  can be calculated in the standard model as a second order process, $(A, Z) \longrightarrow (A, Z + 2) + 2e^- + 2\bar{\nu_e}$. Two neutrino double beta decay ($0\nu\beta\beta$) has been observed in about a dozen nuclei with lifetimes of the order of $ 10^{21}$ yr \cite{Ber12}.  In a number of even-even nuclei, $\beta$-decay is energetically forbidden, while double-beta decay is energetically allowed. Promising candidates are $0\nu\beta\beta$ decay  in $^{48}$Ca, $^{76}$Ge, $^{82}$Se, $^{96}$Zr, $^{100}$Mo, $^{116}$Cd, $^{128}$Te, $^{130}$Te, $^{136}$Xe, and $^{150}$Nd as $2\nu\beta\beta$ have been observed for these nuclei. If  $0\nu\beta\beta$ is observed, it would imply  lepton number violation by two units and additionally that neutrinos have a Majorana mass component. The process would probably be mediated by light and massive Majorana neutrinos. Due to its immense impact on nuclear, particle, and astrophysics, there has been a huge effort to study $0\nu\beta\beta$ decay both experimentally and theoretically \cite{Feld98,Klap04,Due11,Men11,Rod11,KI12,Rob13,Gan13,Ago13,Alb14}. 

\subsection{Supernova radioactivity and light-curve}

During a supernova explosion nucleosynthesis proceeds in the silicon core leading to several radioactive isotopes, including $^{56}$Ni, $^{57}$Ni and $^{44}$Ti, which  have  short half-lives \cite{Die10,Die13}.  The decays of these isotopes can  be directly observed. They are are characterized by either the half-life, $t_{1/2}$, or by the decay time $\tau = t_{1/2}/ \ln 2$. An example is $^{26}$Al, which has a 720 kyr lifetime to decay into $^{26}$Mg.  The decay from the $^{26}$Al 5$^+$ ground state goes to the first two excited states of $^{26}$Mg, both of them  2$^+$ states at 2.938 and 1.809 MeV above its ground state.  97\% of the time the later state is populated, and decays by emitting a 1.809 MeV $\gamma$. The 2.938 MeV state decays to the 1.809 MeV level, ant therefore a small quantity of 1.129 MeV $\gamma$'s are created.  The primary site for  $^{26}$Al production is believed to be Type II and IIb supernovae \cite{Die10}.  For $^{56}$Ni,  $\tau= 8.8$ days, due to the electron capture process $^{56}{\rm Ni}( e^-,\gamma)^{56}{\rm Co}$, with $\gamma$-rays energies of the order of 0.16 - 0.8 MeV. However, $^{56}$Co is not stable either and decays  according to $^{56}{\rm Co}(e^-, \gamma) ^{56}{\rm Fe}$ or $^{56}{\rm Co}\rightarrow \ ^{56}{\rm Fe} +e^+ +\nu_e$, with a branching ratio for the first decay is 80\%  and  20\% for the second, yielding  gamma-ray and positron sharing the energy emission by 96\% and 4\%, respectively. The most intense  gamma-ray lines occur at 0.847 MeV and 1.237 MeV and the mean positron energy is 0.66 MeV.  Recent observations are in agreement with the model of an explosion of a $1.5$ M$_\odot$ WD, providing a proof of the standard hypotheses of Type Ia supernovae explosion  \cite{Chu14}. $^{57}$Ni also decays by electron capture, $^{57}{\rm Ni}(e^-,\gamma)^{57}{\rm Co}$, within a very short time, $\tau= 52$ hours.  Another interesting decay is $^{57}{\rm Co}(e^-, \gamma)^{57}{\rm Fe}$, with $\tau= 390$ days. Produced in a nuclear statistical equilibrium at high densities of ($10^4 - 10^{10}$ g/cm$^3$) and temperatures $(4 - 10) \times 10^9$ K, $^{44}$Ti has a half-life of 59  yr. It decays first to $^{44}$Sc by means of $^{44}{\rm Ti}(e^-,\gamma)^{44}{\rm Sc}$,
followed by a very quick decay time, $\tau = 5.4$ h, to $^{44}{\rm Sc}(e^-,\gamma) ^{44}{\rm Ca}$ or  $^{44}{\rm Sc} \ \rightarrow \ ^{44}{\rm Ca} +e^++\nu_e $ \cite{Clay69,Gre12,WW95,Thie95,Rau02,Tur10}. 
All these radioactive decays yield either $\gamma$-rays or positrons. The $\gamma$-rays Compton-scatter off electrons, loosing about half of their energy to electrons in each collision. Initially, the  $\gamma$-ray energies are in the MeV range, much larger than the atomic electron binding energies, and thus both bound and free electrons contribute to the scattering. The $\gamma$-ray energies decrease until the  photoelectric absorption cross section is larger than the Compton cross section,  at  $E_\gamma \sim 10 - 100$ keV, mostly on iron. When the shock stops the radiation  leaks out on a diffusion time scale, $t_{diff}\simeq {3 R^2 \rho \kappa / \pi^2  c}$,
where $\kappa$ is now the opacity to radiation. The expansion time scale is  $t_{exp} \simeq R/v$ and the opacity due to  Thompson scattering is $\kappa = 0.4$ cm$^2$/g. The two time scales are comparable, $t_{diff}/t_{exp} \sim 1$ for a typical mass of 10 $M_\odot$,  when  the supernova has expanded to $R_{peak}\sim 4\times 10^{15}$ cm.  Only after $t_{peak} = R_{peak}/v \sim 40$ days, the radiation can leak out faster than the ejecta can expand. 

The first information indicating that a core-collapse supernova event has occurred is the burst of neutrinos. A few hours later  electromagnetic radiation is released initially as a ultra-violet flash. The expanding supernova only becomes visible at the optical wavelengths, with a rise and fall of the light curve which is a result  of an increasing surface area and a slow temperature decrease. The peak in the light curve appears when the temperature of the outer layers begin to decrease. A change in the opacity of the outer layer of the exploded star is created by the shock wave, which heats it up to about 100,000 K, leading to hydrogen ionization, which has a high opacity. The radiation from the  star's interior cannot escape, and we thus only observe photons from the surface of the star. The outer parts of the star only cool off after a few weeks to about $4,000-6,000$ K, at which point the ionized hydrogen start recombining into  neutral hydrogen. Being transparent at most wavelengths,  neutral hydrogen forces the opacity to change appreciably at the photosphere of the star.  Then photons from the hotter, inner parts of the hydrogen envelope start to escape. After full hydrogen recombination  in the hydrogen envelope, the light curves are dimmer and generated by a radioactive tail of the $^{56}$ Co to  $^{56}$Fe  conversions.  $^{56}$Ni decays by electron capture in 6.1 d, $^{56}$Co also by electron capture  in 77.3 d, but $^{56}$Fe is stable. Figure \ref{snlight} shows the light curve from the SN1887 \cite{Arn89}. A supernova. It could be beautifully related to 0.85 and 1.24 MeV $\gamma$-lines from $^{56}$Co decay. A lot ($\sim 0.1M_\odot$) of $^{56}$Ni and $^{56}$Fe were formed  in the first moments after the supernova explosion.  The data are also consistent with predictions of a total neutrino count of $10^{58}$ and total energy of $10^{46}$ J \cite{Par09}.  Recent observations of $^{44}$Ti gamma-ray emission lines from SN1987A  have been able to reveal an asymmetric supernovae explosion, open a new and challenging field for supernova explosion models \cite{Bog15}. This is expected because some neutron stars are observed with velocities up to 500 Km/s or greater. But the mechanism leading to this recoil velocity  during the supernovae explosion remains unknown.

\section{Nuclei and nuclear matter in compact stars}\label{nstars}

White dwarfs and neutron stars  are  supported by degenerate fermion gases. A WD star has consumed  most of its hydrogen and helium, leading to  carbon/oxygen  (C/O white dwarfs), neon  (C/O/Ne white dwarfs), silicon, and maybe iron.  Typical WDs have mass smaller than 1.4 $M_\odot$, although some of them are much smaller, e.g. $\sim M_\odot/100$.  The large majority of WDs have $\sim 0.6 M_\odot$ with very little dispersion about this mean. WD radii are of the order of $10^4$ km. Most WD are  a carbon and oxygen mixture in a proportion that is not well known.  A few massive WDs in nova systems contain heavy elements in their spectra, with neon-oxygen-magnesium novae being commonly observed. The surface layers of WDs  vary greatly.   Another common type of WDs are {\it red dwarfs} with $8\%-50\%$ of the solar mass, shining in red due to their low surface temperature of 2500 -- 4000 K. They live  longer than the present age of the Universe,  and are therefore very numerous. {\it Brown dwarfs} are WDs with mass below  $M=0.1M_\odot$. From galactic lensing observations there is evidence of  $0.5$ $ M_\odot$ massive objects  denoted by  {\it Massive Astrophysical Compact Halo Objects} (MACHO).  MACHOs could be white, brown, red dwarfs or also neutron stars, black holes,   or even planets. They might also be associated with dark matter \cite{Alc00}.

Electron pressure is not enough to support the gravitational attraction of stars with $M \sim 1.5 M_{\odot}$. During contraction, when the density increases to $2\times10^{14}$g cm$^{-3}$, matter starts to neutronize via electron capture in nuclei, changing protons into neutrons. A \textit{Neutron Star} (NS) is born with a small radius and very dense matter.   A cubic centimeter within a neutron star weighs 100 million tons.  NS are at the endpoint of stellar evolution and at the limit density that matter can have, with the next step being a black hole. A {\it pulsar} is a rapidly rotating NS, generating high energy radiation ``pulses" due to a misalignment of its rotation axis and its magnetic poles.   We have identified thousands of pulsars and have learned that they are excellent laboratories, because: (a) their density is as large as those of an atomic nucleus; (b) their mass and size generate very strong gravitational fields, second only to those of the black holes, but easier to measure; (c) they can rotate as fast as 700 turns per second implying a rotation of their surface by 36,000 km/s; (d) NS have a larger magnetic field than any other object in the Universe, a million of times stronger than any magnetic field that can be produced in the laboratory; (e) their pulsations are more regular than the precision of some of our best atomic clocks.

\subsection{The structure of neutron stars}

Neutron stars probably consist  of a massive dense core surrounded by a thin crust with a mass $\lesssim 0.01M_\odot$ and a thickness of $\lesssim 1$ km \cite{LP01}. The interface between crust and core occurs at the density  $\approx \rho/2$, where $\rho_0=2.8\times 10^{14}$ g cm$^{-3}$ is the density of saturated nuclear matter. The neutron star crust consists of neutron-rich nuclei, strongly degenerate electrons and, when $\rho \gtrsim 4 \times 10^{11}$ g cm$^{-3}$, of free neutrons dripped from nuclei. The NS core  composition and Equation Of State (EOS) are model dependent, with some constraints set by observations. The EOS are usually separated in components for the outer ($\rho\lesssim 2\rho_0$) and for the inner core ($\rho\gtrsim 2 \rho_0$). The outer core is thought to be composed of a mixture of neutrons, protons and electrons, and appears in all NS's, while the inner core is found  in massive and more compact NS's. The EOS and composition of the inner core is poorly known. It might contain hyperons,  pion or kaon condensates, or quark matter, or even a mixture of these components \cite{Lat12}. 

Pulsars can rotate very fast  because of the conservation of angular momentum at the time of their creation.  Absorption of orbiting matter from a companion star further increases its rotation, which can reach many hundred times per second and lead them to a more oblate form. After this initial stage they slow down as their rotating magnetic fields radiate energy away.  And as they  slow down, their shape becomes more spherical.  The rate with which they slow down is practically constant and very small, only about $-(d\omega/dt)/\omega= 10^{-10} - 10^{-21}$ s/rotation. Often a NS spins up or suffers a {\it glitch}, which is a sudden small increase in its rotation. Glitches are thought to be related to a {\it starquake} when  their stiff crust rupture. Following the starquake, a smaller equatorial radius results, and the rotation increases due to angular momentum conservation. Glitches might also originate from transitions of vortices in the superfluid core from one metastable to a lower energy state  \cite{AI76}.

Due to their large gravitational field, the equations of hydrostatic equilibrium need to be modified to account for the NS gravitational balance and general relativity corrections. It is known as the {\it Tolman--Oppenheimer--Volkoff} (TOV) equation, appropriate for a spherically symmetric body in static gravitational equilibrium, 
\begin{equation}\frac{dp(r)}{dr}=-\frac{G}{r^2}\left[\rho(r)+\frac{p(r)}{c^2}\right]\left[m(r)+4\pi r^3  \frac{p(r)}{c^2}\right]\left[1-\frac{2Gm(r)}{c^2r}\right]^{-1}.
\label{stars:tov}
\end{equation}
where $m(r)$ is the total mass within radius $r$.   The second factors within each square brackets are due to special relativity corrections of order $1/c^2$.  When supplemented by an equation of state (EOS), which links pressure and density, the TOV equation yields the structure of a compact stellar object with isotropic matter in equilibrium.  Neglecting the terms of order $1/c^2$ results in the Newtonian hydrostatic equation \eqref{st:dPdr}. The relativistic correction factors have all a positive effect, meaning that as gravity becomes stronger at a position $r$, relativity has the effect of strengthening the gravity at that point.

The density of matter at the center of a NS is larger than that of an atomic nucleus. For such densities the EOS is still  poorly understood but we believe that we can set an upper limit on the masses of NS in the range $1.6M_\odot \lesssim M_{max} \lesssim 3M_\odot$. A more massive neutron star develops an extremely large central pressure and cannot be supported  against gravitational implosion. The reason is that Eq. \eqref{stars:tov} is quadratic in the pressure at very high pressures and if a star develops a too high central pressure, it will be unable to support itself against the humongous gravitational pull. One integrates the TOV with the boundary conditions $M(0)=0$, $\rho(R)=0$ and $p(R)=0$, where $R$ is the star radius.  Then  $p$, $\rho$  and $m$  can be computed for increasing values of $r$  until one arrives at $p=0$, i.e.  the star boundary.  Model calculations can be represented, e.g., the star mass $M$ versus the central density $\rho_c$  or  versus the radius $R$. We show in Figure \ref{fig:eos}, right, several solutions for $M/M_\odot$ as a function of star radius.  We observe a general feature: the occurrence of a maximum mass for a given value of the central density ($\rho_c \sim 10^{15}$ g/cm$^3$) and a radius of about 10 km. For a non-interacting neutron gas, the maximum mass occurs between 0.7 $M_\odot$ and  3 $M_\odot$, with a stiff EOS \cite{BP12}.

\subsection{The equation of state for neutron stars}

Neutron stars are mostly made of neutrons with a small fraction of electrons and protons. Weak decays transform a neutron into a proton and an electron, ${\rm n} \rightarrow {\rm p} + e^- + \bar\nu_e$, with  the energy $m_n - m_p - m_e$ = 0.778 MeV radiated away by the electron and the neutrino\footnote{Here we use $\hbar=c=1$.}. The Pauli principle prevents the decay when low-energy levels for the  proton are already occupied.  A similar situation  occurs for the electrons which are present in the star to balance the positive charge of the protons. Equal numbers of electrons and protons means that $k_{F,p} = k_{F,e} $. The weak interaction must be in equilibrium and  as many neutrons decay as electrons get captured by means of $p + e^- \rightarrow n + \nu_e$. This equilibrium is expressed in terms of chemical potentials so that $	\mu_n = \mu_p + \mu_e $. The chemical potential for particle $i$ is given by $ \mu_i(k_{F,i}) = \partial E_i/\partial n_i  = (k_{F,i}^2 + m_i^2)^{1/2}$ and the chemical equilibrium for $ i = {\rm n}, {\rm p}$, e,  means that
\begin{equation}(k_{F,n}^2 + m_n^2)^{1/2} - (k_{F,p}^2 + m_p^2)^{1/2} -(k_{F,p}^2 + m_e^2)^{1/2} = 0\, .  \label{eq:kFpconstraint}
\end{equation}
For the NS densities,  $m_e \ll k_{F,n}$, Eq. \eqref{eq:kFpconstraint} yields
\begin{equation}	k_{F,p}(k_{F,n}) 	 \simeq \frac{k_{F,n}^2 + m_n^2 - m_p^2}		{2 (k_{F,n}^2 + m_n^2)^{1/2}}  \, .
\end{equation}

The pressure and energy density within a NS is the sum of those of protons, neutrons and electrons. The liquid drop mass formula for nuclei with $Z$ protons and $N$ neutrons yields for {\it symmetric nuclear matter} ($Z=N$) an equilibrium number  density of $n_0= 0.16$ nucleons/fm$^3$. The corresponding Fermi momentum is $k_{F} = 263$ MeV/$c$ which is small compared with $m_N = 939$ MeV/c$^2$ and justifies a non-relativistic treatment of usual nuclear matter. At the saturation density $n_0$, the mean binding energy per nucleon is $BE =  - 16$ MeV.  The {\it nuclear compressibility}, $K$,   defined by
\begin{equation}
K=\left. k_F^2{\partial(E/A)\over \partial k_F^2}\right|_{k_{F0}}=\left. 9\rho^{2}\frac{\partial^{2} (E/A)}{\partial\rho^{2}}\right|_{\rho_0} . \label{Kcomp}%
\end{equation} 
is a useful quantity for ``measuring" nuclear pressure. This is a quantity which is not so well established but it is probably in the range of 200 to 300 MeV. Another quantity called by {\it symmetry energy}  (discussed below) contributes about 30 MeV of energy (when $Z=0$) above the symmetric matter minimum, occurring at $n_0$ \cite{SR04}.

For {\it symmetric nuclear matter}, $n_n = n_p$ and the total nucleon density $n = n_p + n_n = 2n_n$. The average energy per nucleon is $E(n)/A = \epsilon(n)/n$, including the rest mass, $m_N$. $E(n)/A - m_N$ has a minimum at $n = n_0$ with a value $BE = -16$ MeV, obtained with
\begin{equation}
\frac{d}{d n}\left(\frac{E(n)}{A}\right) = \frac{d}{d n}\left(\frac{\epsilon(n)}{n}\right)= 0  {\ \rm\ at\ } n = n_0 \, . \label{eq:n0constraint}
\end{equation}
The curvature at $n=n_0$ is  
\begin{equation}
K(n) = 9 \frac{d p(n)}{d n} = 9 \left[ n^2 \frac{d^2}{d n^2} \left( \frac{\epsilon}{n} \right)+ 2n \frac{d}{d n} \left( \frac{\epsilon}{n} \right) \right]
\, , \label{eq:Kconstraint}
\end{equation}
where $p(n) = n^2 {d} \left({\epsilon}/{n} \right) /{d n}$, defining the pressure in terms of  the energy density. When $n = n_0$,  $K(n_0)=K$, as given in Eq. \eqref{Kcomp}.

\begin{figure}[t]
\begin{center}
\includegraphics[width=4.in]{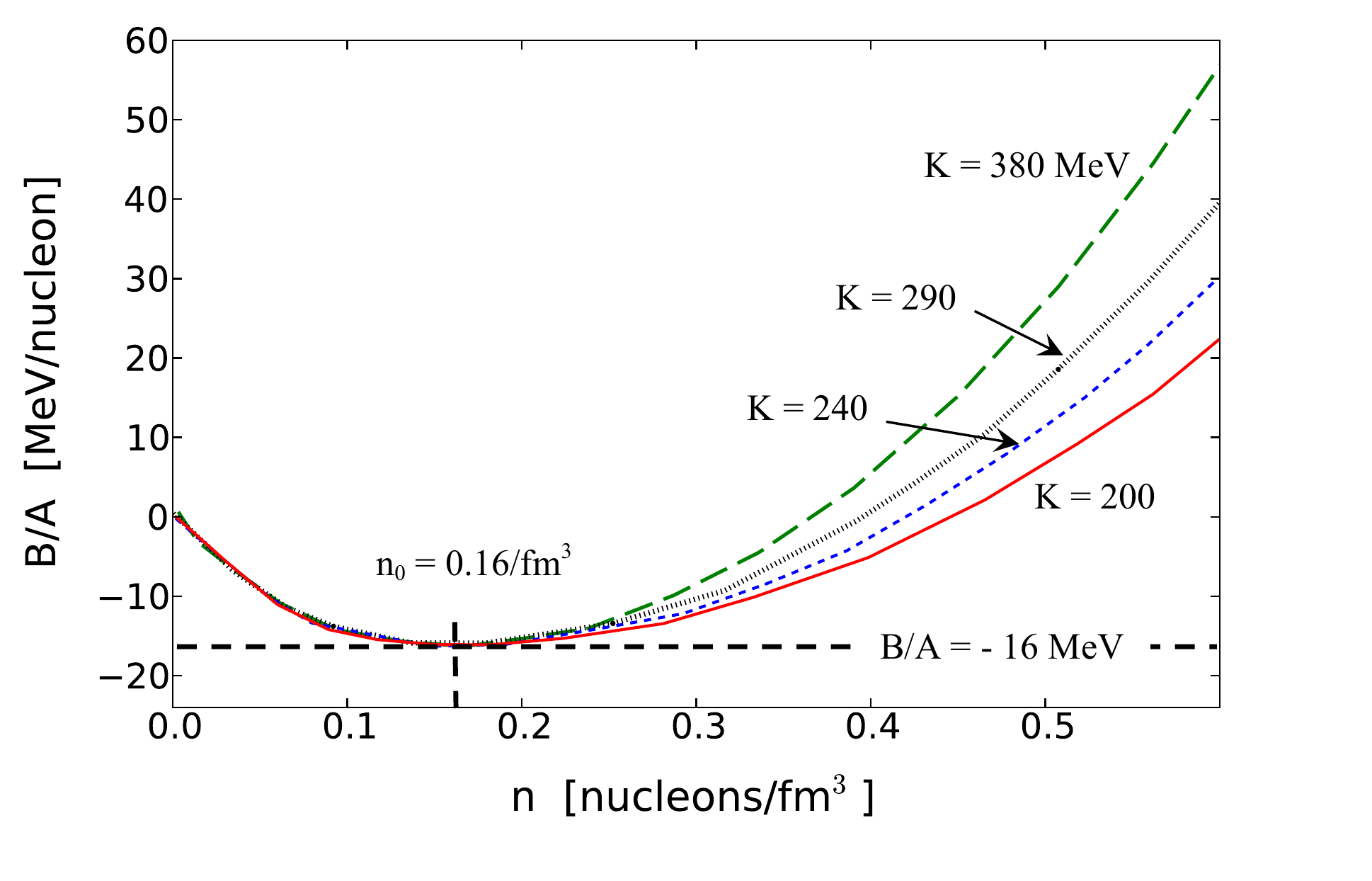}
\includegraphics[width=2.4in]{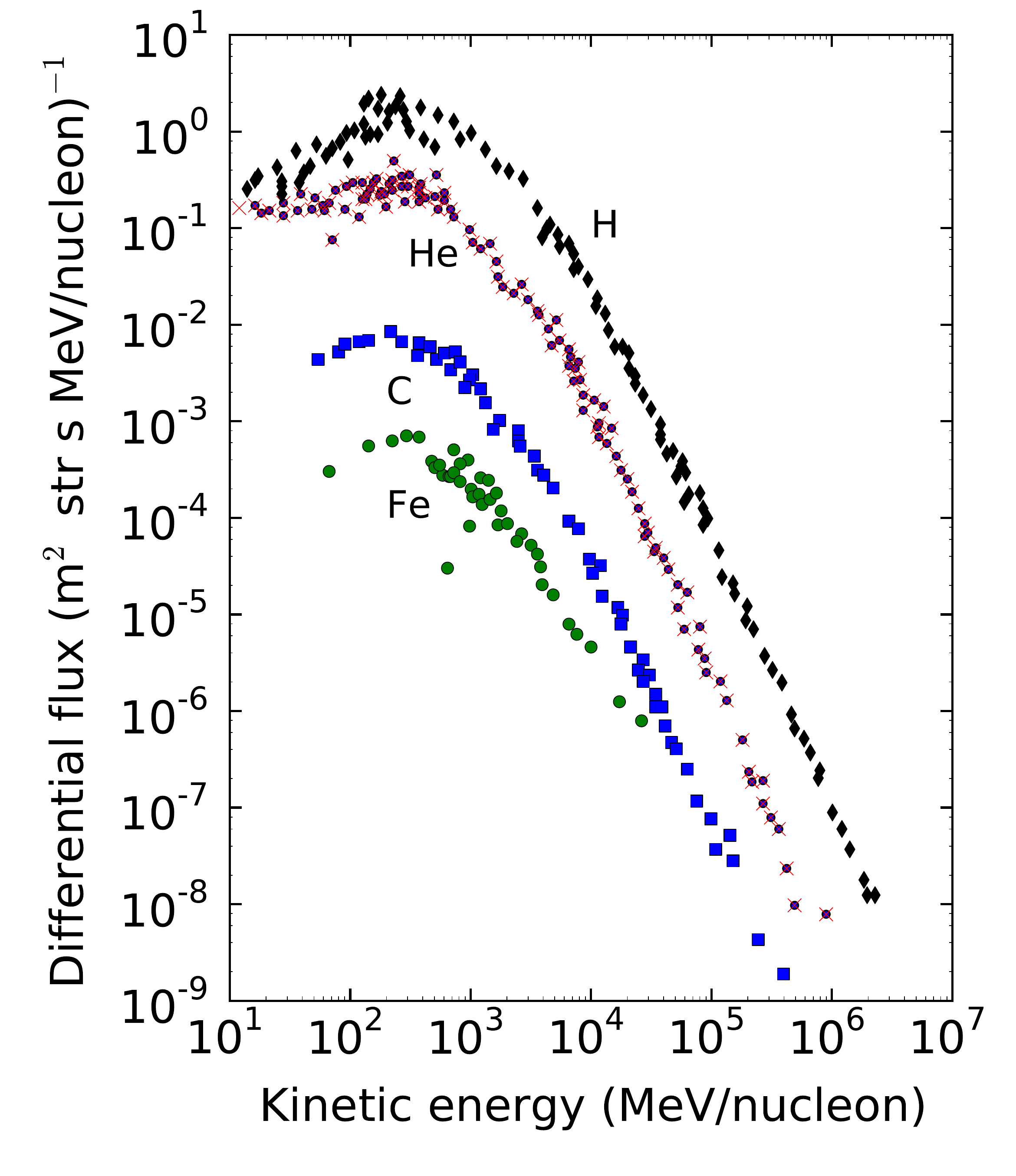}
\caption{\small {\it (Left):} Binding energy per nucleon dependence on the nuclear matter density. The numerical solutions for a few different compressibilities are also shown. {\it (Right):} Main elements of the composition of cosmic rays (data collected from Ref. \cite{Yao06}).}
\label{fig:eos}
\end{center}
\end{figure}

For symmetric nuclear matter, a simple model for the energy density is given by \cite{Pra96}
\begin{equation}
\frac{\epsilon(n)}{n} = m_N + \frac{3}{5} \frac{\hbar^2 k_F^2}{2 m_N} + \frac{A}{2} u + \frac{B}{\sigma + 1} u^\sigma
\, , \label{eq:epsbyn}
\end{equation}
where $u = n/n_0$, and $\sigma$ and $A$ and $B$ constants.   The  second term is the average kinetic energy per nucleon, $\left<E_{F0}\right>$. Using  Eqs.\ (\ref{eq:n0constraint})--(\ref{eq:Kconstraint}), and noting
that $u = 1$ at $n = n_0$, we can solve these equations for the parameters $A$, $B$, and $\sigma$. For a given value of $K$ and and mean value of the kinetic energy, the energy density and pressure are completely determined:
\begin{equation}
p(n) = n^2 \frac{d}{d n} \left( \frac{\epsilon}{n} \right) = n_0 \left[ \frac{2}{3}\left<E_{F0}\right> u^{5/3} +  \frac{A}{2} u^2 +  \frac{B\sigma}{\sigma+1} u^{\sigma+1} \right]\, . \label{eq:pressofnNeqZ0}
\end{equation} 
Several other possible parameterizations of the nuclear matter density are described in the literature \cite{BP12} and a procedure similar to that given above can be used to obtain the EOS. Figure \ref{fig:eos} shows the dependence of the binding energy per nucleon on the nuclear matter density for different EOS commonly found in the literature. The position of the minimum is at $n = n_0 = 0.16$ fm$^{-3}$, with  $BE = -16$ MeV, and a second derivative (curvature) there corresponding to the nuclear compressibility $K$. 

For {\it non-symmetric nuclear matter} one can  represent the proton and neutron densities in  terms of a parameter $\alpha$ so that \cite{Pra96,SR04}
$n_n = {(1 + \alpha)}/2n$, $n_p = {(1 - \alpha)}/2n$  where, for pure neutron matter, $\alpha = 1$.
This definition implies that $\alpha = {(n_n - n_p)}/{n} = {(N-Z)}/{A}$. It is also usual to define the proton fraction in the star as $ x = {n_p}/{n} = {(1 - \alpha)}/{2}$. In the case of symmetric nuclear matter, $\alpha = 0$ (or $x = 1/2$). The contribution of neutrons and protons to the kinetic energy part of $\epsilon$ are
\begin{eqnarray}
\epsilon_{KE}(n,\alpha) = \frac{3}{5} \frac{k_{F,n}^2}{2 m_N}\, n_n + \frac{3}{5} \frac{k_{F,p}^2}{2 m_N}\, n_p  
= n \left<E_F\right> \frac{1}{2} \left[ \left(1+\alpha\right)^{5/3} + \left(1-\alpha\right)^{5/3} \right] \,  ,
\end{eqnarray}
where
\begin{equation}
\left<E_F\right> = \frac{3}{5} \frac{\hbar^2}{2 m_N}\left( \frac{3 \pi^2 n}{2} \right)^{2/3}
\,  \label{eq:EFdef}
\end{equation}
is the average kinetic energy of symmetric nuclear matter at density $n$. When $n=n_0$ one obtains $\left<E_F\right> = 3 \left<E_{F0}\right>/5$
[see Eq.\ (\ref{eq:epsbyn})]. For non-symmetric matter, $\alpha \neq 0$,  and the excess kinetic energy is
\begin{eqnarray}
\Delta\epsilon_{KE}(n,\alpha) &=& \epsilon_{KE}(n,\alpha) - \epsilon_{KE}(n,0) = n \left<E_F\right> \left\{ \frac{1}{2} \left[ 
\left(1+\alpha\right)^{5/3} + \left(1-\alpha\right)^{5/3} \right]- 1 \right\} \nonumber \\
&=& n \left<E_F\right> \left\{ 2^{2/3} \left[ \left(1-x\right)^{5/3} + x^{5/3} \right]- 1 \right\}\, . \label{eq:delepsKE}
\end{eqnarray}
For pure neutron matter, $\alpha = 1$, $\Delta\epsilon_{KE}(n,\alpha) = n \left<E_F\right>\left( 2^{2/3} - 1 \right)$.
To leading order in $\alpha$,
\begin{eqnarray}
\Delta\epsilon_{KE}(n,\alpha) = n \left<E_F\right> \frac{5}{9} \alpha^2 \left( 1 + \frac{\alpha^2}{27} + \cdots \right) =  n \, E_F \, \frac{\alpha^2}{3}\, \left( 1 + \frac{\alpha^2}{27} + \cdots \right)\, .
\end{eqnarray}
We can also assume a quadratic approximation in $\alpha$, so that
\begin{equation}
E(n,\alpha) = E(n,0) + \alpha^2 S(n) \, . \label{eq:Enalphadef}
\end{equation}
The isospin-symmetry breaking is proportional to $\alpha^2$. The function $S(u)$, $u = n/n_0$, only depends on the density of the symmetric nuclear matter.  From the energy density, $\epsilon(n,\alpha) = n_0 u E(n,\alpha)$, we obtain the corresponding pressure,
\begin{equation}
p(n,x) = u \frac{d}{du}\epsilon(n,\alpha) - \epsilon(n,\alpha) = p(n,0) + n_0 \alpha^2 S(n)  \, , \label{eq:presnpdef}
\end{equation}
where $p(n,0)$ is defined by Eq.\ (\ref{eq:pressofnNeqZ0}).
Around $x=0$ 
\begin{equation}
S={1\over 2} \left. {\partial^2 E\over \partial \alpha^2}\right|_{\alpha=0}=J +Lx+{1\over 2} K_{sym}x^2 +\cdots
\end{equation} 
where $J$ is the bulk symmetry energy, $L$ determines the slope, and $K_{sym}$ the curvature of the symmetry energy at saturation, $\rho=\rho_0$. Heavy ion collisions and  has been  a method of choice to study the effects of symmetry energy in nuclei and in nuclear matter (see, e.g., Refs. \cite{BaLi00,BaLi02,Chen05}).

With these EOS  for symmetric or non-symmetric one can  solve the TOV equation for a  NS. Appreciable differences of the solutions for symmetric and asymmetric EOS have been found \cite{Hae07}. There is no experiment on Earth that can reproduce the matter at the high densities and low temperatures occurring inside a NS \cite{Lat12,BP12}. Theorists help in this problem with standard mean field calculations in order to obtain the value $K$ from the in-medium nucleon-nucleon interactions. Microscopic theories such as Hartree-Fock, Relativistic Mean Field Theories, Random Phase Approximation, or other density functional theories,  are used for the purpose \cite{Ben03,Ber09,GC14,Du14,Cha15,Nik11,AB13,Mas15,Zha15,Lian15,Meng06a,Meng06,Lian08,Zhou03,Meng96,Meng98,Zhao11}.    But although some  interactions reproduce many known nuclear data, their predictions  deviate considerably as  the nuclear density  increases \cite{Brow00}. As an example of such studies, we refer to giant monopole resonances which are  compressible modes of the nucleus and  as such should contain information on $K$.  Theoretical studies using the interactions which best fit the experimental data on the excitation of such states have been used to extract the value of $K$ (see, e.g., Ref \cite{AB13}). Another example is the excitation of pigmy resonances in reactions with unstable nuclei which can be used to constrain the symmetry energy dependence $S$ of the EOS \cite{Bru68,Li08}. The determination of neutron skins in nuclei has also been shown to be related to $S$ \cite{TB01,Furn02,Ber07a}. It is worthwhile mentioning that the excitation of giant and pigmy resonances during supernovae explosions have also an impact on predictions for the abundance of heavy elements in the r-process \cite{Go98}.   The experimental study of the nuclear dipole polarizability, related to the photoabsorption cross section  by  $\alpha_D= {\hbar c/ 2\pi^2e^2} \int d\omega {\sigma_{abs}(\omega) / \omega^2} $,  is also another alternative method to obtain the symmetry energy \cite{BHR03,RN10,Piek11,Tam11,Pol12,Krum15}.

\subsection{Quark phase EOS}
 
The matter in the NS core has densities varying between a few times $n_0$ to one order of magnitude higher. Thus, a more detailed knowledge of the EOS is needed for densities $n\gg n_0$ where a description of matter  in terms of leptons and nucleons is not sufficient. For $n\gg n_0$, other particles, such as hyperons and $\Delta$ isobars, may become important and meson condensations may take place. For extremely high densities, nuclear matter should pass by a transition to a quark--gluon phase \cite{Wit84,Bay85,Gle90}.  The cold quark--gluon phase within a NS should not be unique, and other quark--gluon phase could possibly exist at extremely high temperatures and baryonic densities. They probably occurred during the first $10^{-5}$ s  after the Big Bang. The maximum mass of a NS is modified if  the phase transition from hadronic matter to quark matter is accounted for.

A simple EOS including the deconfined quark phase uses the MIT bag model \cite{Cho74}. In this model, the energy density is given as a sum of a non-perturbative energy shift $B$, known as the {\it bag constant}, and the kinetic energy of non-interacting quarks of  mass $m_f$, flavor $f$, and Fermi momentum $k^{(f)}_F =(3\pi^2 n_f)^{1/3}$, where $n_f$ is the  the quark density, i.e.,
 \begin{eqnarray}
\epsilon = B+ \sum_f {3 m_f^4\over 8\pi^2} \left[x_f\sqrt{x_f^2+1}\left(2x_f^2+1\right)^2-\sinh^{-1} x_f\right], \label{st:bag}
\end{eqnarray}
where $x_f=k^{(f)}_F/m_f$. 
Typically, one only  considers  massless $u$ and $d$ quarks, while the mass of the s quark mass is taken as $\sim 150$ MeV. $B$ is interpreted as the difference between the perturbative vacuum energy density  and the energy of the physical vacuum \cite{Cho74}. The inclusion of a perturbative interaction between the quarks amounts to additional terms in the number and energy densities. In the original MIT bag model one uses $B \sim 50$ MeV.fm$^{-3}$, which is  small compared to $\sim 200$ MeV fm$^{-3}$, the value predicted by lattice calculations \cite{Lep00}.

The energy density and  EOS corresponding to Eq. \eqref{st:bag}  leads to stable solutions for compact stars, with small and very dense stars, composed of quark matter. But a model based on Eq.  \eqref{st:bag} is oversimplified. Many other models with mixed phases (hybrid stars) have been developed including quarks, hyperons, quark nuggets,  strange quarks, and so on.  Strangeness reduces the Pauli repulsion because it increases the flavor degeneracy and yields  a smaller charge-to-baryon ratio for strange quark matter compared to ordinary nuclear matter.  The electron chemical potential, $\mu_e$ is appreciably smaller than the quark chemical potential $\mu_q$, and an EOS can be deduced by a power expansion in  $\mu_e/\mu_q$ \cite{JRS06}, leading to
\begin{equation}
p_{\rm QM}=p_0(\mu_q,m_s)-n_q(\mu_q,m_s)\mu_e + {1\over 2} \chi_q(\mu_q,m_s)
\mu_e^2+\ldots
\label{generic_EoS}
\end{equation}
where $p_0$, $n_q$, and $\chi_q$ are calculable functions of $\mu_q$ and of the strange quark mass, $m_s$. The second-order expansion above neglects the electron pressure $p_e\sim\mu_e^4$ and may be used  to model a QCD based EOS for NS. The mass--radius relation for NS built from a mixed or a  pure quark phases varies wildly \cite{JRS06}. 

The bottomline is that NS are rich objects, containing exotic large nuclei or pasta-like nuclei in their crust, superconducting fluids in their interior, strong magnetic fields, and many other unique characteristics which make them a dream object for a theorist's imagination. Observations only set a few constraints such as mass, size, glitches, which are nonetheless difficult to reproduce with standard nuclear physics. But so are many data on nuclear reactions presently available from laboratory experiments on Earth.   A very large number of pulsars ($>2500$) have been identified in  the Galaxy \cite{Man05,Abdo13,Car14}, about 90\% (at least for radio pulsars) of which are single stars and about 250 of them are binary pulsars.  There are now firm observations of neutron stars with masses of about $2M_\odot$ \cite{Ant13}. This sets strict constraints on the EOS for NS. Only a few EOS can reproduce the mass radius relation as displayed in Figure \ref{snlight} and achieve 2 solar masses at small radii.  Those EOS would have   higher pressures at densities above $4 \rho_0$.  A recent review has discussed in details the implications of $2M_\odot$ NS for nuclear physics and astrophysics \cite{LP16}. 

\section{Cosmic rays}\label{cosmicr}

The Earth is bombarded by a nearly constant flux of charged particles originated from elsewhere the cosmos.  They have either galactic or extragalactic origin, with an intensity of the primary cosmic rays of approximately 2 -- 4 particles/s/cm$^2$. It consists basically of nuclei (98\%) and  electrons (2\%). There is a predominance of protons (87\% by mass) and $\alpha$-particles (12\% by mass) and the remaining 1\% includes other nuclei, with a sharply decreasing abundance as their masses increase. The energy distribution and chemical composition of primary cosmic rays is displayed in  Figure \ref{fig:eos}, right. The distribution is nearly independent of energy over the dominant energy range from 10 MeV to several GeV. As seen in the figure,  the main component  are protons, with about 10\% of helium and a much smaller admixture of heavier elements. The chemical distribution of the elements in our solar system differs substantially from that of the cosmic rays with  the most dramatic feature being an enormous enrichment  of the elements Li/Be/B in the cosmic rays.  When compared to  with C, N and O their solar abundances amounts to $\sim 10^{-6}$, whereas  in cosmic rays  this ratio is close to $-0.2$ \cite{Ree70}.

They are richer in even Z elements compared to odd Z and relatively enriched in heavy elements relative to H and He.  It is not displayed in the figure, but numerous elements heavier than iron have been observed with  abundances equal or smaller than $\sim 10^{-5}$. E.g., the C, N and O abundances amounts to $\sim 10^{-6}$, Much of this information was obtained with satellite and spacecraft measurements in recent years. The abundances of even Z nuclei with $30 < Z < 60$ are in good agreement with the corresponding abundances in the solar system, while the abundances for $62 < Z < 80$,  in the platinum-lead region,  are larger compared to that of the solar system by nearly a factor of two. This might be due to contributions from the r-process elements in this mass region. This would be expected because r-processes occur in core-collapse supernovae, which could be a primary source for the acceleration mechanism of low energy cosmic rays and add to the cosmic ray composition in solar abundances.  Nuclei such as  Li, Be and B, have a small percent in cosmic rays, but are several orders of magnitude larger than the their abundance in the whole Universe. That is because these elements are not products of nucleosynthesis in stars, but part of the secondary cosmic ray radiation. In fact, the abundance of several isotopes  pertaining other than the elements Li/Be/B in cosmic rays are different from the universal average. For example, the {\it isotopic ratio} $^{3} $He$/^{4}$He is about 200 times larger in cosmic rays than found elsewhere in the cosmos. There is an increasing observational evidence that the cosmic rays origins are linked to supernova remnants \cite{Acc09,Tav10}, but this hypothesis is far from being settled \cite{Dru12,Bla13,BSY14,Par14,Ahl15}.

\section{Summary and Conclusions}
In the last 80 years we have witnessed an enormous progress in the understanding on how elements are formed in the Universe. The progress in this knowledge is now growing faster due to advances in technologies such as the use of satellites as a tool for astronomical observations. Theoretical simulations are also much better now with advances in computers both in speed, storage, and new algorithms.  There are still many problems that remain and that we have discovered when we  increased our knowledge progressively. The more we dig into these issues the more we learn, but also new questions arise that we did not know before. We have just learned about the  lithium problem in Big Bang nucleosynthesis, we discovered that neutrinos are more complicated than initially thought, but then they also became a powerful tool to unravel many more details of the cosmic evolution. We just started learning about neutrino induced nucleosynthesis. We cannot really reproduce the observed abundance of elements in our Universe. But  we have pretty good ideas on how it has evolved.

Some theories beyond the standard model have been and will continued being tested with robust models that have been developed for nucleosynthesis in diverse scenarios.  But we still need to know how core-collapse supernovae explode, although we have nearly exhausted our imagination on the problem. The solution of such problems could lie in the small details of the physics that we already know but that is hard to implement theoretically or tested experimentally. We are not convinced of what is the site of the r-process and how neutron star mergers may contribute to it. A deeper understanding of nuclear structure and nuclear reactions  is key  to answer these questions and many other discussed in this review. Only recently we have witnessed the correlations in reaction networks involving nuclei far from the stability. New nuclear physics facilities and major efforts in nuclear theory will be crucial to connect the microscopic dynamics of nuclear systems and the big questions in cosmology and stellar physics. Some reactions seem to be beyond direct measurement with present technologies and some nuclear physics problems are heavily based on theory. Being one of the hardest problems in all science, the nuclear physics part of this endeavor will require many more decades of dedicated work.   

\section*{Acknowledgements} 

This work was supported in part by the U.S. DOE grants DE-FG02-08ER41533, the U.S. NSF Grant No. 1415656, and by  Grants-in-Aid for Scientific Research of JSPS, Grants No. 20105004, No. 24340060.

\bibliographystyle{elsart-num}

\end{document}